\documentclass[12pt]{article}
\usepackage{graphicx}
\usepackage{color}

\def\hybrid{\topmargin 0pt      \oddsidemargin 0pt
        \headheight 0pt \headsep 0pt
       \voffset-1cm
        \textwidth 6.25in       
       \textheight 9.5in       
        \marginparwidth 0.0in
        \parskip 5pt plus 1pt   \jot = 1.5ex}
\catcode`\@=11
\def\marginnote#1{}

\newcount\hour
\newcount\minute
\newtoks\amorpm
\hour=\time\divide\hour by60
\minute=\time{\multiply\hour by60 \global\advance\minute by-\hour}
\edef\standardtime{{\ifnum\hour<12 \global\amorpm={am}%
        \else\global\amorpm={pm}\advance\hour by-12 \fi
        \ifnum\hour=0 \hour=12 \fi
        \number\hour:\ifnum\minute<10 0\fi\number\minute\the\amorpm}}
\edef\militarytime{\number\hour:\ifnum\minute<10 0\fi\number\minute}

\def\draftlabel#1{{\@bsphack\if@filesw {\let\thepage\relax
   \xdef\@gtempa{\write\@auxout{\string
      \newlabel{#1}{{\@currentlabel}{\thepage}}}}}\@gtempa
   \if@nobreak \ifvmode\nobreak\fi\fi\fi\@esphack}
        \gdef\@eqnlabel{#1}}
\def\@eqnlabel{}
\def\@vacuum{}
\def\draftmarginnote#1{\marginpar{\raggedright\scriptsize\tt#1}}

\def\draftlabel#1{{\@bsphack\if@filesw {\let\thepage\relax
   \xdef\@gtempa{\write\@auxout{\string
      \newlabel{#1}{{\@currentlabel}{\thepage}}}}}\@gtempa
   \if@nobreak \ifvmode\nobreak\fi\fi\fi\@esphack}
        \gdef\@eqnlabel{#1}}
\def\@eqnlabel{}
\def\@vacuum{}
\def\draftmarginnote#1{\marginpar{\raggedright\scriptsize\tt#1}}

\def\draft{\oddsidemargin -.5truein
        \def\@oddfoot{\sl preliminary draft \hfil
        \rm\thepage\hfil\sl\today\quad\militarytime}
        \let\@evenfoot\@oddfoot \overfullrule 3pt
        \let\label=\draftlabel
        \let\marginnote=\draftmarginnote
   \def\@eqnnum{(\theequation)\rlap{\kern\marginparsep\tt\@eqnlabel}%
\global\let\@eqnlabel\@vacuum}  }

\def\numberbysection{\@addtoreset{equation}{section}
        \def\theequation{\thesection.\arabic{equation}}}

\def\underline#1{\relax\ifmmode\@@underline#1\else
        $\@@underline{\hbox{#1}}$\relax\fi}

\def\titlepage{\@restonecolfalse\if@twocolumn\@restonecoltrue\onecolumn
     \else \newpage \fi \thispagestyle{empty}\c@page\z@
        \def\thefootnote{\fnsymbol{footnote}} }

\def\endtitlepage{\if@restonecol\twocolumn \else  \fi
        \def\thefootnote{\arabic{footnote}}
        \setcounter{footnote}{0}}  
\relax

\numberbysection
\hybrid

\newfont{\Bbb}{msbm10 scaled 1\@ptsize00}
\newcommand{\CC}{\mbox{\Bbb C}}

\newcommand{\ZZ}{\mbox{\Bbb Z}}
\newfont{\Bbbb}{msbm7 scaled 1\@ptsize00}
\newcommand{\z}{\raise-1pt\hbox{$\mbox{\Bbbb Z}$}}

\def\beq{\begin{equation}}
\def\eeq{\end{equation}}
\def\p{\partial}

\def\lbr{\left <}
\def\rbr{\right >}

\def\lvac{\left <0\right |}
\def\rvac{\left |0\right >}

\def\lvacn{\left <n\right |}
\def\rvacn{\left |n\right >}

\def\psistar{\psi^{*}}

\def\normord{ {\scriptstyle {{\bullet}\atop{\bullet}}} }
\def\normordbare{ {\scriptstyle {{ \times}\atop{ \times}}} }
\def\normordboson{ {\scriptstyle {{*}\atop{*}}} }

\begin{document}

\begin{titlepage}

\title{Lectures on nonlinear integrable equations \\and their solutions}

\author{
 A.~Zabrodin\thanks{National Research University Higher School of Economics,
20 Myasnitskaya Ulitsa, Moscow 101000, Russian Federation;
ITEP, 25 B.Cheremushkinskaya, Moscow 117218, Russian Federation;
Skolkovo Institute of Science and Technology, 143026 Moscow, Russian Federation;
e-mail: zabrodin@itep.ru}
}

\date{December 2018}
\maketitle

\vspace{-7cm} \centerline{ \hfill ITEP-TH-38/18}\vspace{7cm}

\begin{abstract}

This is an introductory course on nonlinear integrable partial differential 
and differential-difference equ\-a\-ti\-ons
based on lectures given for students of Moscow Institute of Physics and Technology
and Higher School of Economics. The typical examples of Korteweg-de Vries (KdV),
Kadomtsev-Petviashvili (KP) and Toda lattice 
equations are studied in detail. We give a detailed description
of the Lax representation of these equations and their hierarchies in terms of 
pseudo-differential or pseudo-difference 
operators and also of different classes of the solutions including 
famous soliton solutions. The formulation in terms of tau-function and Hirota bilinear 
differential and difference equations is also discussed. Finally, we give a representation
of tau-functions as vacuum expectation values of certain operators composed of free
fermions. 

\end{abstract}

\end{titlepage}

\vspace{5mm}

\tableofcontents

\newpage

\section{Introductory remarks}

Before sixties of XX century 
the list of problems of classical and quantum physics that admit
exact solution in one or another form
was very short and included just a few examples (some problems on tops
in classical mechanics, Ising model, Heisenberg model)
They seemed to be exotic exceptions. 

In the sixties-seventies the situation drastically changed:  
large families of non-trivial exactly solvable models were discovered in that time and
the basic principles underlying their construction and integrability were understood. 
Herewith it has appeared that many of them unexpectedly emerge in very different 
physical contexts and are directly related to the structure of our world.

\subsubsection*{What types of integrable systems are known}

There are several important types:
\begin{itemize}
\item[--]
Models with a small number of degrees of freedom (integrable cases of tops).
\item[--]
Systems of $N$ interacting particles in one dimension:
the Calogero-Moser model and all its relatives (classical and quantum).
\item[--]
Nonlinear partial differential equations as well as difference equations:
the Kor\-te\-weg-\-de Vries equation (KdV), the nonlinear Schrodinger equation (NLS),
the sine-Gordon equation (SG), the Toda chain and the 2D Toda lattice, the Benjamin-Ono
equation (BO), the Kadomtsev-Petviashvili equation (KP)
and many others.
\item[--]
Models of statistical mechanics on 2D lattice: Ising model, six- and eight-vertex models
and their generalizations.
\item[--]
Integrable models of quantum physics on 1D lattice:
spin chains of XXX, XXZ and XYZ type and their various
generalizations. 
\item[--]
Integrable models of quantum field theory in 
$1+1$ dimensions:
one-dimensional bose-gas with point-like interaction (quantum NLS equation), 
the Thirring model, the sine-Gordon model, ...
\end{itemize}

\noindent
This division is rather conditional and the list is not complete. 
There are deep and beautiful 
connections between the different types of integrable models. For example, 
poles of singular solutions to integrable partial differential equations
move as particles of integrable many-body systems of the Calogero-Moser type. 
Another example is that 
the connection between classical and quantum systems is not
exhausted by their cor\-res\-pon\-den\-ce in the classical limit. It appears that
classical integrable systems emerge in quantum integrable problems even at $\hbar \neq 0$.

\subsubsection*{In what sense one should understand integrability
of a system}

Here are several possible meanings of integrability:
\begin{itemize}
\item[--]
A possibility to integrate the equation (i.e. eliminate
all derivatives).
\item[--]
Existence of a complete set of integrals of motion in involution
(the Liouville integrability).
\item[--]
Presence of a large set of exact solutions and a possibility to express
the answers through known elementary or special functions.
\item[--]
A possibility to reduce the problem to a solution of a finite system of
algebraic or integral equations.
\item[--]
Presence of rich symmetries and interesting algebraic or analytic structures.
\end{itemize}

\noindent
The different types of integrable systems mentioned above provide examples 
to all these meanings of the notion of integrability. 

\subsubsection*{Nonlinear integrable partial differential equations: some more
introductory words}

In these lectures we will be concerned with 
classical integrable models described by nonlinear partial differential equations.
In the first approximation, our world is described by {\it linear equations},
since a response to a small perturbation is usually proportional to the perturbation.
The main content of the standard mathematical physics courses is essentially
the theory of three linear partial differential equations:
the Laplace equation, the heat equation and the wave equation.
Their fundamental role is explained by their exceptional universality.

However, the nonlinear corrections, though small, can lead to significant
outcomes. During last few decades it was understood that
\begin{itemize}
\item[a)]
A correct account of nonlinear effects leads to 
{\it integrable} equations which have the same high degree of
universality as equations of linear mathematical physics;
\item[b)]
These nonlinear equations possess rich hidden symmetries which allow one to
effectively construct large classes of exact solutions
(for example, solutions of soliton type).
\end{itemize}

Each of the nonlinear integrable equations generates an infinite chain 
of compatible ``higher'' equations, {\it the hierarchy}. Besides, all known 
integrable equations (KdV, NLS, SG, Toda chain, KP and many others)
are close relatives. From a general point of view all of them are contained
(as limiting and particular cases, reductions, equivalent forms obtained by 
a change of variables) in 
{\it one} universal equation -- {\it the bilinear difference Hirota
equation}.

\subsubsection*{The aim of these lectures}

These notes are
based on lectures given for students of Moscow Institute of Physics and Technology (MIPT)
and Higher School of Economics (HSE) in 2009-2017. The lectures 
were intended for an audience unfamiliar with the subject and were aimed at an initial 
introduction to it. There are a number of
exercises and problems in the text which are an essential part 
of the course. Exercises are very simple; problems are 
somewhat more complicated.

We start from the KdV equation (section 2) 
which is the most familiar and well-studied example of 
integrable nonlinear equation. After acquaintance with the Lax representation of the KdV 
equation we study its (abelian) 
symmetries and introduce the whole infinite hierarchy of compatible
``higher'' KdV equations. A crucial role in the constructions is played by the common 
solution of the auxiliary linear problems, the Baker-Akhiezer function or $\psi$-function. 
Using properties of the $\psi$-function, the family of soliton solutions is constructed.
We also discuss the non-abelian symmetries of the KdV equation, their stationary 
points and their relation to 
the Painl\'eve equations. 

Section 3 is devoted to the more general infinite hierarchy -- the KP hierarchy. 
We introduce its Lax representation (in terms of pseudo-differential operators) and
zero-curvature (Zakharov-Shabat) representation (in terms of differential operators)
as well as the dressing operator formalism. Again, the role of the $\psi$-function is crucial.
We construct the solutions of the soliton type and discuss other types of solutions. 
Towards the end of this section, we approach the concept of tau-function, the general 
solution to the whole KP hierarchy. We prove the bilinear identity, the formulas 
for the $\psi$-function in terms of tau-function and the bilinear 
generating equation for the KP hierarchy (the Hirota-Miwa equation). 

In section 4 we briefly discuss even more general hierarchy then the KP one -- 
the 2D Toda lattice (2DTL) hierarchy. The commutation representation for it is
constructed with the help of difference and pseudo-difference operators. 
The main objects we introduce here are similar to the case of the KP hierarchy:
the Lax operators, the dressing operators, the $\psi$-function, the tau-function.

In section 5 we present an approach to nonlinear integrable equations from the 
point of view of quantum field theory. Namely, a remarkable fact is that the tau-function
can be realized as vacuum expectation value of certain quantum field operators composed
of free fermions. The equations of the KP hierarchy 
obeyed by tau-function are then basically consequences 
of the Wick's theorem. In the formalism of free fermions, the construction of tau-functions
for the KP, modified KP and 2D Toda hierarchies becomes simple and natural. 

We should note that some standard material is out of the scope of the lectures. In particular, 
we do not even mention the inverse scattering method. Our approach is mostly algebraic.

\subsubsection*{Comments on the literature}

The literature on the subject is enormous. The list of references 
\cite{BBT}--\cite{UT84} includes 
only books and papers that had the greatest impact
on the content of these notes and/or were recommended to students for an additional and 
further reading. We do not cite any literature in the main text. 

\newpage

\section{The KdV equation}

The KdV equation was suggested in 
1895 for description of waves on shallow water. 
Propagation of waves in nonlinear media with dispersion 
in the case of general position is also described by the KdV equation. 

The motivation is as follows.  The wave equation
$u_{tt}=c_0^2 u_{xx}$ has general solution in the form of superposition of waves
propagating to the left and to the right with velocity $c_0$:
$u(x,t)= f(x-c_0t)+g(x+c_0t)$. Let us consider the left-moving wave; it satisfies the
first order equation $u_t =c_0u_x$. Nonlinear effects make the wave velocity 
dependent on the amplitude. In the first order this dependence is linear:
$c_0(u)=c_0 +\alpha u +\ldots$. Therefore, the nonlinearity yields the additional term
$\alpha uu_x$ while dispersion yields the term with third order derivative:
$$
u_t =c_0u_x \,\, \quad \longrightarrow \quad
u_t =c_0u_x +\alpha uu_x + \beta u_{xxx}
$$
(in the presence of the second order derivative
the dynamics becomes dissipative). The coefficients in front of the correction terms
may be small but if 
$u$ is large and/or changes rapidly, these terms become significant. 
In the frame moving with velocity $c_0$ we get the KdV equation
\beq\label{kdv1}
u_t =\alpha uu_x + \beta u_{xxx}.
\eeq

In the limiting case $\alpha =0$ (the linear approximation)
the equation is solved by the Fourier transform.
At $\beta =0$ (the dispersionless approximation) we get the Hopf equation
$u_t =\alpha uu_x$ which (as any equation of the form
$u_t =V(u)u_x$) can be solved by the method of characteristics. 
The general solution is written in an implicit form
$x +\alpha u t = f(u)$ with arbitrary function $f$.
Remarkably, in the general case
$\alpha \neq 0$, $\beta \neq 0$
equation (\ref{kdv1}) can be also integrated but using very different methods. 

Rescaling the variables $x, t, u$, one can get rid of the coefficients
$\alpha$, $\beta$. For reasons which will be more clear later, 
we fix them as follows: $\alpha =3/2$, $\beta =1/4$, so the KdV equation acquires the form
\beq\label{kdv2}
\mbox{\fbox{$\displaystyle{\phantom{\int ^{A}_{B}}
4u_t =6uu_x + u_{xxx}
\phantom{\int ^{A}_{B}}}$}}
\eeq
which we call canonical.

\subsection{An example of exact solution: the traveling wave (one-soliton solution)}

One can try to find solutions to the KdV equation in the form of a traveling wave:
$u(x, t)=f(x+ct)$ with $c>0$. Substituting this ansatz to the equation (\ref{kdv2}), we get 
the ordinary differential equation
$4cf'=6ff'+f'''$ in which one derivative can be eliminated: 
$$
4cf=3f^2 +f'' +C_1.
$$
Multiplying this equation by $f'$, we see that one more derivative can be eliminated
with the result
$$
f'^2=4cf^2-2f^3-2C_1f -C_2,
$$
where $C_1, C_2$ are constants. This first order ordinary differential equation can be 
in\-teg\-ra\-ted:
$$
x-x_0= \int^f \frac{dy}{\sqrt{4cy^2-2y^3-2C_1y -C_2}}.
$$
In the general case $C_1 \neq 0$, $C_2\neq 0$ it is an elliptic integral. It simplifies 
if we assume that the function $f$ with all its derivatives 
tends to zero as $x\to \pm \infty$, then $C_1=C_2=0$ and
the integral can be taken in elementary functions. The result is
$$
2\sqrt{c}\, (x-x_0)=\log \left (\frac{2\sqrt{c}-\sqrt{4c-2f}}{2\sqrt{c}+\sqrt{4c-2f}}
\right )
$$
or 
$$
f(x)=\frac{2c}{\cosh ^2(\sqrt{c}\, (x-x_0))}.
$$
Therefore, the solution to the KdV equation is of the form
\beq\label{kdv3}
u(x, t)=\frac{2c}{\cosh ^2(\sqrt{c}\, (x-x_0+ct))},
\eeq
where $c$ is an arbitrary positive real parameter.
This is the famous one-soliton solution. Note that the soliton excitation propagates 
with velocity which is greater than the velocity of sound (because the solution 
(\ref{kdv3}) is already written in the frame moving with velocity of sound).
The velocity of the soliton is 
pro\-por\-ti\-onal to its amplitude. 

We stress that the existence of traveling 
wave exact solutions is a common feature of all evolution equations of the form
$u_t=F[u, u_x, \ldots ]$ and the possibility to find one-soliton solution is by no means 
a characteristic property of the KdV equation. 
The KdV equation is really distinguished by the property that it has {\it multi-soliton}
solutions. 

\noindent
{\bf Problem.} Find traveling wave solutions of the modified KdV equation 
$4v_t=-6v^2v_x +v_{xxx}$. 

\noindent
{\bf Problem.} Find solutions of the form $\varphi (x,t)=f(x-ct)$ to the sine-Gordon
equation 
$$
\varphi_{tt}-\varphi_{xx}+\frac{m^2}{\beta}\, \sin (\beta \varphi )=0
$$
such that $f(\infty )-f(-\infty )=2\pi / \beta$.

\subsection{The Lax representation}

\paragraph{Proposition.} {\it The KdV equation (\ref{kdv2}) is equivalent to the 
operator relation
\beq\label{Lax1}
\p_t L =[A, \, L],
\eeq
where
$L=\p^2 + u$, $\,\,\, A=\p^3 +\frac{3}{2}\, u \p +\frac{3}{4}\, u_x$,
$\,\,\, \p :=\p /\p x$.}

\vspace{0.3cm}

\noindent
The proof is a direct computation.
Equation (\ref{Lax1}) is called the Lax equation or Lax representation 
(for KdV) while $L$ is called the Lax operator. The Lax equation
can be also written in the form $[\p_t -A, \, L]=0$. 
Note that $L$ is a Hermitean operator
($L^{\dag}=L$) while $A$
is antihermitean ($A^{\dag}=-A$), where the conjugation is defined as
$\p ^{\dag}=-\p$,
$f^{\dag}=f$, $(AB)^{\dag}=B^{\dag}A^{\dag}$.

\noindent
{\bf Exercise.} Prove that the Lax equation for $L$ implies the Lax equation for
$L^n$:
$\p_t L^n =[A, \, L^n]$ for any positive integer $n$.

The Lax equation means that
$
L(t)=U(t)L(0)U^{-1}(t)
$,
where $U$ is some operator with the property $A=\p_t U \, U^{-1}$.
Hence we have the important corollary:
$$
\mbox{{\it Spectrum of $L$ does not depend on time}}
$$
i.e., it is an integral of motion. Accordingly, 
$\det (\lambda \! -\! L)$, $\,\, \mbox{tr}\,
(\lambda - L)^{-1}$ are time-independent. Expanding these quantities in powers of
$\lambda$, one in principle can construct an infinite set of integrals of motion. 
Practically, there are two problems in this way:
a) the meaning of $\det$ and $\mbox{tr}$ for operators in functional space
should be made more precise, b) it is not clear how to 
find local integrals of motion, i.e. integrals such that their densities 
in any point depend only on values of the function $u$ and its derivatives 
with respect to $x$ at this point. 

\noindent
{\bf Remark.} The Lax representation is not unique. For example, the Lax equation
(\ref{Lax1}) is equivalent to the KdV equation in the form
$2u_t +3uu_x +u_{xxx}=0$ for the operators $L=\p^4 +2u\p^2 +u_x \p$,
$A=\p^3 +\frac{3}{2}\, u\p$.

\subsection{Symmetries and conservation laws}

By symmetry of a differential equation 
$\p_t u=K[u]$ we understand an equation
$\p_s u=R[u]$ such that evoltuions in $t$ and $s$
commute: $\p_s K[u] = \p_t R[u]$. (Here $K[u]$ and $R[u]$ are
differential polynomials of $u$, i.e. polynomials of $u$ and its $x$-derivatives.)
This means that any solution $u(x,t)$ 
of the first equation can be extended to a function
$u(x,t,s)$ in such a way that at any fixed $s$ it is a solution to the first 
equation (and at any fixed $t$ it is a solution of the second equation), and
$u(x,t,0)=u(x,t)$. 

If the coefficients of the differential polynomial
$K[u]$ do not depend on $x$ and $t$, the equation
$\p_t u=K[u]$ always has two trivial symmetries: 
shifts of the variables
$x$ and $t$. 

\noindent
{\bf Exercise.} Represent these trivial symmetries
in the differential form 
$\p_s u=R[u]$.

\noindent
{\bf Problem.} Find any non-trivial symmetry 
of the equation 
$\p_t u=uu_x$.

It turns out that the KdV equation has infinitely many non-trivial symmetries. 
They can be found using the technique of pseudo-differential operators. 

\subsubsection{Pseudo-differential operators}

A pseudo-differential operator is a series of the form
$\displaystyle{
\sum_{k=0}^{\infty} v_k \p^{N -k}},
$
where $v_k$ are functions and the operator $\p$ has the following standard
commutation rule with arbitrary function:
$\p f = f' + f \p$. Multiplying both sides of this equality 
by $\p^{-1}$ from the left and from the right, one can understand it as a rule 
of commutation of $\p^{-1}$ with a function: $\p^{-1} f=f\p^{-1} -\p^{-1} f' \p^{-1}$. 
The multiple application of this rule yields: 
$$
\p^{-1} f=f\p^{-1} - f' \p^{-2} + f''\p^{-3} + \ldots
$$
Pseudo-differential operators can be multiplied as Laurent series taking into account
that the symbol $\p$ does not commute with coefficient functions.
For example,
$$
(1+f \p^{-1})(1+g \p^{-1})=1+(f+g)\p^{-1}+ fg \p^{-2}-
fg'\p^{-3}+fg''\p^{-4} +\ldots
$$

\noindent
{\bf Remark.} For brevity we write $\p f$ for composition of the operator of multiplication
by the function $f$ and the differential operator
$\p$. We hope that this will not lead to 
a confusion. The composition is usually written as
$\p \circ f$ but pedantic use of this notation, in our opinion, makes it difficult to read 
formulas. 

\noindent
{\bf Problem.} For any two functions $f, g$ prove the following identities
in the algebra of pseudo-differential operators:
\begin{itemize}
\item[a)]\hspace{4mm}
$(\p -g)^{-1}f =\displaystyle{
\sum _{n=0}^{\infty}(-1)^n f^{(n)}(\p -g)^{-n-1}}$,
\item[b)] \hspace{4mm} $e^{-f}\p^{-1}e^f =
(\p +f')^{-1}$,
\item[c)] \hspace{4mm}
$\displaystyle{
\p^n f = \sum_{k=0}^{n}
\left ( \begin{array}{c}n\\ k\end{array}\right )
f^{(k)} \p^{n-k}\,, \quad n\geq 0}
$,
\item[d)] \hspace{4mm}
$\displaystyle{
\p^{-n} f = \sum_{k\geq 0}(-1)^k
\left ( \begin{array}{c}k\! +\! n\! -\! 1\\ k\end{array}\right )
f^{(k)} \p^{-n-k}\,, \quad n> 0}
$.
\end{itemize}

\noindent
Here $\displaystyle{\left ( \begin{array}{c}n\\ k\end{array}\right )=
\frac{n!}{k! (n-k)!}}$ is the binomial coefficient. Note that formulas
c) and d) can be unified by extending the definition 
of the binomial coefficient to arbitrary complex numbers $n$ as
$$
\displaystyle{\left ( \begin{array}{c}n \\ k
\end{array}\right )=\frac{n(n-1)(n-2)\ldots (n-k+1)}{1\cdot 2 
\cdot 3\cdot\ldots \cdot k}},
$$
then for integer $n<0$ $\displaystyle{
\left ( \begin{array}{c}n\\ k\end{array}\right )=
(-1)^k
\left ( \begin{array}{c}k\! -\! n\! -\! 1\\ k\end{array}\right )}$.
For the proof of c) and d) one may use induction in $n$. 

Given a pseudo-differential operator $P=\displaystyle{
\sum_{k=0}^{\infty} v_k \p^{N -k}}$, we call $N$ the order of it. 
Let
$P_{+}$ be its differential part (i.e. sum of the terms with non-negative
powers of $\p$: $P_+=\displaystyle{
\sum_{k=0}^{N} v_k \p^{N -k}}$), then $P_- =P-P_{+}$ is sum 
of the terms with negative powers.

Let us introduce the important notion of residue of the pseudo-differential operator
$P=\displaystyle{
\sum_{k=0}^{\infty} v_k \p^{N -k}}$: 
$$\mbox{res}\, P =v_{-1}.$$
We note that any operator $P$ can be written putting powers of the symbol
$\p$ both from the right and from the left of coefficient functions:
$$
P=\sum_k v_k \p^k = \sum_k \p^k \tilde v_k.
$$

\noindent
{\bf Exercise.}
Prove that the residue does not depend on the way of writing, i.e.
$\mbox{res} \, P=v_{-1}=\tilde v_{-1}$.
(For other coefficient functions this is in general not the case.) 

\noindent
In what follows we will need an important property of the residue:
for any two pseudo-differential operators
the residue of their commutator is a full derivative.

\noindent
{\bf Lemma.} {\it For any two pseudo-differential operators
$P, Q$
$$
\mbox{res}\, ([P, Q])=\p C,
$$
where $C$ is a differential polynomial
of coefficients of the operators  
$P, Q$, i.e. a linear combination of these coefficients and their $x$-derivatives
of any order.}

\noindent
It is enough to check this statement in the case
$P=f\p^n$, $Q=g\p^{-m}$, where 
$n>m>0$. Using the results of problems c) and d), one can conclude that the residue of the 
commutator is proportional to
$$
fg^{(n-m+1)}+(-1)^{n-m}g f^{(n-m+1)}= \p \left (
\sum_{k=0}^{n-m}(-1)^k f^{(k)}g^{(n-m-k)}\right ),
$$
i.e., it is a full derivative. It is also possible to give an inductive proof
which does not use the results of the exercises: 
assuming that the residue of the commutator
$f\p^n$ и $g\p^{-m}$ is a full derivative for all $n$ and some $m$
(for example, this is easy to check at
$m=1$), to derive from this that the same statement holds for 
$m \to m+1$. 

\noindent
{\bf Problem.} Give a detailed proof of the lemma.

The operation of conjugation (defined as $\p ^{\dag}=-\p$,
$f^{\dag}=f$, $(AB)^{\dag}=B^{\dag}A^{\dag}$) can be extended
to pseudo-differential operators:
\beq\label{conj}
\left (\sum_{k=0}^{\infty} v_k \p^{N -k}\right )^{\dag}=
\sum_{k=0}^{\infty}(-\p )^{N -k}v_k.
\eeq
The following technical lemma, which connects the notions of the operator and usual residue,
turns out to be useful in many cases.

\noindent
{\bf Lemma.} {\it Let $P=\sum_j p_j \p^j$ and $Q=\sum_j q_j \p^j$ be two
pseudo-differential operators, then
\beq\label{bi2}
\mbox{res}_{\p} (PQ^{\dag})
=\mbox{res}_z \left [ (Pe^{xz})\, (Qe^{-xz})\right ],
\eeq 
where by ${\rm res}_{\p}$ and ${\rm res}_{z}$ we denote the operator and usual 
residue (the coefficient of the Laurent series at $z^{-1}$) 
respectively. (It is implied that the rule of acting by 
the operator $\p^{n}$ to the exponential function
$\p^{n}e^{xz}=z^n e^{xz}$ is extended also to negative values of
$n$, i.e., for example, $\p^{-1}e^{xz}=z^{-1} e^{xz}$.)}

\noindent
Indeed,
$$
\mbox{res}_z \left [ (Pe^{xz})\, (Qe^{-xz})\right ]=
\mbox{res}_z \Bigl ( \sum_i p_i z^i \sum_j q_i (-z)^j \Bigr )
=\sum_{i+j=-1}(-1)^j p_i q_j,
$$
and the same expression is obtained after writing what is $\mbox{res}_{\p} (PQ^{\dag})$:
$$
\mbox{res}_{\p} (PQ^{\dag}) =\mbox{res}_{\p}\Bigl (
\sum_{i,j}p_i \p^i (-\p )^j q_j \Bigr )
=\sum_{i+j=-1}(-1)^j p_i q_j.
$$

The pseudo-differential operators allow one to take the square root of
$L=\p^2 +u$:
\beq\label{bi3}
(\p^2 +u)^{1/2}=\p + \frac{u}{2}\, \p^{-1}-
\frac{u_x}{4}\, \p^{-2} + \frac{u_{xx}\! -\! u^2}{8}\,
\p^{-3} \, - \frac{u_{xxx}\! -\! 6uu_x}{16}\, \p^{-4}\, 
+\ldots
\eeq
Note that if the symbol 
$\p$ commutes with $u$
(for example, if $u=\mbox{const}$), then derivatives in the right hand side vanish
and one obtains the usual Laurent series for the function $\sqrt{\p^2 +u}$. 
One can also define any half-integer powers of the operator
$L$: $L^{n/2}= (L^{1/2})^n$. It is easy to see that all of them
commute with $L$.

\noindent
{\bf Problem} (unsolved). Prove that the ``even'' coefficients $u_{2k}$, $k\geq 1$
in the expansion $\displaystyle{(\p^2 +u)^{1/2}=\p + \sum_{m\geq 1}u_k \p^{-m}}$
are full derivatives (see (\ref{bi3})).

\subsubsection{The KdV hierarchy}

It is easy to check that
$
A = (L^{3/2})_{+}$, 
$\frac{3}{2}uu_x +\frac{1}{4}u_{xxx}=
2\p \, \mbox{res}\,L^{3/2}
$, so the Lax equation can be written as
$\p_t L =[(L^{3/2})_{+}, \, L]$, and the KdV equation acquires the form
$\p_t u = 2\p \, \mbox{res}\,L^{3/2}$. In these terms, all its symmetries are written
in a unified way: instead of power $3/2$ it is enough to take arbitrary
other half-integer power of the operator
$L$.

\paragraph{Proposition.} {\it The Lax equations
\beq\label{Laxk}
\p_{t_{k}}L=[A_{k}, \, L]\,, \quad \quad
A_{k}=(L^{k/2})_{+}
\eeq
for any odd $k\geq 1$ generate the equations
\beq\label{resk}
\p_{t_{k}}u=2\p \, \mbox{{\rm res}}L^{k/2},
\eeq
which are symmetries of the KdV equation.
}

\noindent
The set of equations (\ref{resk}) is called the KdV hierarchy.
The variables $t_k$ are called times. 
Any equation of the form
$\p_t u= \sum_j c_j \, \p \, \mbox{{\rm res}}L^{j/2}$,
where $c_j$ are arbitrary constants, also belongs to the hierarchy.
The KdV equation itself is obtained at
$k=3$ if we identify
$t_3=t$. At $k=1$ we have $u_{t_1}=u_x$; this alow us to identify
$t_1$ with $x+c$. Here are the first three equations of the hierarchy:
$$
\begin{array}{l}
\, u_{t_1}=u_x,
\\ \\
4u_{t_3}=6uu_x +u_{xxx},
\\ \\
16u_{t_5}=30 u^2 u_x +20u_x u_{xx}+10 uu_{xxx}+u_{xxxxx}.
\end{array}
$$

For the proof of the proposition we should check two facts: 
first, that in any Lax equation the right hand side is actually 
the operator of multiplication by a function 
(and that it has the form $2\p \, \mbox{{\rm res}}\, L^{k/2}$), and, second,
that $\p_{t_{k}}\mbox{res} \,L^{3/2}
=\p_{t_{3}}\mbox{{\rm res}}\, L^{k/2}$.
The first fact follows from the equality
\beq\label{L+L-}
[(L^{k/2})_{+}, \, L] =
-\, [(L^{k/2})_{-}, \, L]
\eeq
(the left hand side is a purely differential operator while the right hand side 
contains only non-positive powers of $\p$, hence we have the operator 
of multiplication by a function in both sides). The second statement holds true
even in a more general form
$\p_{t_{k}}\mbox{res} \,L^{m/2}
=\p_{t_{m}}\mbox{{\rm res}}\, L^{k/2}$ for any pair of times
$t_k , t_m$. It is proved by the following chain of equalities:
$$
\begin{array}{c}
\p_{t_{k}}\mbox{{\rm res}}\, L^{m/2} =
\mbox{{\rm res}}\, ( \p_{t_{k}}L^{m/2})=
\mbox{{\rm res}}
\left ([(L^{k/2})_{+}, L^{m/2}]\right )
\\ \\
=\mbox{{\rm res}}
\left ([(L^{k/2})_{+}, (L^{m/2})_-]\right ) =
\mbox{{\rm res}}
\left ([L^{k/2}, (L^{m/2})_-]\right )
\\ \\
=\mbox{{\rm res}}
\left ([(L^{m/2})_{+}, L^{k/2}]\right )=
\mbox{{\rm res}}\, ( \p_{t_{m}}L^{k/2})=
\p_{t_{m}}\mbox{{\rm res}}\, L^{k/2}.
\end{array}
$$
Here we use the fact that
$\mbox{{\rm res}}
\left ([(L^{k/2})_{+}, (L^{m/2})_+]\right ) =
\mbox{{\rm res}}
\left ([(L^{k/2})_{-}, (L^{m/2})_-]\right ) =0$.
Besides, at the second step we implicitly assumed that the Lax equation for 
$L$ implies the Lax equation for its half-integer powers. 
Strictly speaking, this is not obvious and requires a proof. 
Clearly, it is enough to check this for the pseudo-differential operator 
$L^{1/2}$. 

\noindent
{\bf Lemma.} {\it It follows from the Lax equation $\p_{t}L=[A , L]$ 
that $\p_{t}L^{1/2}=[A , L^{1/2}]$.}

\noindent
For brevity denote $L^{1/2}={\cal L}$, then from
$\p_{t}{\cal L}^2=[A , {\cal L}^2]$ we find 
$(\dot {\cal L}-[A, {\cal L}]){\cal L}
+{\cal L}(\dot {\cal L}-[A, {\cal L}])=0$, where dot denotes the
$t$-derivative. Using the same argument as above (see (\ref{L+L-})), 
it is easy to see that the operator
$\dot {\cal L}-[A, {\cal L}]:= P_- = p_1 \p^{-1}+
p_2 \p^{-2}+\ldots $ contains only negative powers of
$\p$. Since ${\cal L}$ has the form
${\cal L}=\p + O(\p^{-1})$, from $P_- {\cal L}+
{\cal L}P_- =0$ it follows that $p_1 =0$, i.e. that $P_-$ is actually
of the form $P_- = \tilde P_- \p^{-1}$. Substituting it in this form into
the equality mentioned above, we obtain  
$\tilde P_- \tilde {\cal L}+{\cal L}\tilde P_- =0$, where
$\tilde {\cal L} =\p^{-1}{\cal L}\p = \p + O(\p^{-1})$, 
hence it follows in the same way that $p_2 =0$. Repeating this process, we find that
all coefficients of the operator $P_- = 
\dot {\cal L}-[A, {\cal L}]$ are equal to 0, i.e., 
$\dot {\cal L}=[A, {\cal L}]$, what is the statement of the lemma.

Hence the KdV equation has infinitely many {\it commuting} symmetries, i.e.
such that any two symmetries from this set are symmetries for each other. 
It turns out that the KdV equation has a larger set of symmetries. It includes 
also those which do not possess the commutativity property 
(so-called additional or non-abelian 
symmetries).
They will be discussed later in these notes.

\subsubsection{Integrals of motion}

The infinite number of symmetries implies the existence of an infinite set
of integrals of motion (conservation laws).

\paragraph{Proposition.} {\it The quantities
\beq\label{Laxk1}
I_{j}=\int \mbox{{\rm res}}\, L^{j/2}\, dx
\eeq
at $j\geq 1$ are integrals of motion for all equations
of the KdV hierarchy.
}

\noindent
(Obviously, $I_{2l}=0$, so only integrals with odd indices are non-trivial.)
The proof is based on the Lax representation:
$$
\p_{t_n} I_{j}=\int \mbox{{\rm res}}\, (\p_{t_n} L^{j/2})dx
=\int \mbox{{\rm res}} \left ([(L^{n/2})_+ , L^{j/2}]\right )dx.
$$
Now we recall that $\mbox{{\rm res}}\, [P,Q]$ for any two pseudo-differential operators
$P,Q$ is a full derivative. Therefore, in the case of rapidly decreasing or periodic 
solutions we obtain $\p_{t_n} I_{j}=0$.

\subsubsection{The Gelfand-Dickey coefficients and their properties}

The densities of integrals of motion
$R_j \equiv \mbox{{\rm res}} \, L^{j/2}$ are called the Gelfand-Dickey co\-ef\-fi\-ci\-ents.
There is a simple recurrence formula for them. 
In order to derive it, we write
$
(L^{j/2})_- = R_j \p^{-1}+S_j \p^{-2}+T_j \p^{-3}+\ldots
$,
then
$$
\begin{array}{c}
R_{j+2}=\mbox{{\rm res}} \, L^{\frac{j}{2} +1}=
\mbox{{\rm res}} \left ((\p^2 +u) L^{j/2}\right )
\\ \\
\phantom{aaaaa}=\, \mbox{{\rm res}} \left (\p^2
(R_j \p^{-1}+S_j \p^{-2}+T_j \p^{-3})+uR_j \p^{-1}\right )
\\ \\
=\, R_{j}'' +uR_j +2S_{j}' +T_j.
\end{array}
$$
Now calculate
$$
\begin{array}{c}
-[(L^{j/2})_{-}, \, L]=-[R_j \p^{-1}+S_j \p^{-2}+T_j \p^{-3}+\ldots \, ,\,
\p^2 +u]
\\ \\
=\, 2R_{j}' +(R_{j}'' +2S_{j}')\p^{-1} +
(u'R_j + S_{j}'' +2 T_{j}')\p^{-2} +\ldots
\end{array}
$$
and use equality (\ref{L+L-}). One can see from it that the operator
$[(L^{j/2})_{-}, \, L]$ does not contain negative powers of
$\p$, i.e., the following identities hold:
$$
\begin{array}{l}
R_{j}'' +2S_{j}'=0,
\\ \\
u'R_j + S_{j}'' +2 T_{j}'=0.
\end{array}
$$
Expressing $S_j$, $T_j$ through $R_j$, we find the recurrence relation 
from the formula for
$R_{j+2}$ written above:
\beq\label{recurr1}
4R_{j+2}'=R_{j}''' +4u R_{j}' +2u' R_j.
\eeq
One can rewrite it in the form
\beq\label{recurr2}
4R_{j+2}'=(\p^2 +4u +2u'\p^{-1})R_{j}'
\eeq
or
\beq\label{recurr3}
4R_{j+2}=(\p^2 +2u +2\p^{-1}u\p )R_{j}.
\eeq
It is convenient to start with $j=-1$
putting $R_{-1}=1$. The recurrence relation connects 
$R_j$ with odd indices; all $R_{2l}$ are equal to $0$.
The generating function
\beq\label{Rz}
R(z)=\sum_{n=-1}^{\infty}R_n z^{-n-2}
=z^{-1}+R_1 z^{-3}+\ldots
\eeq
allows one to represent the recurrence relation in the form of the 
linear differential equation
\beq\label{Rz1}
R'''(z)+4(u-z^2)R'(z)+2u' R(z)=0.
\eeq

\noindent{\bf Problem.}
Prove that $R(z)$ satisfies the nonlinear differential equation
\beq\label{nlind}
2R(z)R''(z)-(R'(z))^2 +4(u-z^2)R^2(z)+4 =0.
\eeq

\noindent{\bf Remark.}
The function $R(z)$ has the meaning of diagonal of the kernel
of the operator 
$(\p^2 +u -z^2)^{-1}$, i.e., $R(z)=R(z;x,x)$, where
$R(z;x,x')$ is the kernel of the resolvent of the operator $\p^2 +u$.
Thus the generating function of integrals of motion
$\int R(z)dx = \int R(z;x,x)dx$ should be understood as
$\mbox{tr}\, (\p^2 \! + \! u \! -\! z^2)^{-1}$, which is in agreement with
conservation of this quantity by virtue of the Lax equation.

The operator $\Lambda =\p^2 +2u +2\p^{-1}u\p$
in the right hand side of (\ref{recurr3}) has a name. It is called the
recursion operator. It allows one to formally solve the recurrence relation writing
\beq\label{recurr4}
R_{2j-1}= 2^{-2j}\Lambda ^{j} \cdot 1\,, \quad j=0,1,2, \ldots
\eeq
Here are the first few $R_j$'s:
$$
\begin{array}{l}
\phantom{1}2R_{1}=u,
\\ \\
\phantom{1}8R_3 = 3u^2 + u'',
\\ \\
32R_5 = 10u^3 +5u'^{\,2}+10 uu''+ u''''.
\end{array}
$$
We remind the reader that the higher KdV equations have the form
$\p_{t_{j}}u=2R_{j}'$.

\noindent{\bf Problem.}
Prove that
\beq\label{L-adic}
A_m = (L^{m/2})_+ =
\sum_{j=0}^{\frac{m-1}{2}} \left (R_{2j-1}\p -
\frac{1}{2}R'_{2j-1}\right )
L^{\frac{m-1}{2}-j}\,, \quad m=1,3,5 \ldots
\eeq
(Hint: first prove the recurrence relation 
$A_{m+2}=A_m L +R_m \p -\frac{1}{2}R'_{m}$.)

The coefficients $R_j$ have an important property:
for any $j,l$ the product $R_j R'_l$ is a full derivative, i.e.
there exists a differential polynomial $P_{j,l}$
such that
\beq\label{RRP}
R_j R'_l =P'_{j,l}.
\eeq
This statement can be proved by induction with the help of the recurrence relation
(\ref{recurr1}). Namely, assume that this property holds for some (odd)  
$j$ for all $l$ (for example, this is obvious
at $j=-1$); then the recurrence relation implies that this property
holds for all 
$l$ and the next odd $j$.

\noindent
{\bf Problem.} Prove the property (\ref{RRP}).

\noindent
In particular, this property guaranties that the operator $\p^{-1}$ in 
(\ref{recurr3}) is always applied to a full derivative of a differential polynomial:
$u\p R_j =
2R_1 R'_j = 2P_{1,j}'$, and so the result of its action is the differential polynomial 
$2P_{1,j}$. 

There is also a recurrence relation of a different type for the Gelfand-Dickey
co\-ef\-fi\-ci\-ents:
\beq\label{recurr5}
\frac{\delta}{\delta u}\int R_m \, dx = \frac{m}{2}\, R_{m-2},
\eeq
where in the left hand side we have the variational derivative of
$m$th integral of motion
$I_m = \int R_m dx$. This relation immediately 
follows from the fact that under 
$\int \mbox{res}$ the variation of the operator 
$L$ elevated to the power $m/2$ can be handled as variation 
of the usual power function:
$$
\int \mbox{res}\, (\delta (L^{\frac{m}{2}}))dx =
\frac{m}{2} \int \mbox{res}\,
(L^{\frac{m-2}{2}}\, \delta L)dx.
$$
Since $\delta L =\delta u$, the relation (\ref{recurr5})
is now obvious.

\noindent
{\bf Problem.} Give a detailed derivation of relation (\ref{recurr5}).

\subsection{Hamiltonian formulation}

The relation (\ref{recurr5}) allows one to write
$m$th equation of the KdV hierarchy in the form 
\beq\label{ham}
\p_{t_m}u= \frac{4}{m+2}\, \frac{d}{dx} 
\frac{\delta I_{m+2}}{\delta u}.
\eeq
Clearly, this equation has the standard Hamiltonian 
form $\p_{t_m}u =\{u, H_m\}$ with the Hamiltonian
$H_m =\frac{4}{m+2}\, I_{m+2}$ and the Poisson bracket defined on 
functionals ${\cal F}$, ${\cal G}$ of $u$ in the following way:
\beq\label{br1}
\left \{{\cal F}, {\cal G}\right \}=
\int \frac{\delta {\cal F}}{\delta u}\, \frac{d}{dx}\,
\frac{\delta {\cal G}}{\delta u}\, dx.
\eeq
It is easy to see that the integrals $I_m$ are in involution:
$$
\{I_m , I_n\}=\int \frac{\delta I_m}{\delta u}\, \frac{d}{dx}\,
\frac{\delta I_n}{\delta u}\, dx =
\frac{mn}{4}\int R_{m-2}R'_{n-2}dx =0
$$
by virtue of (\ref{RRP}).

\noindent
{\bf Problem.} Check that the bracket (\ref{br1}) satisfies the Jacobi identity.

If in the $m$th KdV equation $\p_{t_m}u=2\p R_m$ one first expresses
$R_m$ through $R_{m-2}$ by means of the recurrence operator 
($4R_{m}=\Lambda R_{m-2}$), and then use relation 
(\ref{recurr5}), one obtains another Hamiltonian 
formulation of the same equation: 
\beq\label{ham2}
\p_{t_m}u= \frac{1}{m}\, \frac{d}{dx} \, \Lambda \,
\frac{\delta I_{m}}{\delta u}\,,
\quad \quad \frac{d}{dx} \, \Lambda = \p^3 +4u\p +2 u'.
\eeq
Now the Hamiltonian is $\frac{1}{m}I_m$, and the Poisson bracket is given by
\beq\label{br2}
\left \{{\cal F}, {\cal G}\right \}_2=
\int \frac{\delta {\cal F}}{\delta u}\, \frac{d}{dx}\, \Lambda \,
\frac{\delta {\cal G}}{\delta u}\, dx.
\eeq
It is easy to check that it is antisymmetric, but the proof of the Jacobi identity
for it is non-trivial.

The bracket (\ref{br1}) (which can be naturally denoted as 
$\{\phantom{|}, \phantom{|}\}_1$) and the bracket (\ref{br2}) define respectively
{\it first and second Hamiltonian structures} of the KdV equation.
It is easy to see that the integrals of motion 
$I_m$ are in involtuion with respect to the both brackets. 
It can be also verified that these brackets are
{\it compatible}, i.e., any their linear combination
$\{\phantom{|}, \phantom{|}\}_1 +\lambda \{\phantom{|}, \phantom{|}\}_2$
with constant $\lambda$ is also a Poisson bracket.



\subsection{Auxiliary linear problems and $\psi$-function}

\subsubsection{The Baker-Akhiezer function}

The Lax equation (\ref{Lax1}) is the compatibility condition
of overdetermined system of linear differential equations
(auxiliary linear problems)
\beq\label{lin1}
\left \{\begin{array}{l}
L\psi =z^2 \psi ,
\\ \\
\p_t \psi =A\psi .
\end{array}\right.
\eeq
Indeed, taking the $t$-derivative of the first equation and substituting the second one,
one gets $\Bigl (\p_t L+[L,A]\Bigr )\psi =0$. 
The compatibility means existence of a large set of common solutions. It then follows that 
the operator $\p_t L+[L,A]$ should be equal to 0. 

The solutions $\psi$
can be found in the form of a series in $z$:
\beq\label{psi}
\psi = e^{zx+z^3t}\left ( 1+ \frac{\xi _1}{z}+\frac{\xi _2}{z^2}
+ \ldots \right ),
\eeq
where $\xi _i$ depend only on $x$ (and on $t$). The functions $\xi_i$ can be
expressed through $u$ by substitution of the series for $\psi$
into the equation $(\p^2 \!+\! u)\psi =z^2 \psi$. For example,
\beq\label{lin1a}
\begin{array}{l}
2\xi _1' +u=0,
\\ \\
2\xi _2'+\xi_{1}^{''}+\xi _1 u=0
\end{array}
\eeq
and so on (the recurrence relations for $i\geq 2$ 
are $2\xi _i'+\xi_{i-1}^{''}+\xi _{i-1} u=0$).

In a similar way, any higher KdV equation is the compatibility condition
of the linear problems
\beq\label{lin2}
\left \{\begin{array}{l}
L\psi =z^2 \psi ,
\\ \\
\p_{t_m} \psi =A_m\psi 
\end{array}\right.
\eeq
with the same operator $L$ and $A_m = (L^{m/2})_{+}$.
Their common solution in this case has the form
\beq\label{psi2}
\psi = e^{zx+z^3t_{3}+z^5t_{5}+\ldots}
\left ( 1+ \frac{\xi _1}{z}+\frac{\xi _2}{z^2}
+ \ldots \right ).
\eeq
So far it is only some formal series. 

The function $\psi$ regarded as a function of
the ``spectral parameter''
$z$ is called the Baker-Akhiexer function. Strictly speaking, it should be called
formal Baker-Akhiexer function because it is not yet a function but only a formal series. 
However, we will not emphasize this distinction. 
The Baker-Akhiexer function plays a fundamental role in the theory of the KdV equation
(and other soliton equations). It serves as a basic tool for construction of exact 
solutions. Namely, the integration scheme of the KdV equation will consist in 
constructing a family of solutions for
$\psi$. They will be already well-defined functions of $z$ whose expansion
around $\infty$ will be of the form (\ref{psi2}) ). The desired solution
$u$ will be then found using the formula $u=-2\p_x \xi _1$.

\noindent
{\bf Exercise.} Prove that
$\p_{t_n}\xi_1 = -R_n$, 
where $R_n$ is the Gelfand-Dickey coefficient.

\subsubsection{Integrals of motion from the $\psi$-function}

The Baker-Akhiezer function allows one to find an infinite set of
integrals of motion. 

\noindent
{\bf Proposition.}
{\it The function
\beq\label{chi}
\chi := \p \log \psi - z =\sum_{j=1}^{\infty}
\chi _j z^{-j}
\eeq
is a generating function of densities of integrals of motion, i.e.
$
\p_t \int \chi _j dx=0
$
for all $j\geq 1$.
}

\noindent
Indeed, the function $\chi$ satisfies the equation of the
Riccati type
$$
\chi ^{2}+\chi_x +2z\chi +u =0,
$$
hence its coefficients can be recursively expressed through
$u$, $u_x$, $u_{xx}$, $\ldots$,
for example: $\chi_1 =-\frac{1}{2}\, u$, $\chi_2 =\frac{1}{4}\, u_x$.
On the other hand, 
$$
\p_t \chi =\p \p_t \log \psi =\p \left (
\frac{\p_t \psi}{\psi}\right )=
\p \left (
\frac{A\psi}{\psi}\right ).
$$
But the function $A\psi /\psi$
is expressed through $\chi , u$ and their $x$-derivatives
(because
$\p^3 \psi /\psi$ is expressed through
$\chi$ and its derivatives). Therefore, the expansion coefficients 
of this expression are differential polynomials of $u$.
The derivative in the right hand side implies that
$\p_t \int \chi  dx=0$ (for rapidly decreasing and periodic functions
$u$).

Note that the non-trivial integrals of motion only come from 
$\chi_j$ with odd indices. All $\chi_{2n}$ are full derivatives. 
Indeed, writing the Riccati equations for 
$\chi (\pm z)$ and subtracting them, we get
$$
(\chi (z) \! +\! \chi (-z))(\chi (z)\! -\! \chi (-z))
+(\chi (z)\! -\! \chi (-z))_x +2z(\chi (z) \! +\! \chi (-z))=0,
$$
hence
$
\chi (z) \! +\! \chi (-z)=-\p \log \Bigl (
\chi (z)\! -\! \chi (-z) \! +\! 2z\Bigr )
$
is the full derivative.

A natural question is how the integrals
$\displaystyle{\int \chi _j \, dx}$ are connected with the integrals
$I_j$ given by equation (\ref{Laxk}).
Let $\psi (z)$ be a solution of the equation
$(\p^2 +u)\psi =z^2 \psi$, then the second solution
is $\psi (-z)$, and their Wronskian
\beq\label{Wr}
\psi (-z)\psi _x (z) -\psi (z)\psi _x (-z):=W(z)
\eeq
does not depend on $x$.
Dividing both sides by $\psi (z)\psi (-z)$, we get
\beq\label{chi1}
2z+ \chi (z)\! -\! \chi (-z) =\frac{W(z)}{\psi (z)\psi (-z)},
\eeq
where in the left hand side we have the generating function of densities 
of non-trivial integrals
$\chi_j$ (with odd indices).
Next, it is not difficult to see that $\psi (z)\psi (-z)$
satisfies the same third order equation (\ref{Rz1}) as
$R(z)$. Since both functions have the same structure of expansions in powers of
$z$, they may differ only by a $x$-independent common factor which can be found
from the limit $x\to \infty$
(assuming that $u\to 0$ as $x\to \infty$). Taking into account that
$\chi (z)\to 0$ as $x\to \infty$, (\ref{chi1})
implies that $\psi (z)\psi (-z)$ tends to $W(z)/(2z)$ and, therefore, 
\beq\label{Rchi}
R(z)=\frac{2\psi (z)\psi (-z)}{W(z)}.
\eeq
We see that the generating functions of densities of the integrals
$\int \chi _j dx$ and $I_j$ at odd $j$
are connected by the relation
$$
\left (z+\frac{\chi (z)\! -\! \chi (-z)}{2}\right )R(z)=1.
$$

Expanding both sides of (\ref{Rchi}) in powers of $z$, we can write:
\beq\label{Rchi1}
R_l =2\, \mbox{res}_{z=\infty}
\left (\frac{z^{l+1}}{W(z)}\, \psi (z) \psi (-z)\right ),
\eeq
where the residue is understood as the coefficient in front of
$z^{-1}$. Note that this equality is a consequence of a more general
relation
\beq\label{Lchi}
\left (L^{m/2}\right )_{-}=2\, \mbox{res}_{z=\infty}
\left (\frac{z^{m+1}}{W(z)}\, \psi (z) \p^{-1}\psi (-z)\right ).
\eeq
We will not prove it here (it follows from even more general relation
which is proved in the section on the KP hierarchy).

\subsubsection{The mKdV equation}

Let us show how the $\psi$-function allows one to pass from the KdV equation
to the so-called modified KdV equation (mKdV). 
Put $v=\p\log \psi$, then equations
$\psi_{xx} +u \psi =\lambda \psi$ and $\psi_t =
\psi_{xxx} +\frac{3}{2}u\psi_x +\frac{3}{4}u_x \psi$ 
(the spectral parameter is denoted here by 
$\lambda$) can be written in the form
\beq\label{miura}
u=\lambda - v^2 -v_x
\eeq
and
\beq\label{miura1}
v_t=\p_x \left ( \frac{\psi_{xxx}+\frac{3}{2}u\psi_x +
\frac{3}{4}u_x\psi}{\psi}\right ).
\eeq
The relation (\ref{miura}) is called the Miura transformation
(usually with $\lambda =0$). Using the Miura transformation 
and obvious identities 
$$
\frac{\psi_{xx}}{\psi}=v_x +v^2\,, 
\quad \quad
\frac{\psi_{xxx}}{\psi}=v_{xx}+3vv_x +v^3,
$$
one can represent the relation (\ref{miura1}) as an
equation for $v$: $v_t =\frac{1}{4}\,
\p_x (v_{xx}-2v^3 +6 \lambda v)$ or
\beq\label{miura2}
4v_t =-6v^2v_x +v_{xxx}+6\lambda v_x,
\eeq
which is called the mKdV equation (usually without the
last term).

\noindent
{\bf Exercise.} Let $u$ and $v$ be connected by the Miura transformation
(\ref{miura}). Show that
\beq\label{miura3}
-(4u_t -6uu_x -u_{xxx})=
(\p_x +2v)(4v_t +6v^2 v_x -v_{xxx}-6\lambda v_x ).
\eeq

\subsection{Construction of solutions to the KdV equation
with the help of the
$\psi$-function}

\subsubsection{The basic lemma}

Let us present simple but important lemma of a technical nature
on which the cons\-truc\-tion of exact solutions is based.

\noindent
{\bf Lemma.}
{\it For the function $\psi$ of the form (\ref{psi})
the following formal equalities hold:
\beq\label{lemma1}
\begin{array}{r}
(\p^2 -z^2 +u)\psi =O(z^{-1})e^{zx +z^3t},
\phantom{aaaaaaaaa}
\\ \\
(\p_t +\frac{1}{2}\, \p^{3}-\frac{3}{2}\, z^2 \p
+ \frac{3}{4}\,u_x )
\psi =O(z^{-1})e^{zx +z^3t},\phantom{aaaaaa}
\end{array}
\eeq
where the function $u=u(x,t)$ can be found form vanishing 
of the coefficients at non-negative powers of 
$z$:
$u=-2 \xi _{1,x}$.}

\noindent
The proof is a direct verification. The meaning and use of this 
statement is demonstrated below. It allows one to construct exact solutions
to the KdV equation using methods of linear algebra. 
Assume that the space of
$\psi$-functions of the form
$(\ref{psi})$ (defined by imposing certain requirements on
analytic properties of these functions as functions of the complex variable 
$z$) is {\it one-dimensional}, i.e. there is only one such function
up to multiplication by a constant. Assume also that the operators in the
left hand sides of the formal equalities preserve this space.
Then the form of the right hand sides implies that they are equal to zero
identically and not only up to the terms
$O(z^{-1})e^{zx +z^3t}$. In its turn, this means that 
the function 
$\psi$ for all $z$ is a common solution to the pair of linear problems 
(\ref{lin1}). The compatibility of these linear problems
iplies the KdV equation for 
$u=-2 \p\xi _{1}$.

\subsubsection{One-soliton solution}

We begin with the simplest example. Let us consider the space of functions
$\psi =\psi (z)$ which are meromorphic everywhere except infinity and such that
\begin{itemize}
\item[a)]
The function $\psi e^{-zx -z^3t}$ is regular at $z=\infty$;
\item[b)]
The function $\psi$ has the only simple pole at $z=0$ and holomorphic everywhere else
except $\infty$;
\item[c)]
In some point $p \in \CC$ the relation
$
\psi (p)=\psi (-p)
$ holds for all $x,t$.
\end{itemize}
The variables $x,t$ and the point $p$ are regarded here as
fixed parameters. Clearly, such functions form a linear space.
It is easy to see that if the parameters are in general position, then the dimension
of this space is equal to 1, i.e., tere is only one such function (up to multiplication by
a constant). Indeed, such $\psi$ has the form
$$
\psi =e^{zx +z^3t} \left (b_0 + \frac{b_1}{z}\right )
$$
with some $b_0$, $b_1$ but condition c) fixes the ratio
$b_1 /b_0$, and thus only the common factor remains arbitrary. 

Next, it is not difficult to convince oneself that the operators in
the left hand sides of (\ref{lemma1}) preserve the 
linear space of functions defined above. 
It is enough to check that for any choice of the function 
$u$
$$
\begin{array}{r}
(\p^2 -z^2+u)\psi =O(1)e^{zx +z^3t},
\phantom{aaaaaaaaaaa}
\\ \\
(\p_t +\frac{1}{2}\, \p^{3}-\frac{3}{2}\, z^2 \p
+ \frac{3}{4}\,u_x )
\psi =O(1)e^{zx +z^3t}\phantom{aaaaaa}
\end{array}
$$
and that the left hand sides at
$z=p$ and $z=-p$ are the same. The first is checked by a direct calculation and the
second is obvious from the fact that the left hand sides contain $z^2$ only.
Thus in the right hand sides we have functions from the same linear space;
denote them by $\psi_1$ and
$\psi_2$.
According to the lemma, at $u=-2 \xi _{1,x}$ the right hand sides behave actually
as $O(1/z)e^{zx +z^3t}$.
This means that the functions
$\psi_1 e^{-zx -z^3t}$ and $\psi_2 e^{-zx -z^3t}$
vanish at infinity. The uniqueness then implies
$\psi_1 =\psi_2 \equiv 0$, i.e.,
$$
\begin{array}{l}
(\p^2 +u -z^2)\psi =0,
\\ \\
(\p_t +\frac{1}{2}\, \p^{3}-\frac{3}{2}\, z^2 \p
+ \frac{3}{4}\,u_x )
\psi =0.
\end{array}
$$
Substituting $z^2 \psi$ from the first equality to the second one,
we come to the pair of linear problems
(\ref{lin1}) together with the explicitly found family of common solutions. 
Their compatibility guarantees that $u=-2 \xi _{1,x}$
is a solution to the KdV equation.

It remains to find the solution explicitly. 
For the function
$$
\psi =e^{zx +z^3t} \left (1 + \frac{\xi_1}{z}\right )
$$
the condition $\psi (p)=\psi (-p)$ is equivalent to the linear equation
$e^{2px +2p^3t}(p+\xi_1)=p-\xi_1$ for $\xi_1$, hence
$
\xi_1 =-p\, \tanh (px +p^3 t)
$
and, therefore, 
\beq\label{solit1}
\mbox{\fbox{$\displaystyle{\phantom{\int ^{A}_{B}}
u(x,t)=\frac{2p^2}{\cosh ^2 (px +p^3 t)}
\phantom{\int ^{A}_{B}}}$}}
\eeq
(the same formula as (\ref{kdv3}) after identification $c=p^2$).
Note that $\xi_1 =-\p_x \log \cosh (px +p^3 t)$ and hence 
equation (\ref{solit1}) can be written in the form
$$
u(x,t)=2\p_{x}^{2}\log \cosh (px +p^3 t)
$$
The one-soliton solution of the whole KdV hierarchy is given by the same formula
in which instead of $px +p^3 t$ under $\cosh$ we have
$px +p^3 t_3 +p^5 t_5 +\ldots$

\noindent
{\bf Remark.} Instead of functions with a pole at $0$ one can consider functions
with a pole at an arbitrary point
$a\in \CC$ and with a more general condition
$\psi (p)=\alpha \psi (-p)$, where $\alpha$ is an arbitrary nonzero constant.

\noindent
{\bf Problem.} Show that the function $\psi =
e^{zx +z^3t} \left (1 + \frac{\xi_1}{z-a}\right )$ with the condition
$\psi (p)=\alpha \psi (-p)$ leads to the one-soliton solution
of the similar form (\ref{solit1}) which differs from it only by a shift
$x \to x +x_0$ (here $x_0$ is generally speaking a complex number)
and express $x_0$ through $a$ and $\alpha$.

From the point of view of the Schrodinger equation with the potential
$-u$,
i.e., $-\p^2 \psi -u\psi =E\psi$, the one-soliton solution is remarkable in that
the corresponding potential well has only one bound state with energy $E=-p^2$
and the states belonging to the continuous spectrum with energy $E=-z^2$
at purely imaginary $z$ have zero reflection coefficient. The propagation
through the potential well brings only the phase shift equal to
$\mbox{arg}\, \frac{z-p}{z+p}$.

\subsubsection{Multisoliton solutions}

Consider the linear space of functions
$\psi =\psi (z)$ which are meromorphic everywhere except infinity and such that
\begin{itemize}
\item[a)]
The function $\psi e^{-zx -z^3t}$ is regular at $z=\infty$;
\item[b)]
The function $\psi$ has no more than $N$ poles (counted with multiplicities)
at some marked points of the complex plane and holomorphic everywhere else in the complex
plane except at $\infty$;
\item[c)]
At $N$ distinct points $p_j \in \CC$ the relations
$\psi (p_j)=\alpha _j\psi (-p_j)$, $j=1, 2, \ldots , N$, hold.
\end{itemize}
For simplicity consider the case when all the poles are concentrated at the point
$z=0$, i.e., there is a pole of multiplicity not greater than $N$ in this point and
no other poles. Then $\psi$ can be represented in the form
$$
\psi =e^{zx +z^3t} \left (b_0 + \frac{b_1}{z}
+\frac{b_2}{z^2} +\ldots +\frac{b_N}{z^N}\right ).
$$
The space of such functions has dimension $N+1$.
Similarly to the case of just one pole, 
$N$ linear conditions $\psi (p_j)=\alpha _j\psi (-p_j)$
for the coefficients $b_j$ make this space one-dimensional. 
The operators in the left hand sides of (\ref{lemma1}) preserve it.
Therefore, if we normalize
$\psi$ by the condition that the coefficient in front of
$z^0$ is $1$, as in (\ref{psi}),
$$
\psi =e^{zx +z^3t} \left (1 + \frac{\xi_1}{z}
+\frac{\xi_2}{z^2} +\ldots +\frac{\xi_N}{z^N}\right ),
$$
then $u=-2 \xi _{1,x}$
is going to be a solution of the KdV equation.

Let us find this solution explicitly. The conditions
$\psi (p_j)=\alpha _j\psi (-p_j)$ are equivalent to the following
system of linear equations for $\xi_j$:
$$
\sum _{j=1}^{N}M_{ij}\, \xi _j = - M_{i0},
$$
where $M_{ij}$ has the form
$
M_{ij}=p_{i}^{-j}e^{p_ix+p_{i}^{3}t}-\alpha _i
(-p_{i})^{-j}e^{-p_ix-p_{i}^{3}t}
$.
The Kramer's rule yields
$$
\xi_1 = -\,\frac{\det M_{ij}^{(0)}}{\det M_{ij}}\,,
\quad \quad i,j =1,2,\ldots , N,
$$
where the matrix $M_{ij}^{(0)}$ differs from $M_{ij}$ by the change of the first column
$M_{i1}$ to $M_{i0}$ ($i$ numbers the rows).
Since $\p_x M_{ij}=M_{i, j-1}$, we have $\det \! M_{ij}^{(0)} =
\p_x \det \! M_{ij}$, and $\xi_1 = -\p_x \log \det M_{ij}$, so that
$
u=2\p_x^2 \log \det M_{ij}
$. We have thus obtained the family of solutions
\beq\label{uN}
\mbox{\fbox{$\displaystyle{\phantom{\int ^{A}_{B}}
u=2\p_{x}^{2}\log \tau ,
\phantom{\int ^{A}_{B}}}$}}
\eeq
where
\beq\label{tauN}
\tau=
\det_{1\leq i,j \leq N} \left (
p_{i}^{-j}e^{p_ix +p_i^3 t}
-\alpha_i (-p_i)^{-j}e^{-p_ix -p_i^3 t}
\right ).
\eeq
The solution of the whole hierarchy is given by the same formula with the change
$p_ix+p_i^3t$ $\to$ 
$p_ix +p_i^3 t_3 + p_i^5 t_5+\ldots$ \hspace{1mm} Note that all
$\alpha _i$'s can be ``hidden'' in suitably choisen 
initial values of the times $t_j$, so from the point of view of the 
hierarchy the true parameters of the solution are only
the $p_i$'s.

This solution is called the $N$-soliton solution and $p_i$ are called
momenta of the solitons. 
The function $\tau= \tau (x, t_3, t_5, \ldots )$ is called the
{\it tau-function}. It plays a fundamental role
not only in the theory of the KdV equation but also 
in the theory of all other integrable equations. It turns out that any exact solution
of the KdV hierarchy (not only the $N$-soliton solution) can be represented as
(multiplied by 2) second loragithmic derivative of the determinant of some matrix or operator
(in general case infinite-dimensional). This determinant is the tau-function. 

Since the solution is expressed through the second logarithmic derivative of 
$\tau$, the tau-functions which differ from each other only by a factor of the form
$Ce^{ax}$, with constant $C$ and $a$ are equivalent. 

It can be shown that the potentials in the Schrodinger equation corresponding to
the $N$-soliton solutions with real 
$p_i$ are reflectionless and have exactly $N$ bound states with energies
$E_i=-p_i^2$. In the quantum-mechanical interpretation, 
the conditions $\psi (p_j)=\alpha _j\psi (-p_j)$ mean that at the points of the discrete
spectrum there is only one linearly independent eigenfunction of the Schrodinger operator.

\noindent
{\bf Problem.} Find the phase shift of the wave function
for scattering on the potential corresponding to the 2-soliton solution.

Let us point out two other useful forms of the soliton tau-function. 
One of them is the following determinant of the $N\times N$ matrix:
\beq\label{tauN1}
\tau =\det_{1\leq i,j \leq N} \left (\delta_{ij}+
\frac{2\beta_i p_i}{p_i +p_j}\, e^{2p_ix +2p_i^3 t_3 +2p_i^5 t_5 +\ldots}
\right ).
\eeq
Here $\beta_i$ are some parameters (which are similar to $\alpha_i$). They also
can be eliminated by a suitable shift of times.

\noindent
{\bf Problem.} Prove the equivalence of the determinant representations
(\ref{tauN}) and (\ref{tauN1}).

\noindent
Expanding the determinant (\ref{tauN1}), we get the following formula:
\beq\label{tauN2}
\tau =\sum_{\{\epsilon_1 , \ldots , \epsilon_N\}\in \z_{2}^{N}}
\,\,\, \prod_{i<j}^{N}\left (\frac{p_i -p_j}{p_i+p_j}
\right )^{2\epsilon_i \epsilon_j}\prod_{k=1}^{N}
\left (\beta_k
e^{2p_kx +2p_k^3 t_3  +\ldots}\right )^{\epsilon_k}.
\eeq
Here the sum is taken over all sets of $N$ numbers
$\epsilon_i$ taking values $0,1$, so that the sum contains
$2^N$ terms. For example, at $N=2$ we have:
$$
\tau = 1+ \beta_1 e^{2p_1 x}+ \beta_2 e^{2p_2 x}+
\beta_1 \beta_2
\left (\frac{p_1 -p_2}{p_1+p_2}
\right )^{2}e^{2(p_1+ p_2) x},
$$
where only the terms containing $x$ in the exponential functions are left for
simplicity. 

The multisoliton solutions are stationary points for higher commuting symmetries.

\noindent
{\bf Proposition.} {\it Any $N$-soliton solution satisfies the ordinary
differential equation
(a higher stationary KdV)
\beq\label{stat}
\sum_i c_i R_i [u] =0,
\eeq
where $R_i$ are the Gelfand-Dickey coefficients, 
with some choice of the constants $c_i$.}

\noindent
For example, the one-soliton solution (\ref{solit1})
satisfies the equation $R_3-p^2 R_1=0$ or
$3u^2 +u_{xx}-4p^2 u=0$. This is obvious from the fact that 
the solution depends on the combination 
$x+p^2 t$ and, therefore, 
$u_t = p^2 u_x$. The general proof can be carried out using 
the explicit formulas
(\ref{uN}), (\ref{tauN}) and the fact that in the rapidly decreasing case
equation (\ref{stat}) is equivalent to 
$\sum_i c_i \p_{t_i}\tau =0$. It is convenient to use 
the expression (\ref{tauN2}) for the tau-function. 

\noindent
{\bf Problem.} Prove that the 2-soliton solution satisfies the equation
$R_5[u]+c_3R_3[u]+c_1R_1[u]=0$ and find $c_3, c_1$.

\noindent
{\bf Remark.} Equations (\ref{stat}) (called Novikov's equations)
besides rapidly decreasing so\-lu\-ti\-ons of the soliton type have a large family
of periodic and quasiperiodic solutions. The corresponding potentials in the 
Schrodinger operators are distinguished by the fact that these operators
have only a finite number of forbidden (unstable) bands in the spectrum. 
These solutions are expressed through the Riemann theta-functions.

\subsection{Commutation representation of the KdV hierarchy by 
$2\! \times \! 2$ matrices}

There is an alternative commutation representation of the KdV hierarchy
which is realized in usual $2\times 2$ matrices depending on an additional
complex parameter (it is called the spectral parameter). 
It is independent of the technique of pseudo-differential operators and in some cases
is more convenient.

\subsubsection{Zero curvature representation}

The second order equation $(\p^2 +u)\psi =z^2 \psi$ can be rewritten as a vector
first order equation
\beq\label{mm1}
\p \Psi =U_1 (\lambda ) \Psi ,
\eeq
where $\lambda =z^2$ is the spectral parameter,
$$
\Psi =\left ( \begin{array}{c}
\psi_1 \\ \psi_2 \end{array} \right ),
\quad 
U_1 (\lambda )=\left ( \begin{array}{cc}
0 & 1 \\ \lambda \! -\! u & 0 \end{array} \right )
$$
and $\psi_1 =\psi$, $\psi_2 = \psi ' =\p\psi$. In order to find a similar
representation for the equation $\p_{t_m}\psi =A_m \psi$, we use the formula
(\ref{L-adic}) and write:
$$
\p_{t_m}\psi = A_m \psi = 
\sum_{j=0}^{\frac{m-1}{2}} \left (R_{2j-1}\p -
\frac{1}{2}R'_{2j-1}\right )
\lambda ^{\frac{m-1}{2}-j}\psi =
-\frac{1}{2}R'_{m-2}(\lambda )\psi +R_{m-2}(\lambda )\psi '
$$
where the standard notation
$$
R_m (\lambda ):= \sum_{j=0}^{\frac{m+1}{2}}R_{2j-1}
\lambda ^{\frac{m+1}{2}-j}, \quad m=-1, 1, 3, \ldots
$$
for ``incomplete generating functions'' of the Gelfand-Dickey coefficients is introduced. 
For example, 
$$
\begin{array}{l}
R_{-1}(\lambda )=1, \\ \\
R_{1}(\lambda )= \lambda +R_1, \\ \\
R_{3}(\lambda )= \lambda ^2  +\lambda R_1 + R_3 \quad
\mbox{and so on.}
\end{array}
$$
The recurrence relation has the form $R_{m+2}(\lambda )=
\lambda R_m(\lambda )+R_{m+2}$.
It is also easy to find that
$$
\p_{t_m}\psi ' = 
\left [ (\lambda -u)R_{m-2}(\lambda ) -
\frac{1}{2} R''_{m-2}(\lambda )\right ] \psi
+\frac{1}{2} R'_{m-2}(\lambda )\psi '.
$$
Unifying this with the previously found formula for 
$\p_{t_m}\psi$, it is possible to represent the equation $\p_{t_m}\psi
=A_m \psi$ in the vector form similar to (\ref{mm1}):
\beq\label{mm2}
\p_{t_m}\Psi =U_m (\lambda ) \Psi ,
\eeq
where
$$
U_m (\lambda )=\left ( \begin{array}{cc}
-\frac{1}{2}R'_{m-2}(\lambda ) & R_{m-2}(\lambda ) 
\\  & \\
(\lambda \! -\! u)R_{m-2}(\lambda ) 
\! -\! \frac{1}{2}R''_{m-2}(\lambda )& 
\,\,\, \frac{1}{2}R'_{m-2}(\lambda ) \end{array} \right ).
$$
Note that at
$m=1$ this equation coincides with (\ref{mm1}).
At $m=3$ we have
$$
U_3 (\lambda )=\frac{1}{4}\left ( \begin{array}{cc}
-u' & 4\lambda + 2u
\\  & \\
4\lambda ^2 \! -\! 2\lambda u \! -\! (2u^2 +u'')& 
u' \end{array} \right ).
$$
In the general case $U_m (\lambda )$ is a matrix polynomial
of $\lambda$ of degree $\frac{1}{2} (m+1)$.
We have rewritten the auxiliary linear problems as matrix equations
of first order. 

The compatibility condition of the problems (\ref{mm1}) and
(\ref{mm2}) can be written in the form $[\p_{t_m}-
U_m (\lambda ), \, \p_{t_1}-U_1 (\lambda )]=0$ or
\beq\label{mm3}
\p_{t_m}U_1 (\lambda ) -\p_{t_1}U_m (\lambda ) 
+[U_1 (\lambda ), U_m (\lambda )] =0
\eeq
for all $\lambda$, which yields the $m$th equation of the KdV hierarchy.

\noindent
{\bf Exercise.} Check this statement by a direct 
calculation. 

\noindent
The representation of the KdV hierarchy in the form
(\ref{mm3}) is called the zero curvature (or Zakharov-Shabat) representation. 
Note that the more general relations of the same type
\beq\label{mm4}
\p_{t_m}U_n (\lambda ) -\p_{t_n}U_m (\lambda ) 
+[U_n (\lambda ), U_m (\lambda )] =0
\eeq
hold true for all $m,n \geq 1$.

\subsubsection{Zero curvature representation, 
another gauge}

The linear problems (\ref{mm2}) can be 
``gauge transformed'': $$\Psi \rightarrow {\cal W}\Psi, \quad
U_m (\lambda ) \rightarrow {\cal W}U_m (\lambda ){\cal W}^{-1},$$ where the matrix
${\cal W}$ does not depend on all $t_k$'s (but can depend on
$\lambda$). Such trans\-for\-ma\-ti\-on means that when passing from the second order equation
$(\p^2 +u)\psi =z^2 \psi$ to the first order vector equation
the components of the vector 
$\Psi$ are not $\psi$ and $\psi '$ but their linear combinations with
coefficients which may depend on $z$. 

For example, consider the choice
$\tilde \psi_1 =\psi$, $\tilde \psi_2 = \psi ' -z\psi$.
The new vector $\tilde \Psi$ is connected with $\Psi$ 
as follows:
$$
\left ( \begin{array}{c}
\psi_1 \\ \psi_2 \end{array} \right )=
\left ( \begin{array}{cc}
1& 0 \\ z& 1
\end{array} \right )
\left ( \begin{array}{c}
\tilde \psi_1 \\ \tilde \psi_2 \end{array} \right ).
$$
Ten the lineaar problems acquire the form
\beq\label{mm5a}
\p_{t_m}\tilde \Psi = \tilde U_m (z) \tilde \Psi \, ,
\quad m=1,3, \ldots
\eeq
where $z$ plays the role of the spectral parameter and 
$$
\tilde U_m (z)=\left ( \begin{array}{cc}
1& 0 \\ -z& 1
\end{array} \right )
U_m (z^2)
\left ( \begin{array}{cc}
1& 0 \\ z& 1
\end{array} \right )
$$
are matrix polynomials of $z$ of degree $m$:
$$
\tilde U_m (z)=\left ( \begin{array}{cc}
zR_{m-2}(z^2)-\frac{1}{2}R'_{m-2}(z^2) & R_{m-2}(z^2) 
\\  & \\
zR'_{m-2}(z^2)\! -\! uR_{m-2}(z^2)\! -\! \frac{1}{2}R''_{m-2}(z^2) & 
\,\,\, -zR_{m-2}(z^2)+\frac{1}{2}R'_{m-2}(z^2) \end{array} \right ).
$$
Here are the first two matrices:
\beq\label{mm6}
\hspace{-6cm} \tilde U_1 (z)=\left ( \begin{array}{cc}
z& 1 \\ -u& -z
\end{array} \right )
\eeq
\beq\label{mm7}
\tilde U_3 (z)=\frac{1}{4}
\left ( \begin{array}{cc}
4z^3 +2uz -u' & 4z^2 +2u \\ & \\
- 4uz^2 \! +\! 2u'z \! -\! (2u^2 +u'') & \,\,\, 
-4z^3 \! -\! 2uz \! +\! u'
\end{array} \right ).
\eeq

This commutation representation suggests possible generalization
of the KdV theory to  other evolution equations:
the integrable equation should be the compatibility con\-di\-ti\-on
of the linear problems (\ref{mm5a}) with some matrices $V_j (z)$
which are rational functions of the spectral parameter $z$. 
In particular, a minimal generalization of the KdV theory
corresponds to the choice
$$
V_1 (z)=\left ( \begin{array}{cc}
z& v \\ -u& -z
\end{array} \right ),
$$
where $u,v$ are some functions included in the equation (at 
$v=1$ we come back to KdV). The form of the other matrices 
should be chosen in such a way that the corresponding zero curvature equation
be equivalent to some evolution partial differential equations for
$u$ and $v$. This is the way to obtain the mKdV equation,
the nonlinear Schrodinger equation, the sine-Gordon equation and others.

\subsubsection{Spectral curve}

The zero curvature representation becomes especially useful in the case when
the solution is stationary with respect to some higher flow or their combination.
Take, for example,
$\p_t u := \displaystyle{\sum_m} c_m \p_{t_m}u =0$ with some constants
$c_m$, then $\p_t \Psi = U(\lambda )\Psi$ with the matrix
$U(\lambda )=\displaystyle{\sum_m} c_m U_m (\lambda )$.
Taking linear combination of equations (\ref{mm4}) with the coefficients
$c_m$ and imposing the condition
$\p_{t}U_n (\lambda )=0$, we get the Lax type equation for
$U(\lambda )$:
\beq\label{mm5}
\p_{t_n}U (\lambda ) 
=[U_n (\lambda ), U(\lambda )].
\eeq
This equation implies that the characteristic polynomial of the matrix
$U(\lambda )$ does not depend on the times $t_n$ for all
$\lambda$. In other words, the algebraic curve defined by the equation
$\det (\mu + U(\lambda ))=0$ is an integral of motion for all equations
of the hierarchy. This curve is called the spectral curve because it is closely
connected with the spectrum of the Schrodinger operator
$\p^2 +u$.
Since $\mbox{tr}\, U(\lambda )=0$, the equation of the spectral curve has the form
$\mu^2 +\det  U(\lambda )=0$. 
For example, if $U(\lambda )=
U_3 (\lambda )$, then the spectral curve is the elliptic curve
$$
\mu^2 = \lambda^3 -\frac{3u^2 +u''}{4}\, \lambda -
\frac{4u^3 -u'^{2} +2uu''}{16}.
$$

\noindent
{\bf Problem.} Verify directly that this spectral curve
is an integral of motion for all equations of the hierarchy, i.e. that
$\p_{t_k}(3u^2 +u'')=\p_{t_k}(4u^3 -u'^{2} +2uu'')=0$ for all $t_k$.

\noindent
In the general case $U(\lambda )=
c_m U_m (\lambda )+ c_{m-2}U_{m-2}(\lambda )+
\ldots + c_1 U_1 (\lambda )$, $c_m \neq 0$, the curve is hyperelliptic. The equation
of the curve is of the form $\mu^2 = P_m (\lambda )$, where 
$P_m(\lambda )$ is a polynomial, $\mbox{deg}\, P_m (\lambda )=m$.

\subsection{Nonabelian symmetries}

\subsubsection{The Galilean transformation and the similarity transformation}

We start with a very simple statement.

\noindent
{\bf Proposition.} {\it The KdV equation preserves its form 
under the transformations
\beq\label{ns1}
\left \{
\begin{array}{l}
u\rightarrow \,\, \lambda^{-2}u -2a \lambda^{-1},
\\ \\
x\rightarrow \,\, \lambda x +3a \lambda^2 t,
\\ \\
t\rightarrow \,\, \lambda^3 t
\end{array}\right.
\eeq
with arbitrary constants $\lambda, a$.}

\noindent
In other words, if $u(x,t)$ satisfies the KdV equation
(\ref{kdv2}), then $\tilde u$ defined as a function of
$\tilde x , \tilde t$
by the equalities
$$
\left \{
\begin{array}{l}
\tilde u = \,\, \lambda^{-2}u -2a \lambda^{-1},
\\ \\
\tilde x = \,\, \lambda x +3a \lambda^2 t,
\\ \\
\tilde t = \,\, \lambda^3 t
\end{array}\right.
$$
satisfies the same equation
$4\tilde u_{\tilde t}=6\tilde u \tilde u_{\tilde x}+
\tilde u_{\tilde x\tilde x\tilde x}$. Even simpler, one can say that
the transformation
$$
u(x,t)\rightarrow \,\,
\lambda^{-2}u\left (\lambda^{-1}x\! -\! 3a\lambda^{-2}t, \,
\lambda^{-3}t\right ) -2a\lambda^{-1}
$$
sends a solution to another solution. 
This can be checked by a direct substitution.

The transformation with $\lambda =1, \, a\neq 0$, i.e.
$$
u(x,t)\rightarrow \,\, u(x-3at, \, t)-2a
$$
is called the Galilean transformation.
In the case when $\lambda \neq 1, \, a= 0$, i.e.
$$
u(x,t)\rightarrow \,\, \lambda^{-2}u\left (\lambda^{-1}x, \,
\lambda^{-3}t\right )
$$
we have a similarity transformation. For example, 
the Galilean transformation of 
the one-soliton solution
(\ref{solit1}) is
$$
u= \frac{2p^2}{\mbox{cosh}^2 (px +(p^3 \! -\! 3ap)t)} - 2a
$$
while the similarity transformation of this solution is
$$
u= \frac{2(p/\lambda )^2}{\mbox{cosh}^2 (
(p/\lambda )x +(p/\lambda )^3t)}.
$$
In the latter case the form of the solution is preserved with
$p \rightarrow p/\lambda$.

In the infinitesimal form the Galilean and similarity transformations
have the form
$$
\begin{array}{l}
u\rightarrow u+ \left (\frac{3}{2}\, tu_x +1\right )\varepsilon ,
\\ \\
u\rightarrow u+ \left (3tu_t +xu_x +2u\right )\varepsilon ,
\end{array}
$$
where $\varepsilon$ is the small parameter of the transformation.
One may introduce the cor\-res\-pon\-ding ``times''
$s_{-1}$, $s_1$ and represent the infinitesimal transformations in the form of
the differential equations
\beq\label{ns2}
\begin{array}{l}
\displaystyle{
\frac{\p u}{\p s_{-1}}=\frac{3}{2}\, tu_x +1},
\\ \\
\displaystyle{
\frac{\p u}{\p s_{1}}=\frac{3}{4}\, t
\left (6uu_x +u_{xxx}\right )
+xu_x +2u}.
\end{array}
\eeq
They are symmetries of the KdV equation in the sense of section 2.3. 
This fact can be checked directly by calculation of the derivatives
$\p_{s_{i}}(\p_t u)$ and $\p_t(\p_{s_{i}}u)$ for $i=-1,1$.
However, equations (\ref{ns2}) are not symmetries for each other:
the derivatives
$\p_{s_{-1}}(\p_{s_{1}}u)$ and $\p_{s_{1}}(\p_{s_{-1}}u)$
are not equal. This is why these symmetries are called nonabelian. 

\noindent
{\bf Exercise.} Calculate $\p_{s_{-1}}(\p_{s_{1}}u)-
\p_{s_{1}}(\p_{s_{-1}}u)$.

\noindent

The characteristic feature of nonabelian symmetries is the form of the 
right hand sides of (\ref{ns2}): they are non differential polynimials
of $u$ but contain $x,t$ explicitly. In this respect, they differ from the 
previously discussed commuting symmetries. 

\subsubsection{Infinite series of nonabelian symmetries}

It turns out that the KdV equation has an infinite series of nonabelian
symmetries. 
They can be compactly written in the form
\beq\label{ns3}
\frac{\p u}{\p s_{m}}= 2^{-\frac{m+1}{2}} \p_x \Lambda^{\frac{m+1}{2}}
\! \left (\frac{3}{2}\, t u +x\right ),
\quad m=-1,1,3, \ldots
\eeq
where $\Lambda =\p^2 +2u +2\p^{-1}u\p$ is the recursion operator.
It is easy to check that the first two symmetries coincide with
(\ref{ns2}). The other ones do not have such a simple form. 
The symmetries 
(\ref{ns3}) can be extended to symmetries of the whole hierarchy
if instead of 
$\frac{3}{2}\, tu +x$ in the right hand side one substitutes
$$
S:= \sum_{n\geq 1} nt_n R_{n-2}=x+3t_3 R_1 +5t_5 R_3 +\ldots
$$

\subsubsection{Stationary points of nonabelian symmetries: examples}

It is interesting to find solutions which are stationary points
of nonabelian symmetries. For example, the solution invariant with respect to
the Galilean transformation in which all
$t_j$ starting from
$t_5$ are equal to $0$ should satisfy the condition 
$\p u/\p s_{-1}=\frac{3}{2}\, tu_x +1=0$, hence
\beq\label{ns4}
u(x,t)=-\, \frac{2x}{3t}.
\eeq
This is probably the simplest solution of the KdV equation which is not constant 
in each of the variables. 
If one puts all higher times equal to $0$
starting from $t_7$ while $t_5\neq 0$ (it is convenient to put
$t_5=2/5$), then the stationarity condition $\p u/\p s_{-1}=0$ acquires the form 
$3u^2 +u_{xx}+6tu +4x=0$. In this case the solution can not be expressed through
elementary functions. The best one can do is to express it through a solution
of the Painl\'eve I. 
Let $f(x)$ be a solution of Painl\'eve I
$$
3f^2 +f_{xx}+4x=0,
$$
then an easy calculation shows that
$$
u(x,t)=f\left (x-\frac{3}{4} \, t^2\right )-t
$$
satisfies the stationarity condition and the KdV equation. 
Note that the Painl\'eve I equ\-a\-ti\-on (more precisely, its
$x$-derivative) can be represented as the commutation relation
$[L, A]=1$. This is so-called ``string equation'', which became popular
in the beginning of nineties of the last century in connection with the attempts
to construct the theory of 2D quantum gravity based on the model of random matrices.

\subsubsection{Digression on Painl\'eve equations}

The appearance of the Painl\'eve equation in connection with the KdV equation
illustrates a rather general fact: ordinary differential equations that are 
obtained as reductions of integrable partial differential equations also have
certain ``good'' properties which distinguish them from all others. Namely, 
they have so-called Painl\'eve property. 

To formulate it, we should introduce the notion of critical point of 
solution to ordinary differential equation: a singular point is called critical
if it is not a pole of arbitrary integer order. In other words, critical points are
ramification points (algebraic and logarithmic) and essential singularities.  
Consider the full set of solutions to a differential equation; their critical points
can be divided in two groups: those which depend only on the equation itself
and does not depend on the choice of solution (immovable singularities) and those which
depend on the integration constants (movable singularities). A differential equation
has the Painl\'eve property if all its solutions have only immovable critical points. 

As it follows from the theory of linear ordinary differential equations,
all linear equations have the Painl\'eve property. 
Simple examples show that nonlinear equations may or may not possess the
Painl\'eve property ($y'=y^2$ and 
$y'=y^3$ respectively). For equations of first and second order which a linear
in the highest derivative all equations possessing the Painl\'eve property are known. 
For first order equations of the form
$y'=F(y,x)$ the answer is simple (Fuchs, 1884): only generalized Riccati equations
$$
y'=f_2(x)y^2 + f_1(x)y + f_0(x)
$$
do not have movable critical points. 

For second order equations of the form
$y''=G(y, y', x)$ the answer was obtained in the beginning
of XX century in the works of Painl\'eve, Fuchs and Gambier. 
There are 50 equations of this form which do not have movable
critical points. They can be reduced either to equations integrable in
known elementary or special functions or to one of 6 canonical equations
which in general can not be integrated in known functions.  
These six equations are now called Painl\'eve equations:
$$
\begin{array}{ll}
{\rm P}_{\rm I}:& \quad y'' =6y^2 +x
\\& \\
{\rm P}_{\rm II}:& \quad y''=2y^3 +xy +\alpha
\\ &\\
{\rm P}_{\rm III}:& \quad \displaystyle{
y''=\frac{y'^{2}}{y}-\frac{y'}{x}+x^{-1}(\alpha y^2 +\beta )
+\gamma y^3 +\frac{\delta}{y}}
\\ &\\
{\rm P}_{\rm IV}:& \quad \displaystyle{
y''=\frac{y'^{2}}{2y}+\frac{3}{2}\, y^3+4xy^2+2(x^2-\alpha )y
+\frac{\beta}{y}}
\\ &\\
{\rm P}_{\rm V}:& \quad \displaystyle{
y''=\left (\frac{1}{2y}+\frac{1}{y-1}\right )y'^{2}-\frac{y'}{x}
+\frac{y(y-1)^2}{x^2}\left (\alpha +\frac{\beta}{y^2}
+\frac{\gamma x}{(y-1)^2}+\frac{\delta x^2(y+1)}{(y-1)^3}\right )}
\\ &\\
{\rm P}_{\rm VI}:& \quad \displaystyle{
y''=\frac{1}{2}\left (\frac{1}{y}+\frac{1}{y-1}+\frac{1}{y-x}\right )
y'^{2}-\left (\frac{1}{x}+\frac{1}{x-1}+\frac{1}{y-x}\right )
y'}
\\ &\\
&\displaystyle{
\hspace{1cm}+\,\, \frac{y(y-1)(y-x)}{x^2(x-1)^2}
\left (\alpha +\frac{\beta x}{y^2}+\frac{\gamma (x-1)}{(y-1)^2}
+\frac{\delta x(x-1)}{y-x)^2}\right )}
\end{array}
$$
There is an extensive literature devoted to analysis of these equations and 
properties of their solutions. There are still open questions in the theory of
Painl\'eve equations. Besides, the Painl\'eve equations, as other fundamental 
objects, sometimes appear in rather unexpected contexts. 

In the context of integrable partial differential equations the Painl\'eve
equations emerge as self-similar reductions of two-dimensional equations.

\newpage

\section{The KP hierarchy}

The theory of the KdV equation allows generalizations in different directions.
One of them is to take the Lax operator to be an $N$th order differential operator
rather than a second order operator. For example, at 
$N=3$ one can consider the operator $L=\p^3 + u\p + w$
with two independent coefficients $u,w$ which depend on
$x$ and all the times $t_j$, with the dynamics being determined by the Lax equations
$\p_{t_j}L=[(L^{j/3})_{+}, \, L]$. It can be shown that 
each of the Lax equations defines a well-defined system of evolution equations
for the functions
$u,w$, and the equations obtained in this way are symmetries for each other.
At arbitrary $N>2$ the situation is similar, with the only difference that
the genarators of the flows are 
$(L^{j/N})_{+}$, and the equations are written for $N-1$
unknown function. All these hierarchies
(generalized KdV hierarchies of order $N$) can be embedded in one big hierarchy
(in some sense ``the biggest'' one) the definition of which already does not depend
on the number $N$. It is called the Kadomtsev-Petviashvili (KP) hierarchy. 

To make such embedding possible and natural, it is necessary to
``equalize in rights'' the Lax operators for different 
$N$, i.e. represent them as elements of one common algebra. 
The way how to do this was suggested mainly in the works of the Japanese 
school (Sato, Jimbo, Miwa). The idea is to work not with the Lax operators
themselves but with roots of $N$th degree from them. All of them are 
pseudo-differential operators of first order and satisfy the same Lax equations.
After that one may forget about their origin and extend the Lax equations to
first order pseudo-differential operators of general form.

\subsection{The Lax equations}

From now on by the Lax operator the {\it pseudo-differential} operator
of the form
\beq\label{kp1}
L= \p + u_1 \p^{-1} +u_2 \p^{-2} +\ldots
\eeq
will be understood. The coefficients
$u_i$ are in general independent functions of $x$ and the times
$t_j$ with integer $j\geq 1$.
The operator $L$ is subject to the Lax equations
\beq\label{kp2}
\p_{t_j}L=[(L^j)_{+}, \, L]\,, \quad \quad j=1,2,3, \ldots
\eeq
Each of them defines an infinite system of evolution equations for the infinite
set of functions $u_i$:
$\p_{t_j}u_i ={\cal P}_{ij} (\{u_l\})$, where ${\cal P}_{ij}(\{u_l\})$ are
differential polynomials of $u_l$. For example, comparing of the coefficients
in front of $\p^{-1}$ in both sides yields the equations 
\beq\label{kp6}
\p_{t_j}u_1 =\p_x \, \mbox{res}\, L^j
\eeq
which are similar in this form to the equations of the KdV hierarchy. However,
these equations are not closed because the right hand sides contain also the functions
$u_2, u_3$, $\ldots$. A closed system includes also equations for
$\p_{t_j}u_2$, $\p_{t_j}u_3$ and so on, which are obtained by comparing the coefficients
in front of higher degrees of the operator $\p^{-1}$.

The Lax equation at $j=1$ tells us that $\p_{t_1}L=[\p , L]$, or
$\p_{t_1}u_i =\p_x u_i$, which allows one to identify $t_1$ with $x$.
In other words, the evolution in the time $t_1$ is simply a shift of
the argument $x$: $u_i (x)\rightarrow u_i(x+t_1)$.

\noindent
{\bf Remark.} The KdV hierarchy is obtained after imposing the condition
that $L^2$ is a purely differential operator, i.e. does not contain negative powers of
$\p$: $(L^2)_{-}=0$ or $(L^2)_{+}=L^2$
(in the section devoted to the KdV equation 
it was this differential operator
$L^2$ which was denoted by $L$ and was called the Lax operator). This condition
makes the functions $u_i$ dependent; there is only one independent function among them:
$u=2u_1$, and all other
$u_i$ with $i\geq 2$ are expressed as 
differential polynomials of $u$.
Since $(L^{2m})_{+}=L^{2m}$ for all positive integer $m$, 
$(L^{2m})_{+}$ commutes with $L$
and all flows with even
orders are trivial in this case.

\subsection{Zero curvature representation}

There is another (equivalent) representation of the KP hierarchy
in which the Lax operator does not participate explicitly. In this approach 
the main objects are
differential operators
$(L^j)_+$.
For brevity we denote $A_j =(L^j)_+$.
For example, $A_1 =\p$,
$A_2 =\p^2 +2u_1$.

\noindent
{\bf Exercise.} Find $A_3 = (L^3)_+$.

\noindent
{\bf Proposition.} {\it The Lax equations (\ref{kp2}) imply
the equations
\beq\label{kp3}
\p_{t_m}A_n - \p_{t_n}A_m -[A_m, A_n]=0
\eeq
for all $m,n \geq 1$.}

\noindent
For the proof we first note that the Lax equations imply
$\p_{t_m}L^n=[A_m , L^n]$ for all $n$, then
\beq\label{kp5}
\begin{array}{ll}
&\p_{t_m}(L^n)_+ -\p_{t_n}(L^m)_+ -[A_m, A_n]
\\ &\\
=& \Bigl (\p_{t_m}L^n -\p_{t_n}L^m -[A_m, A_n] \Bigr )_+
\\ &\\
=& \Bigl ([A_m, L^n] -[A_n, L^m] -[A_m, A_n] \Bigr )_+
\\ &\\
=& \Bigl ([A_m,\, L^n \! -\! A_n] -[A_n, L^m] \Bigr)_+
\\ &\\
=&\Bigl ([(L^m)_+, (L^n)_-] -[(L^n)_+, L^m] \Bigr )_+
\\ &\\
=&\Bigl ([L^m, (L^n)_-] +[L^m , (L^n)_+] \Bigr )_+ \, =\,
\Bigl ( [L^m , L^n ] \Bigr )_+ \, = \, 0.
\end{array}
\eeq
The converse is also true: from the full set of equations
(\ref{kp3}) the Lax equations follow. Clearly, equation
(\ref{kp3}) is equivalent to the commutation relation
$[\p_{t_m}-A_m , \, \p_{t_n}-A_n]=0$.

\noindent
{\bf Exercise.} Check that equations (\ref{kp3}) 
can be rewritten in the form
\beq\label{kp3-}
\p_{t_m}(L^n)_- -\p_{t_n}(L^m)_- +
[(L^m)_- , \, (L^n)_-]=0.
\eeq

By analogy with (\ref{mm4}), the representation of the KP hierarchy
in the form (\ref{kp3}) or (\ref{kp3-})
is called the zero curvature (or Zakharov-Shabat) representation.
Each of the zero curvature equations generates a closed system of a finite number
of equations for a finite number of unknown functions which, however, does not have
the evolution form. It contains derivatives with respect to three times: 
$x=t_1$, $t_m$, $t_n$.
(Unlike the matrix zero curvature equation (\ref{mm3}),
equations (\ref{kp3}) at $n=1$ are identities, so in non-trivial equations 
three times rather than two will participate.) 
If $n>m$, then we have a system of $n-1$ equations
for unknown functions $u_1, u_2, \ldots , u_{n-1}$.
These systems are called equations of the KP hierarchy. 

The simplest non-trivial example corresponds to the choice
$m=2$, $n=3$.
Denote $t_1 =x$, $t_2 =y$, $t_3 =t$, $u=2u_1$, $w=u_2$,
then from (\ref{kp3})
we obtain the system
$$
\left \{ \begin{array}{l}
4w_x=3u_{y}+3u_{xx},
\\ \\
u_{t}-w_{y}=\frac{3}{2}\, uu_x +u_{xxx}-w_{xx}.
\end{array}
\right.
$$
Excluding $w$, one obtains the closed equation for
$u$:
\beq\label{kp0}
3u_{yy}=\Bigl (4u_t -6uu_x -u_{xxx}\Bigr )_x,
\eeq
which is called the KP equation. It was suggested in 1970. 
In general case the system obtained from (\ref{kp3}) can not be reduced
to a single equation.

\noindent
{\bf Remark.}
In physical literature equation (\ref{kp0}) is called
KP2; the equation KP1 is obtained by the change $y\to iy$.
Properties of solutions and physical applications in these two cases are very
different. We will not discuss them since we are mainly interested in algebraic
structures.

\subsection{Symmetries and conservation laws}

Similarly to the KdV case, all equations of the KP hierarchy are symmetries for
each other. In other words, all vector fields
$\p_{t_j}$ commute and the quantities $u_i$
are functions not only on $x$ but also on all 
the times $t_j$ simultaneously.
This can be easily verified:
$$
\begin{array}{ll}
&\p_{t_m}(\p_{t_n} L)-\p_{t_n}(\p_{t_m} L)
\\ &\\
=&\p_{t_m}[A_n , L]-\p_{t_n}[A_m , L]
\\ &\\
=&[\p_{t_m}A_n -\p_{t_n}A_m,\, L]+
[A_n, \p_{t_m} L]-[A_m, \p_{t_n} L]
\\ &\\
=&[[A_m , A_n], L]+[A_n, [A_m, L]]-[A_m, [A_n, L]]\, =0.
\end{array}
$$
When passing to the last line, we have used the zero curvature equation. 
The expression in the last line vanishes identically after opening brackets
in the commutators.

One of the consequences of commutativity of the vector fields
$\p_{t_j}$ are the relations
\beq\label{kp4}
\p_{t_m}\mbox{{\rm res}}\, L^{n}=\p_{t_n}\mbox{{\rm res}}\, L^{m}
\eeq
which literally generalize the corresponding formulas for KdV. 
One can prove them independently by a calculation similar to
(\ref{kp5}) under the sign $\mbox{res}$ rather than $(\ldots )_+$:
$$
\p_{t_m}\mbox{{\rm res}}\, L^{n}-\p_{t_n}\mbox{{\rm res}}\, L^{m}
=\mbox{res}\, \Bigl (\p_{t_m}L^n -\p_{t_n}L^m -[A_m, A_n]\Bigr )
\quad \mbox{and so on.}
$$
Besides, these relations immediately follow from (\ref{kp3-}).

The integrals of motion are constructed in the same way as in the KdV case.
Now all the integrals are non-trivial (not only those with odd numbers). 

\noindent
{\bf Proposition.} {\it The quantities
\beq\label{kpLaxk}
I_{j}=\int \mbox{{\rm res}}\, L^{j}\, dx
\eeq
at $j\geq 1$ are integrals of motion for all equations of the KP
hierarchy: $\p_{t_n}I_j=0$.
}

\noindent
The proof is again based on the Lax equations:
$$
\p_{t_n} I_{j}=\int \mbox{{\rm res}}\, (\p_{t_n} L^{j})dx
=\int \mbox{{\rm res}} \left ([A_n, L^{j}]\right )dx.
$$
Since $\mbox{{\rm res}}\, [P,Q]$ is a full derivative for any $P,Q$, 
we conclude that in the rapidly decreasing and periodic cases
$\p_{t_n} I_{j}=0$.

Note that equation (\ref{kp4}) implies that densities of integrals of motion
are time de\-ri\-va\-ti\-ves of one and the same function $v$:
$\mbox{{\rm res}}\, L^{j}=\p_{t_j}v$. The fact that the residue
of commutator is a full derivative in $x=t_1$ implies that 
$v$ can be represented as derivative of some function
$f$:
$$
\p_{t_m}\mbox{res}\, L^n =\mbox{res} \, [ (L^m)_+ , \, L^n]
=\p_{t_m}\p_{t_n}v=\p_{t_1} \p_{t_m}\p_{t_n}f,
$$
and also
$$
\mbox{res}\, L^n =\p_{t_1} \p_{t_n}f.
$$
The function $f$ will play an important role. 
We will see that it is logarithm of the tau-function. 

\subsection{Dressing operator}

The Lax and zero curvature representations admit a nice reformulation
in terms of so-called 
{\it dressing operator} $K$. This operator could be introduced also for KdV
but the construction becomes really useful for the KP hierarchy which is free of
any constraints like $(L^N)_-=0$.

The dressing operator is the pseudo-differential 
operator of the form
$$
K= 1+ \xi _1 \p^{-1}+\xi _2 \p^{-2} +\ldots
$$
such that
\beq\label{kp7}
L=K\p K^{-1}.
\eeq
This equality is said to be the 
``dressing'' of the operator $\p$ by the operator $K$. The operator 
$L$ is the result of the dressing. Obviously,
$L^m = K \p^m K^{-1}$. Note that 
$K$ is defined up to multiplication from the right to an operator of the form
$1+\sum _{k\geq 1}a_k \p^{-k}$ with constant coefficients $a_k$.

\noindent
{\bf Proposition.} {\it Assume that the dressing operator satisfies the equations
\beq\label{kp8}
\p_{t_n}K = -(K\p^n K^{-1})_- \, K\,,
\eeq
then $L=K\p K^{-1}$ satisfies the Lax equations of the KP
hierarchy. 
}

\noindent
{\bf Exercise.} Prove this proposition by a direct
calculation and check that 
(\ref{kp8}) can be written in the form
$$
\p_{t_n}K = (L^n)_+ K -K \p^n.
$$

\noindent
By a similar calculation one can prove that the vector fields
$\p_{t_j}$ defined by
(\ref{kp8}) commute not only when they act to coefficients of the operator
$L$ but also when they act to coefficients of the operator $K$
(which, generally speaking, are not differential polynomials 
of $u_i$): $\p_{t_m}(\p_{t_n}K)=\p_{t_n}(\p_{t_m}K)$.

Since 
$$
K(\p_{t_m}\! -\! \p^m )K^{-1} =\p_{t_m} \! -\! (\p_{t_m}K)K^{-1}
\! -\! L^m = \p_{t_m} \! +\!  (K\p^m K^{-1})_{-}\! -\! L^m =\p_{t_m}-A_m,
$$
the Lax and zero curvature equations can be represented in the form
$$
\begin{array}{l}
K[\p_{t_m} -\p^m, \, \p]K^{-1}=0,
\\ \\
K[\p_{t_m} -\p^m, \, \p_{t_n} -\p^n]K^{-1}=0.
\end{array}
$$
One can say that they are obtained by dressing
the obvious relations
$[\p_{t_m} -\p^m, \, \p]=0$ and
$[\p_{t_m} -\p^m, \, \p_{t_n} -\p^n]=0$.

\subsection{Linear problems and Baker-Akhiezer function}

The zero curvature equation (\ref{kp3}) is the compatibility condition
of the linear problems
\beq\label{kplin1}
\left \{\begin{array}{l}
\p_{t_m} \psi =A_m\psi ,
\\ \\
\p_{t_n} \psi =A_n\psi .
\end{array}\right.
\eeq
As before, compatibility means existence of a large set of common solutions. 
The solution can be again found as a series in a spectral parameter
$z$, which now does not enter the linear problems explicitly. 
For brevity we denote
\beq\label{kpxi}
\xi ({\bf t}, z)=xz + t_2z^2  + t_3 z^3  + \ldots
\eeq
Let us find the solution of (\ref{kplin1}) in the form
\beq\label{kppsi}
\psi = \left ( 1+ \frac{\xi _1}{z}+\frac{\xi _2}{z^2}
+ \ldots \right )e^{\xi ({\bf t}, z)},
\eeq
where the coefficients $\xi _i$ depend only on $x$ (and on $t_j$).
One can add to the system the equation $L\psi = z\psi$,
which contains the spectral parameter explicitly.

It is easy to see that common solutions to the system
(\ref{kplin1}) are constructed by application of the dressing operator
$K$ to the function
$e^{\xi ({\bf t}, z)}$:
\beq\label{psiK}
\psi = K e^{\xi ({\bf t}, z)}=\left (1+\xi_1 \p^{-1}+
\xi_2 \p^{-2}+\ldots \right )e^{\xi ({\bf t}, z)}
\eeq
($\p^{-1}$ acts to the exponential function according to the rule
$\p^{-1}e^{xz}=z^{-1}e^{xz}$).
Indeed,
$$
\begin{array}{lll}
\p_{t_m}\psi &=& z^m \! Ke^{\xi ({\bf t},z)}+\p_{t_m}\! 
K e^{\xi ({\bf t},z)}=
(K\p^m -(K\p^m K^{-1})_- \, K)e^{\xi ({\bf t},z)}
\\ && \\
&=&(K\p^m K^{-1} -(K\p^m K^{-1})_-)Ke^{\xi ({\bf t},z)}=
(L^m-(L^m)_-)\psi =A_m\psi .
\end{array}
$$

Together with the dressing operator $K$ it is useful to consider the formally 
conjugate operator
$
K^{\dag}=1-\p^{-1}\xi _1 +\p^{-2}\xi _2 -\ldots
$
and construct the {\it adjoint Baker-Akhiezer function}
\beq\label{psiKa}
\psi ^{*} = (K^{\dag})^{-1} e^{-\xi ({\bf t}, z)}.
\eeq
It has the form
$$
\psi ^{*}= \left ( 
1+ \frac{\xi ^* _1}{z}+\frac{\xi ^* _2}{z^2}
+ \ldots \right )e^{-\xi ({\bf t}, z)}
$$
(here the star does {\it not} mean the complex
conjugation!) and satisfies the system of compatible linear problems
\beq\label{kplin1adj}
\left \{\begin{array}{l}
\p_{t_m} \psi ^{*}=-A_m^{\dag}\psi ^{*},
\\ \\
L^{\dag}\psi ^{*} =\, z\psi ^{*}.
\end{array}\right.
\eeq

\noindent
{\bf Problem.} Prove that in the KdV case ($(L^2)_- =0$)
$\psi ^* (z)=2z\psi (-z)/W(z)$, where $W$ is the Wronskian of the functions
$\psi (z)$ and $\psi (-z)$ (see (\ref{Wr})).

Let us prove a formula for $(L^m)_-$ through $\psi$ and $\psi ^{*}$: 
\beq\label{ff1}
(L^m)_- = {\rm res}_z \! \left (z^m \psi (z)\, \p^{-1}
\psi ^{*}(z)\right ).
\eeq
Here $\p^{-1}\psi ^{*}(z)$ in the right hans side is understood as composition 
of operators (and not as the result of action of
$\p^{-1}$ to $\psi ^{*}(z)$). Since we will deal with both ordinary and operator
residues, let us denote them by
${\rm res}_{z}$ and ${\rm res}_{\p}$ respectively
(${\rm res}_{z}$ is the coefficient in front of $z^{-1}$).
It is obvious that
$$
(L^m)_- = \sum_{l\geq 0}{\rm res}_{\p} \!
\left (L^m \p^l\right )\p^{-l-1}
=\sum_{l\geq 0}{\rm res}_{\p} \!
\left (K\p^m K^{-1}\p^l \right )\p^{-l-1}.
$$
Now, in order to transform the operator residue into
the ordinary one, we use the previously proved lemma which states that
for any two pseudo-differential operators
$P$, $Q$ the relation
$\mbox{res}_z \left [ (Pe^{xz})\, (Qe^{-xz})\right ]=
\mbox{res}_{\p} (PQ^{\dag})$ holds. We write, continuing the equality:
$$
(L^m)_- = \sum_{l\geq 0}{\rm res}_{z} \!
\left [ \left (K\p^m e^{\xi (t,z)}\right )
\left ( (-\p)^l  (K^{\dag})^{-1}e^{-\xi (t,z)}\right )
\right ] \p^{-l-1}
$$
(Here the operator $(-\p)^l$ acts to what stands from the right of it.) 
Rewriting the right hand side in terms of the Baker-Akhiezer function and its adjoint,
we will have: 
$$
(L^m)_- = \sum_{l\geq 0}{\rm res}_{z} \!
\left [(L^m \psi (z))(\p^l \psi ^{*}(z))\right ] \!
(-1)^l \p^{-l-1}
= \sum_{l\geq 0}{\rm res}_{z} \!
\left [z^m \psi (z)\, \p^l \psi ^{*}\! (z)\right ] \!
(-1)^l \p^{-l-1}.
$$
Finally, using the commutation relation
$\displaystyle{\p^{-1}\! f=\sum_{l\geq 0}(-1)^l f^{(l)}\p^{-l-1}}$,
we arrive at (\ref{ff1}). The reduction to KdV gives equation (\ref{Lchi}).

The key for construction of solutions to the KP equation is the following technical lemma
(we use the notation
$t_1 =x$, $t_2 =y$, $t_3 =t$, $u=2u_1$). 

\noindent
{\bf Lemma} {\it (simple but important)}.
{\it For the function $\psi$ of the form
\beq\label{psi1}
\psi = e^{zx+z^2 y+z^3t}
\left ( 1+ \frac{\xi _1}{z}+\frac{\xi _2}{z^2}
+ \ldots \right )
\eeq
the formal equalities
\beq\label{lemma2}
\begin{array}{r}
(-\p_y +\p^2 +u)\psi =O(z^{-1})e^{zx+z^2 y+z^3t},
\phantom{aaaaaaaaa}
\\ \\
(-\p_t +\p^{3}+ \frac{3}{2}u\p +w)
\psi =O(z^{-1})e^{zx+z^2 y+z^3t}\phantom{aaaaaa}
\end{array}
\eeq
hold, where the functions $u,w$ are found from the condition that coefficients
at non-negative powers of $z$ vanish:
\beq\label{uw}
\begin{array}{l}
u=-2 \xi _{1,x},
\\ \\
w= 3\xi_1 \xi _{1,x}-3\xi _{1,xx}-3\xi _{2,x}.
\end{array}
\eeq
}

\noindent
The meaning of this lemma is similar to the corresponding
statement about the 
$\psi$-function of the KdV equation. 
The proof is a direct calculation. 

This lemma can be generalized in two directions:
a) one may allow the function 
(in fact the formal series) $\psi$ to have, on the background of the essential
singularity, a pole at 
$\infty$ of order $n$, б) the lemma can be extended to the whole hierarchy.

\noindent
{\bf Lemma} {\it (generalization of the previous one)}.
{\it For the function $\psi$ of the form
\beq\label{psi101}
\psi = z^n Ke^{\xi ({\bf t},z)}=
\left ( 1+ \frac{\xi _1}{z}+\frac{\xi _2}{z^2}
+ \ldots \right )z^ne^{\xi ({\bf t},z)}
\eeq
built with the help of the dressing operator 
$K=1+\xi_1 \p^{-1}+\xi_2 \p^{-2}+\ldots  \,$,
the formal equalities
\beq\label{lemma201}
(\p_{t_k} -A_k)\psi =O(z^{n-1})e^{\xi ({\bf t},z)}
\eeq
hold, where the coefficient functions of the differential operators
$A_k$ are differential po\-ly\-no\-mi\-als of
$\xi_1, \xi_2, \ldots \,$. Their explicit form is determined from the equalities
$A_k= (K\p^k K^{-1})_+$.
}

\noindent
The proof which uses the technique of dressing operators is very simple. 
We have:
$$
(\p_{t_m} -A_m)\psi =\Bigl ((\p_{t_m} K)K^{-1} +z^m -A_m\Bigr )\psi
=\Bigl ((\p_{t_m} K)K^{-1} +L^m -(L^m)_+\Bigr )\psi .
$$
It is clear that the operator
$(\p_{t_m} K)K^{-1} \! +\! 
L^m \! -\! (L^m)_+=(\p_{t_m} K)K^{-1} +(L^m)_-$
has the form $O(\p^{-1})$, so acting to the exponential function it gives
the factor
$O(z^{-1})$.

\subsection{Solutions of the KP hierarchy}

\subsubsection{Soliton solutions}

The construction of soliton solutions is similar to the corresponding
construction for KdV. 
Consider the linear space of functions
$\psi =\psi (z)$ meromorphic everywhere except at infinity and such that
\begin{itemize}
\item[a)]
The function $\psi e^{-zx -z^2 y -z^3t}$ is regular at $z=\infty$;
\item[b)]
The function $\psi$ has not more than $N$ poles (counted with multiplicities)
at some marked points of the complex plane and holomorphic everywhere else except 
$\infty$;
\item[c)]
For $N$ pairs of distinct points $p_j , q_j \in \CC$ the relations
$\psi (p_j)=\alpha _j\psi (q_j)$, $j=1, 2, \ldots , N$ hold.
\end{itemize}
Similarly to the KdV case, this space is one-dimensional and the operators
in the left hand sides of
(\ref{lemma2}) preserve it. For simplicity we consider the case when all poles 
are concentrated at
$z=0$. The function $\psi$ can be found in the form
$$
\psi =e^{\xi ({\bf t},z)} \left (1 + \frac{\xi_1}{z}
+\frac{\xi_2}{z^2} +\ldots +\frac{\xi_N}{z^N}\right ),
$$
then $u=-2 \xi _{1,x}$
is a solution to the KP equation.

Let us find this solution explicitly. The conditions
$\psi (p_j)=\alpha _j\psi (q_j)$ are equivalent to the following
system of linear equations for $\xi_j$:
$$
\sum _{j=1}^{N}M_{ij}\, \xi _j = - M_{i0},
$$
where
$
M_{ij}=p_{i}^{-j}e^{\xi ({\bf t},p_i )}-\alpha _i
q_{i}^{-j}e^{\xi ({\bf t},q_i )}
$.
The Kramer's rule yields
$$
\xi_1 = -\,\frac{\det M_{ij}^{(0)}}{\det M_{ij}}\,,
\quad \quad i,j =1,2,\ldots , N,
$$
where the matrix $M_{ij}^{(0)}$ differs from $M_{ij}$ by the change
of the first column $M_{i1}$ to $M_{i0}$.
Since $\p_x M_{ij}=M_{i, j-1}$, we have $\det \! M_{ij}^{(0)} =
\p_x \det \! M_{ij}$, and $\xi_1 = -\p_x \log \det M_{ij}$, so that
$
u=2\p_x^2 \log \det M_{ij}
$. We have obtained a family of solutions which are expressed by the same formula
(\ref{uN}) with the tau-function
\beq\label{kptauN}
\tau=
\det_{1\leq i,j \leq N} \left (
p_{i}^{-j}e^{\xi ({\bf t}, p_i )}
-\alpha_i q_i^{-j}e^{\xi ({\bf t}, q_i )}
\right ).
\eeq
As before, we have got solutions of the whole hierarchy. 
All $\alpha _i$'s can be again ``hidden'' in initial values
of the times $t_j$, so the real parameters are only
$p_i$ and $q_i$ ($2N$ parameters).

At $N=1$ we obtain one-soliton solution:
$\tau = p^{-1}e^{\xi ({\bf t},p)}-\alpha q^{-1}e^{\xi ({\bf t},q)}$,
$$
u=-\, \frac{2pq(p-q)^2 e^{\xi ({\bf t},p)+
\xi ({\bf t},q)}}{(qe^{\xi ({\bf t},p)}-
\alpha pe^{\xi ({\bf t},q)})^2}=-\, 
\frac{(p-q)^2}{2\sinh ^2 \left (\frac{1}{2}
(\xi ({\bf t},p)-\xi ({\bf t},q)+\varphi )\right )},
$$
where $\varphi =\log\left (\frac{q}{\alpha p}\right )$.
Putting $\alpha = -q/p$, $t_4=t_5=\ldots =0$, we write this solution in the form 
\beq\label{solitonKP}
u(x,y,t)=\frac{(p-q)^2}{2\cosh ^2 \left (
\frac{1}{2}(p\! -\! q)x +\frac{1}{2}(p^2\! -\! q^2)y 
+\frac{1}{2}(p^3\! -\! q^3)t\right )}.
\eeq
In the case $q=-p$ the dependence on $y=t_2$ (and on all even times) disappears, 
and this formula reproduces the one-soliton solution of the KdV equation
(\ref{solit1}). Note that
$u(x,y,t)$ defined by (\ref{solitonKP}) exponentially decreases in the plane
$(x,y)$ in all directions except the direction along the line
$x+(p+q)y=0$, and so it can not be regarded as a 2D soliton in the physical sense. 
For this reason the solutions found in this section are sometimes called
soliton-like.  

We point out the equivalent determinant representation of the multi-soliton 
tau-function:
\beq\label{kptauN1}
\tau =\det_{1\leq i,j \leq N} \left (\delta_{ij}+
\frac{\beta_i (p_i -q_i)}{p_i -q_j}\,
e^{\xi ({\bf t},p_i )-\xi ({\bf t},q_i )}
\right ).
\eeq
Expanding the determinant, using the known expression for the Cauchy determinant
$$
\det_{i,j=1, \ldots , N}\frac{1}{p_i-q_j}=
\frac{\prod\limits_{1\leq i<j\leq N}
(p_i-p_j)(q_j-q_i)}{\prod\limits_{1\leq i,j\leq N}(p_i-q_j)}
$$
(which can be proved by via comparing analytic properties of rational functions
in the both sides), we get: 
\beq\label{kptauN2}
\tau =\sum_{\{\epsilon_1 , \ldots , \epsilon_N\}\in \z_{2}^{N}}
\,\,\, \prod_{i<j}^{N}
\left (\frac{(p_i -p_j)(q_j-q_i)}{(p_i-q_j)(p_j-q_i)}
\right )^{\epsilon_i \epsilon_j}\prod_{k=1}^{N}
\left (\beta_k
e^{\xi ({\bf t},p_k )-\xi ({\bf t},q_k)}\right )^{\epsilon_k}.
\eeq
The summation goes over all sets of
$N$ numbers
$\epsilon_i =0,1$, so that the sum contains
$2^N$ terms.
For example, at $N=2$ we have:
$$
\tau = 1+ \beta_1 e^{(p_1 -q_1)x}+ \beta_2 e^{(p_2-q_2) x}+
\beta_1 \beta_2
\frac{(p_1 -p_2)(q_2-q_1)}{(p_1-q_2)(p_2-q_1)}
e^{(p_1+ p_2-q_1-q_2) x},
$$
where for simplicity only the terms containing $x$ are left.

\subsubsection{Soliton-like solutions of general form and solutions
depending on func\-ti\-o\-nal parameters}

The construction described above can be extended to a much larger class 
of solutions. 
For this it is enough to notice that all arguments go through if 
instead of $N$ linear conditions $\psi (p_i)=\alpha _i \psi (q_i)$ 
for coefficients of the $\psi$-function one imposes more general conditions
$\sum_{l}A_{il} \psi (p_{i}^{(l)})=0$ with some
(generally speaking, rectangular) matrix $A$.

A more precise formulation is as follows.  
\begin{itemize}
\item[a)]
Fix some points $p_{m}^{(j)}\in \CC$ which are assumed to be distinct. 
It is convenient to regard them as divided in  
${\cal N}_c$ ``clusters'', so that $j=1, 2, \ldots , {\cal N}_c$ 
numbers the clusters, and $m$ numbers points inside the clusters. Let the number
of pints in $j$th cluster be $M_j>1$. Formally the case 
$M_j=1$ is also possible but not very interesting. To each cluster
(with the number $j$) assign a rectangular matrix $A^{(j)}_{\alpha , m}$,
where $m=1, \ldots , M_j$ and $\alpha =1, \ldots , \mu_j$ 
with some $1\leq \mu_j \leq M_j$. Put 
$\displaystyle{N=\sum_{j=1}^{{\cal N}_c}\mu_j}$. This $N$ has the meaning 
of the number of solitons in the solution. 
\item[b)]
Fix $N$ distinct points $z_1, z_2, \ldots , z_N\in \CC$ 
not coinciding with $p_{m}^{(j)}$ and consider the Baker-Akhiezer function
of the form
$$
\psi ({\bf t},z)= \left ( 1 +\sum_{k=1}^N \frac{w_k({\bf t})}{z-z_k}\right )
e^{\xi ({\bf t}, z)},
$$
with $N$ linear conditions
$$
\sum_{m=1}^{M_i}A^{(j)}_{\alpha , m}\psi ({\bf t}, p_{m}^{(j)})=0
$$
(where $j=1, \ldots , {\cal N}_c, \quad \alpha = 1, \ldots , \mu_j$).
\end{itemize}
\noindent
It is easy to see that the lemma can be applied to this more general situation,
and we obtain a large class of solutions which are sometimes also called soliton-like.
If the number of points in the clusters is greater that 2, these solutions do not
have analogues in the KdV theory. 

One may generalize this even further and impose the conditions of the form
$$\int \rho _i (p) \psi (p)d\mu (p)=0$$ with some measure 
$d\mu$ and functions $\rho_i(p)$ in the complex plane. 
More precisely, consider the linear space
of functions $\psi =\psi (z)$ with the same analytic properties
a), b) as above but instead of condition c) impose
\begin{itemize}
\item[c$'$)]
For $N$ different fixed functions $\rho _i(z)$ and some measure
$d\mu (z)$ in the complex plane
the relations
$$
\int \rho _i(z)\psi (z) d\mu (z) =0\,, \quad \quad
i=1, 2, \ldots , N
$$
hold. 
\end{itemize}
It is easy to see that this space is one-dimensional in the general case
and the operators in the right hand sides of
(\ref{lemma2}) preserve it. The procedure again reduces to solving a linear system
with the result
\beq\label{functau}
\tau ({\bf t})= \det_{1\leq i,j\leq N}\int z^{-j}\rho_i(z)
e^{\xi ({\bf t},z)}d\mu (z).
\eeq
It is assumed that the measure is such that the integral converges. 
The functions $\rho_i(z)$ (and the measure $d\mu$) are functional parameters
of the solution. Note that such solutions exist only for the KP hierarchy.
In the KdV case the general linear condition of the type
c$'$) is broken under the action of the operators in the left
hand sides of (\ref{lemma1}).

It is clear that the soliton-like solutions correspond to the case when 
$\rho _i (z)$ is a linear combination of  
$\delta$-functions concentrated in the points $p_m^{(j)}$.

\subsubsection{Rational solutions}

Tending the points in each cluster to each other and choosing 
$A^{(j)}_{\alpha , m}$ in a speciial way, in the limit one can obtain solutions
whose tau-function is a polynomial of $x$
and all $t_j$, and $u$ is thus a rational function. 
Such solutions are called rational. They can be constructed in the same way as 
soliton-like solutions if one considers the linear space of functions
$\psi =\psi (z)$ with the same analytic properties
a) and b) and 
\begin{itemize}
\item[c$''$)]
In $N$ distinct points $p_i$ of the complex plane
the relations
$$
\sum_{m=0}^{M_i}a_{im}\p_z^m \psi (z) \Bigr |_{z=p_i}
=0\,, \quad \quad
i=1, 2, \ldots , N
$$
hold for all $t_j$. 
\end{itemize}
Here $a_{im}$ are some constants (parameters of the solution together with the points
$p_i$). In general some points $p_i$ may coincide which means that more than one
condition is imposed in one point. 

A physically meaningful example of such solution can be obtained by putting
$q_1 =p_1 +\epsilon$, $q_2 =p_2 +\epsilon$, 
$\beta_1 =\beta_2 =-1$ in the two-soliton solution and tending
$\epsilon \to 0$. Setting $p_1 =p$, $p_2 =-\bar p$ and taking
$y$ to be purely imaginary (i.e. redefining $y\to iy$ thus passing from KP2 to KP1),
we obtain in the limit
$$
\tau = |x+2ipy +3p^2 t|^2 +\frac{1}{(p+\bar p)^2}.
$$
This tau-function describes the motion of a bell-shaped excitation in the plane
$(x,y)$. This is a two-dimensional soliton 
in the physical sense. The equation KP2 does not have solutions of this type.

\noindent
{\bf Problem.} Find the linear conditions for the $\psi$-function of the form c$''$) 
for this solution.

Let us give an explicit description of rational solutions which are obtained
after imposing the condition c$''$) to the Baker-Akhiezer function
\begin{equation}\label{rat1}
\psi ({\bf t},z)=z^N e^{\xi ({\bf t},z)}\left (1+\frac{\xi_1({\bf t})}{z}
+\ldots + \frac{\xi_N({\bf t})}{z^N}\right )
\end{equation}
(in this case it is convenient to place all poles at $\infty$).
The conditons c$''$) are equivalent to the system of linear equations
\begin{equation}\label{rat3}
A_{i}(N,{\bf t})+\sum_{k=1}^{N}A_{i}(N-k,{\bf t})\xi_k ({\bf t})=0,
\end{equation}
where
\begin{equation}\label{A}
\left. A_{i}(n,{\bf t})=\sum_{m=0}^{M_i}a_{im}
\p_{z}^{m}
\left (z^{n}e^{\xi ({\bf t},z)}\right ) \right |_{z=p_i}.
\end{equation}
These functions are polynomials of 
$t_j$ multiplied by 
exponential factors $e^{\xi ({\bf t}, p_i)}$.
Solving the system by the Kramer's rule, 
we write the answer for $\psi$:
\begin{equation}\label{rat5}
\psi ({\bf t},z)= e^{\xi ({\bf t},z )}\, (\tau ({\bf t}))^{-1}
\left |\begin{array}{cccc}
z^N&z^{N-1}&\ldots & 1\\
A_{1}(N,{\bf t})& A_{1}(N\! -\! 1,{\bf t})& \ldots &
A_{1}(0,{\bf t})\\
\vdots & \vdots &\ddots & \vdots \\
A_{N}(N,{\bf t})&A_{N}(N\! -\! 1,{\bf t})& \ldots
&A_{N}(0,{\bf t})
\end{array}\right |,
\end{equation}
where
\beq\label{rat5a}
\tau ({\bf t}) = \left |\begin{array}{ccc}
A_{1}(N-1,{\bf t})&\ldots & A_{1}(0,{\bf t})\\
\vdots &\ddots & \vdots \\
A_{N}(N-1,{\bf t})& \ldots & A_{N}(0,{\bf t})
\end{array}\right | =\det_{1\leq i,j\leq N} A_i(N-j, {\bf t}).
\eeq
The result of the next problem implies that 
$\tau$ is the tau-function for this class of solutions and 
$u=-2\xi_1'=2\p^2 \log \tau$.

\noindent
{\bf Problem.} Check that $\xi_1 ({\bf t})=-\p_{t_1} \log \tau ({\bf t})$
(hint: use the identity
$\p_{t_1}A_i(n,{\bf t})=A_i(n+1,{\bf t})$).

A further degeneration of rational solutions can be obtained by
merging the points $p_i$ and tending them to $0$. 
The solutions remain rational and admit an explicit description 
in terms of Schur polynomials.

Let us define the polynomials $h_j = h_j({\bf t})$ of the times
$t_i$ using the expansion
\beq\label{schur}
e^{\xi ({\bf t},z)}=\sum_{l=0}^{\infty} h_l ({\bf t})z^l.
\eeq
These are the simplest Schur polynomials. 
For example, $h_0 =1$, $h_1 =t_1$, $h_2 = t_2 + \frac{1}{2}t_1^2$,
$h_3 = t_3 + \frac{1}{2}t_1 t_2 +\frac{1}{6}t_1^3\, $ and so on. 
Here is the general formula:
\beq\label{schur1}
h_k ({\bf t})=\sum_{k_1+2k_2+\ldots =k}\frac{t_1^{k_1}}{k_1!}\, 
\frac{t_2^{k_2}}{k_2!}\ldots \,\,\, = \,\,\,
\sum_{l=1}^{k}\frac{1}{l!}
\sum_{{k_1, \ldots , k_l\geq 1}\atop{k_1+\ldots +k_l=k}}
\!\!\! t_{k_1}\ldots t_{k_l}.
\eeq
It is convenient to put $h_{k}({\bf t})=0$ at $k<0$.

\noindent
{\bf Problem.} Prove the identities
\beq\label{schur2}
\hspace{-2.5cm} \p_{t_j}h_k({\bf t})=h_{k-j}({\bf t}),
\eeq
\beq\label{schur3}
\sum_{j=1}^{n}jt_j h_{n-j}({\bf t})=nh_n({\bf t}), \quad n\geq 1.
\eeq
It turns out that any polynomial $h_j({\bf t})$ is the tau-function 
of the KP hierarchy, i.e. $u ({\bf t})=2\p^2_{t_1} \log h_j ({\bf t})$ is a solution.
The general Schur polynomials can be defined using Young diagrams
$\lambda$ of arbitrary form. A Young diagram is a set of positive integer numbers
$\lambda _1 , \lambda _2 , \ldots \, , \lambda _n$ such that
$\lambda _1 \geq \lambda_2 \geq \ldots \geq \lambda_n$
($\lambda_i$ are lengths of rows of the diagram $\lambda$). 
The Schur polynomial corresponding to the Young diagram $\lambda$ is
\beq\label{schur4}
s_{\lambda}({\bf t})=\det_{1\leq i,j\leq n} h_{\lambda_i -i+j}({\bf t}).
\eeq
The polynomial
$h_j$ corresponds to the diagram consisting of just one row of length
$j$.

All Schur polynomials $s_{\lambda}({\bf t})$ are tau-functions
of the KP hierarchy. 
They are maximally degenerate solutions. Their characterization 
with the help of the Baker-Akhiezer function 
$\psi$ of the form (\ref{rat1}) is as follows. 
Fix a sequence of non-negative integer numbers 
$n_i$ such that $n_1>n_2>\ldots >n_{N-1}\geq 0$
and impose $N$ conditions
$$
\p_z^{n_i} \psi ({\bf t},z)\Bigr |_{z=0}=0.
$$
They are equivalent to the following system of linear equations for the coefficients
$\xi_j$:
\beq\label{linsyst}
\sum_{j=1}^{N} h_{n_i +j -N}({\bf t})\, \xi_j({\bf t})=-
h_{n_i -N}({\bf t})\,, \quad i=1, \ldots , N.
\eeq
Instead of $n_i$ introduce the sequence $\lambda_i$
according to the formula
$
\lambda_i =n_i+i -N.
$
Obviously, 
$\lambda _1 \geq \lambda_2 \geq \ldots \geq \lambda_N\geq 0$,
so that they define a Young diagram.  
In terms of the numbers $\lambda_i$ the system is rewritten as
$$
\sum_{j=1}^{N} h_{\lambda_i -i+j}({\bf t})\, \xi_j({\bf t})=-
h_{\lambda_i-i }({\bf t})\,, \quad i=1, \ldots , N.
$$
The Kramer's rule yields
\beq\label{linsyst1}
\xi_1 ({\bf t})=-\, \frac{\left |
\begin{array}{llll}
h_{\lambda_1-1} & h_{\lambda_1+1} & \ldots & h_{\lambda_1-1+N}
\\
h_{\lambda_2-2} & h_{\lambda_2} & \ldots & h_{\lambda_2-2+N}
\\
\ldots & \ldots & \ldots & \ldots
\\
h_{\lambda_N-N} & h_{\lambda_N-N+2} & \ldots & h_{\lambda_N}
\end{array}\right |}{\left |
\begin{array}{llll}
h_{\lambda_1} & h_{\lambda_1+1} & \ldots & h_{\lambda_1-1+N}
\\
h_{\lambda_2-1} & h_{\lambda_2} & \ldots & h_{\lambda_2-2+N}
\\
\ldots & \ldots & \ldots & \ldots
\\
h_{\lambda_N-N+1} & h_{\lambda_N-N+2} & \ldots & h_{\lambda_N}
\end{array}\right |}\,
=\, -\p_x \log \det_{i,j=1, \ldots , N}h_{\lambda_i-i+j}({\bf t}).
\eeq

\noindent
{\bf Problem.} Give a detailed derivation of formulas (\ref{linsyst}) and
(\ref{linsyst1}).

\subsubsection{Dynamics of poles of rational solutions to the KP equation
and the Calogero-Moser system of particles}

As we have seen, the tau-function for rational solutions of the KP equation
is a polynomial in 
$x=t_1$:
$$
\tau = C\prod_{j=1}^{N}(x-x_j ({\bf t}))
$$
Its roots $x_j$ (assumed to be distinct) are poles of the function
\beq\label{dyn1}
u({\bf t})=
2\p_x^2 
\log \tau ({\bf t})=-\sum_{j=1}^{N}\frac{2}{(x-x_j ({\bf t}))^2}\,.
\eeq
They are functions of $t_2, t_3, \ldots \, $.
(The construction of rational solutions given above implies that these solutions
are rational functions of all the times.) The problem about dynamics of 
poles of rational solutions to the KP equation was solvd by Krichever in 1978. 
It turns out that the dynamical equations for 
$x_j$ coincide with equations of motion of the integrable $N$-body system 
on the line with pairwise interaction. The interaction potential
is proportional to the inverse square of the distance between particles.
This is the (rational) Calogero-Moser system. 

The method suggested by Krichever consists in the substitution of the pole
expression
(\ref{dyn1}) for $u$ not into the nonlinear equation 
but into the auxiliary linear problem for the Baker-Akhiezer function $\psi$.
This allows one to separate the variables from the very beginning and to obtain the
Lax representation for the Calogero-Moser system. 
For this, one should derive or guess the corresponding ''pole ansatz'' for
$\psi$.

Let us describe the main points of the derivation of equations of motion for
$x_j$. For simplicity we will follow only the dynamics with respect to the time
$t_2$, which in this section will be denoted by $t$. All higher times will be put
equal to 0. 
The linear equation for $\psi$ has the form
\beq\label{dyn2}
\p_t \psi = \p_x^2 \psi +u \psi ,
\eeq
where $u$ is the sum of pole terms (\ref{dyn1}).
The function $\psi$ will be found in the form
\beq\label{CM4}
\psi = e^{xz+t z^2}\left (c_0 (z) +
\sum_{i=1}^{N}\frac{c_i(z,t)}{x-x_i (t)}\right )
\eeq
with some $x$-independent coefficients $c_i$.
The fact that it should have simple poles at the points
$x_i$ follows from (\ref{rat5}) (the denominator of this expression is
the tau-function).
Substituting $\psi$ in this form into (\ref{dyn2}), we obtain
$$
e^{-xz-t z^2}\bigl (\p_t -\p_x^2\bigr )\left [ e^{xz+t z^2}\Bigl (c_0  +
\sum_{i=1}^{N}\frac{c_i}{x-x_i}\Bigr )\right ]
+2 \Bigl (\sum_{i=1}^{N}\frac{1}{(x-x_i)^2}\Bigr )\Bigl (c_0  +
\sum_{i=1}^{N}\frac{c_i}{x-x_i}\Bigr )=0.
$$
The left hand side is a rational function of
$x$ with first and second order poles at $x=x_i$ (possible poles of third order
cancel identically), which is equal to 0 at infinity.
Therefore, it is enough to cancel all the poles. 
Equating the coefficients at each pole to zero, we obtain the following
system of
$2N$ linear equations for the coefficients 
$c_1, \ldots , c_N$:
$$
\left \{
\begin{array}{l}\displaystyle{
(\dot x_i +2z)c_i +2\sum_{k\neq i}\frac{c_k}{x_i-x_k}=-2c_0
\quad \mbox{(cancellation of second order poles)}},
\\ \\
\displaystyle{\dot c_i  +2c_i \sum_{k\neq i}\frac{1}{(x_i-x_k)^2}
-2\sum_{k\neq i}\frac{c_k}{(x_i-x_k)^2}=0
\quad \mbox{(cancellation of first order poles),}}
\end{array}
\right.
$$
where $i=1, \ldots , N$. The coefficient $c_0$ can be put equal to a constant
(for example, 1), since it affects only the common factor of the
$\psi$-function.
These equations can be compactly written in the matrix form:
\beq\label{CM5}
\left \{
\begin{array}{l}
({\cal L}-2zI ){\bf c}=c_0 (z){\bf 1},
\\ \\
{\bf \dot c}={\cal M}{\bf c},\end{array}
\right.
\eeq
where $I$ is the unity matrix, 
${\bf c}=(c_1, c_2, \ldots , c_N)^{T}$,
${\bf 1} =(1, 1, \ldots , 1)^{T}$ are $N$-component
vectors and $N \! \times \!N$ matrices ${\cal L}={\cal L}(t)$, ${\cal M}={\cal M}(t)$ 
have the form
\beq\label{CM6}
{\cal L}_{ik}= -2p_i \delta_{ik}-
2\frac{1-\delta_{ik}}{x_i-x_k}\,, \quad \quad
p_i:=\frac{1}{2}\, \dot x_i,
\eeq
\beq\label{CM7}
{\cal M}_{ik}=-\delta_{ik}\sum_{j\neq i}\frac{2}{(x_i-x_j)^2}+
\frac{2(1-\delta_{ik})}{(x_i-x_k)^2}.
\eeq
The system (\ref{CM5}) is overdetermined. 
The compatibility condition is
\beq\label{CM8}
\dot {\cal L}=[{\cal M}, \, {\cal L}].
\eeq
A calulation shows that the non-diagonal elements of the matrices
on the left and right hand sides equal identically while equality
of diagonal elements yields the equations
\beq\label{CM9}
\ddot x_i = -8\sum_{j\neq i} \frac{1}{(x_i-x_j)^3}.
\eeq
These are equations of motion of the Calogero-Moser system.
The matrices ${\cal L}, {\cal M}$ form the Lax pair; ${\cal L}$ is the
Lax matrix for this system. 
As it is seen from
(\ref{CM8}), the time evolution preserves the spectrum of the Lax matrix.
The coefficients ${\cal J}_k$ of its characteristic po\-ly\-no\-mial
\beq\label{char-pol}
\det \Bigl (2zI -{\cal L}(t)\Bigr )=\sum_{k=0}^{n}{\cal J}_k z^{n-k}
\eeq
are integrals of motion.

Introduce the matrix ${\cal X}=\mbox{diag}\, (x_1, x_2, \ldots , x_N)$.
It is easy to check that the matrices ${\cal X}, {\cal L}$ satisfy the commutation
relation
\beq\label{comm}
[{\cal X}, \, {\cal L}]=I -{\bf 1}\otimes {\bf 1}^{T}
\eeq
(here ${\bf 1}\otimes {\bf 1}^{\sf t}$ is the $N\! \times \! N$ matrix
of rank $1$ with all elements equal to $1$).

The function $\psi$ (\ref{CM4}) is found from the solution
of the linear equatios 
(\ref{CM5}) in the following form:
\beq\label{cc1}
\psi =c_0(z)\, e^{xz+tz^2}\Bigl (
1-{\bf 1}^{T}(x -{\cal X}(t))^{-1}(z -{\cal L}(t))^{-1}{\bf 1}\Bigr ).
\eeq

The equations of motion of the Calogero-Moser system 
cam be written in the Ha\-mil\-to\-ni\-an form
\beq\label{CM10}
\left (\begin{array}{l}\dot x_i \\ \dot p_i \end{array} \right )=
\left (\begin{array}{r} \p_{p_i} {\cal H}_2 \\
- \p_{x_i} {\cal H}_2 \end{array} \right )
\eeq
with the canonical variables $p_i, x_i$ and the Hamiltonian
\beq\label{CM11}
{\cal H}_2= \frac{1}{4}\, \mbox{tr}\, {\cal L}^2 = \sum_i p_i^2 -
\sum_{i<j}\frac{2}{(x_i-x_j)^2}.
\eeq
The connection of the pole dynamics with the Calogero-Moser system
can be extended to the whole KP hierarchy
(Shiota's result). Namely, the higher times dynamics is given by the 
Hamitonian equations
\beq\label{CM12}
\left (\begin{array}{l}\p_{t_k}x_i \\ \p_{t_k} p_i \end{array} \right )=
\left (\begin{array}{r} \p_{p_i} {\cal H}_k \\
- \p_{x_i} {\cal H}_k \end{array} \right ),
\quad \quad {\cal H}_k =2^{-k}\mbox{tr}\, {\cal L}^k,
\eeq
where ${\cal H}_k$ are higher integrals of motion (Hamiltonians)
of the Calogero-Moser system.  
It can be shown that they are in involution. This is in agreement with
commutativity of the KP flows. 

Note that in the case of KdV the correspondence with the Calogero-Moser system
means that the particles in the system with the Hamiltonian
${\cal H}_2$ do not move, i.e. they are in equilibrium with respect to
this dynamics while the $t_3$-dynamics is generated by the Hamiltonian 
${\cal H}_3$. The set of all such equilibrium positions is called 
{\it locus}. For rational solutions the locus is not empty only if
$N=n(n+1)/2$ with positive integer $n$. 

\noindent
{\bf Problem.} Solve the KdV equation $4u_t=6uu_x+u_{xxx}$ with the 
initial condition $u(x, 0)=-6/x^2$. 

Finally, we give an explicit determinant formula for the tau-function.
Let ${\cal X}_0 ={\cal X}(0)$ be the diagonal matrix
${\cal X}_0=\mbox{diag}(x_1(0), x_2(0), \ldots , x_n(0))$
and let ${\cal L}_0$ be the Lax matrix (\ref{CM6}) taken at
${\bf t}=0$. It can be shown that the tau-function
is given by the formula
\beq\label{CM13}
\tau ({\bf t})=
\det_{N\times N} \Bigl ( xI -{\cal X}_0 +\sum_{k\geq 1}kt_k {\cal L}_0^{k-1}\Bigr ).
\eeq

\subsubsection{Trigonometric solutions}

Now we come back to the non-degenerate soliton-like solutions. 
Let all
$p_i$, $q_i$ be real. Then the constructed solutions
$u(x)$ exponentially decrease as 
$x\to \pm \infty$ along the real axis and oscillate alng the 
imaginary axis. At $N>1$ the solution in general does not have any definite
period along the imaginary axis
(because the numbers $p_i -q_i$ and $p_j -q_j$ are in general incommensurate).
Nevertheless, among $N$-soliton solutions of the KP equation
there is an $N$-parametric family of solutions such that they have
a period $2\pi L$ along the imaginary axis. For this it is enough for the parameters
$p_i, q_i$ to be connected by the constraints
$q_i = p_i + 2\pi/L$ which make the number of free parameters equal to
$N$ (not counting the period itself).
Equation (\ref{kptauN2}) then implies that the tau-function
as a function of $x$ will be a polynomial of
$e^{2\pi x/L}$ of degree $N$. Multiplying it by a inessential common factor,
one can represent it in the form
\beq\label{factor}
\tau = C\prod_{j=1}^{N}\mbox{sinh}\, \frac{\pi (x-x_j (t))}{L},
\eeq
where the zeros $x_j$depend on all the times starting from $t_2$:
$x_j = x_j (t_2, t_3, \ldots )$. For $u$ we get the pole expansion
\beq\label{factor1}
u=-2\sum_{j=1}^{N} \frac{(\pi /L)^2}{\mbox{sinh}^2 \, \pi (x-x_j({\bf t}))/L}.
\eeq
We call solutions of this form trigonometric. 

The remarkable connection with the Calogero-Moser model discussed above
is ge\-ne\-ra\-li\-zed to the trigonometric solutions 
(and even to elliptic solutions, see below).
If the function of the form
(\ref{factor1}) satisfies the KP equation, then its poles
$x_j$ as functions of $t_2$ move according to the equations of motion
of the system of $N$ particles on the line with the Hamiltonian
\beq\label{CMtrig}
H_2=\sum_{j=1}^{N}p_j^2 - 2\sum_{i<j}^{N}
\frac{(\pi /L)^2}{\mbox{sinh}^2 \, \pi (x_i-x_j)/L}
\eeq
which is the trigonometric version of the Calogero-Moser system (the Sutherland
model).

\noindent
{\bf Problem.} Solve the KdV equation $4u_t=6uu_x+u_{xxx}$ with the 
initial condition
$$
u(x, 0)=\frac{6p^2}{\cosh ^2 px}.
$$

\subsubsection{Elliptic solutions}

Elliptic (double periodic in the complex plane) solutions 
to the KP equation were studied by Krichever in 1980 who showed 
that their
poles move according to the equations of motion of the Calogero-Moser system with
the elliptic potential. 

The double periodicity means the existence of two periods $2\omega , \, 2\omega ' \in \CC$
such that ${\rm Im} (\omega '/ \omega )>0$: $u(x+2\omega )=u(x)$, 
$u(x+2\omega ' )=u(x)$. 
For such solutions the tau-function is an 
``elliptic quasi-polynomial'' in the variable $x$:
\beq\label{ell1}
\tau = e^{Q(x, t_2, t_3, \ldots )}\prod_{i=1}^{N}\sigma (x-x_i ({\bf t}))
\eeq
where $Q(x, t_2, t_3, \ldots )$ is a quadratic form in the times $t_i$ and
$$
\sigma (x)=\sigma (x |\, \omega , \omega ')=
x\prod_{s\neq 0}\Bigl (1-\frac{x}{s}\Bigr )\, e^{\frac{x}{s}+\frac{x^2}{2s^2}},
\quad s=2\omega m+2\omega ' m' \quad \mbox{with integer $m, m'$},
$$ 
is the Weierstrass 
$\sigma$-function with the quasi-periods $2\omega$, $2\omega '$.
It is connected with the Weierstrass 
$\zeta$- and $\wp$-functions by the formulas $$\zeta (x)=\sigma '(x)/\sigma (x), \quad
\wp (x)=-\zeta '(x)=-\p_x^2\log \sigma (x).$$
We set $Q=cx^2+ \ldots$ with some constant $c$.
Correspondingly, 
the function $u=2\p_x^2\log \tau$ is an 
elliptic function with double poles at the points $x_i$:
\beq\label{ell2}
u=-2\sum_{i=1}^{N}\wp (x-x_i) +4c.
\eeq
The poles depend on the times $t_2$, $t_3$.

According to Krichever's method, the basic tool for 
studying $t_2$-dynamics of poles is the auxiliary linear problem 
\beq\label{ell3}
\p_{t_2} \psi =\p_x^2\psi +u \psi .
\eeq
Since the coefficient function $u$ is double-periodic, 
one can find double-Bloch solutions $\psi (x)$, i.e., solutions such that 
$\psi (x+2\omega )=b \psi (x)$, $\psi (x+2\omega ' )=b' \psi (x)$
with some Bloch multipliers $b, b'$.
The ansatz for the $\psi$-function is
\beq\label{ell4}
\psi = e^{xz+t_2z^2}\sum_{i=1}^N c_i \Phi (x-x_i, \lambda ),
\eeq
where the coefficients $c_i$ do not depend on $x$.
Here the function
\beq\label{Phi}
\Phi (x, \lambda )=\frac{\sigma (x+\lambda )}{\sigma (\lambda )\sigma (x)}\,
e^{-\zeta (\lambda )x}
\eeq
has a simple pole
at $x=0$ ($\zeta$ is the Weierstrass $\zeta$-function). 
The expansion of $\Phi$ as $x\to 0$ is
$$
\begin{array}{l}
\Phi (x, \lambda )=x^{-1}-\frac{1}{2}\, \wp (\lambda )x 
- \frac{1}{6}\, \wp '(\lambda ) x^2 +\ldots , \qquad 
x\to 0.
\end{array}
$$
The parameter $\lambda$ is a 
spectral parameter. 
Using the quasiperiodicity properties of the function $\Phi$,
$$
\Phi (x+2\omega , \lambda )=e^{2(\zeta (\omega )\lambda - \zeta (\lambda )\omega )}
\Phi (x, \lambda ),
$$
$$
\Phi (x+2\omega ' , \lambda )=e^{2(\zeta (\omega ' )\lambda - \zeta (\lambda )\omega ' )}
\Phi (x, \lambda ),
$$
one concludes that $\psi$ given by (\ref{ell4}) 
is indeed a double-Bloch function with Bloch mul\-ti\-pli\-ers
$$b=e^{2(\omega z + \zeta (\omega )\lambda - \zeta (\lambda )\omega )}, \qquad
b '=e^{2(\omega ' z + \zeta (\omega ' )\lambda - \zeta (\lambda )\omega ' )}.$$
We will often suppress the second argument of $\Phi$ writing simply 
$\Phi (x)=\Phi (x, \lambda )$. 
We will also need the $x$-derivatives
$\Phi '(x, \lambda )=\p_x \Phi (x, \lambda )$, 
$\Phi ''(x, \lambda )=\p^2_x \Phi (x, \lambda )$.

Substituting (\ref{ell4}) into (\ref{ell3}) with $u$ given by (\ref{ell2}), we get:
$$
-\sum_i \dot c_i \Phi (x-x_i)+\sum_i c_i \dot x_i \Phi '(x-x_i)+2z
\sum_i c_i \Phi '(x-x_i)+\sum_i c_i \Phi ''(x-x_i)
$$
$$
-\, 2\left (\sum_i \wp (x-x_i)\right ) \left ( \sum_k c_k \Phi (x-x_k)\right )
+4c \sum_i c_i \Phi (x-x_i)=0,
$$
where dot means the $t_2$-derivative. Different terms of this expression have poles at $x=x_i$. 
The highest poles are of third order but it is easy to see that they cancel identically.
The conditions of cancellation of
second and first order poles have the form
\beq\label{ell5}
c_i\dot x_i=-2zc_i -2\sum_{j\neq i}c_j \Phi (x_i-x_j),
\eeq
\beq\label{ell6}
\dot c_i=(4c +\wp (\lambda )) c_i -2\sum_{j\neq i}c_j \Phi '(x_i-x_j)-2c_i \sum_{j\neq i}
\wp (x_i-x_j).
\eeq
Introducing $N\! \times \! N$ matrices
\beq\label{ell7}
{\cal L}_{ij}=-\delta_{ij}\dot x_i -2(1-\delta_{ij})\Phi (x_i-x_j),
\eeq
\beq\label{ell8}
{\cal M}_{ij}=\delta_{ij}(\wp (\lambda )+4c)-2\delta_{ij}
\sum_{k\neq i}\wp (x_i-x_k) -2(1-\delta_{ij})\Phi ' (x_i-x_j),
\eeq
we can write the above conditions as a system of linear equations for the vector
${\bf c}=(c_1, \ldots , c_N)^T$:
\beq\label{ell9}
\left \{ \begin{array}{l}
{\cal L}{\bf c}=2z{\bf c}
\\ \\
\dot {\bf c}={\cal M} {\bf c}.
\end{array} \right.
\eeq
The compatibility condition is
\beq\label{ell12}
\dot {\cal L}+[{\cal L}, {\cal M}] =0.
\eeq
Using certain identities for the $\Phi$-function, one can see that it is equivalent to
the equations of motion
\beq\label{int1}
\ddot x_i=4\sum_{k\neq i} \wp ' (x_i-x_k).
\eeq
The Hamiltonian is
\beq\label{ell13}
{\cal H}_2=\sum_i p_i^2 -2\sum_{i<j}\wp (x_i-x_j).
\eeq
This is the elliptic version of the Calogero-Moser model. We have obtained it 
together with its Lax representation (\ref{ell12}) with the matrices ${\cal L}$,
${\cal M}$ depending on the spectral parameter $\lambda$. 

\subsubsection{Algebro-geometric (finite gap) solutions}

The soliton-like solutions are degenerate cases of a more general family
of solutions which are associated with algebraic curves (Riemann surfaces)
according to Krichever's construction.
These solutions are called algebro-geometric. In general they are quasi-periodic in 
all the times. 

The main building block of the algebro-geometric solutions is the Riemann 
theta-function given by the absolutely convergent series
\beq\label{ag1}
\Theta (\vec z \, |\, T)=\sum_{\vec n \in \z ^g}\exp \Bigl (\pi i (T\vec n, \vec n)+
2\pi i (\vec n, \vec z)\Bigr ).
\eeq
Here $\vec z=(z_1, \ldots , z_g)^T$ is a complex vector, the brackets denote the
Euclidean scalar product $\displaystyle{(\vec x, \vec y)=\sum_{i=1}^g x_iy_i}$ and
$T$ is a complex $g\! \times \! g$ symmetric matrix with positively definite imaginary part
(called the Riemann matrix). 

The tau-function of the algebro-geometric solutions
can be found in the form
\beq\label{ag2}
\tau ({\bf t})= e^{Q({\bf t})}\Theta \Bigl (\vec Z_0+ 
x\vec U_1 +t_2 \vec U_2 +t_3 \vec U_3 +\ldots \Bigl | \, T\Bigr ),
\eeq
where $Q({\bf t})$ is a quadratic form of the times $t_k$ and 
$\vec Z_0$, $\vec U_k$ are some $g$-dimensional constant vectors. However, 
in general (\ref{ag2}) is not a solution (i.e. it is not the tau-function) 
if the vectors $U_k$ and the Riemann 
matrix $T$ are arbitrary. For (\ref{ag2}) to be the tau-function these parameters
have to obey some constraints. 
The necessary relations between the parameters can be in principle 
found by a direct
substitution of the expression (\ref{ag2}) into the KP equation. In practice 
this method effectively works for small $g$ only. 

In general the relations between the parameters can be implemented if one starts with
a smooth algebraic curve (a Riemann surface) ${\cal C}$
of genus $g$ with a marked point which we call $\infty$. Topologically the surface of
genus $g$ is a sphere with $g$ handles. In the space spanned by closed cycles 
one can choose a basis of cycles (closed contours) $a_1, \ldots , a_g$ and
$b_1, \ldots , b_g$ with the following intersections:
$$
a_i \circ a_j =b_i \circ b_j =0, \quad a_i \circ b_j =\delta_{ij}, \quad
i,j =1, \ldots , g.
$$
As is known, there are $g$ linear independent holomorphic differentials 
$\Omega_k$ on the curve
of genus $g$. We normalize them by the condition
$
\oint_{a_i}\Omega_k =\delta_{ik}.
$
Then the $b$-periods form the matrix of periods
\beq\label{ag3}
T_{ik}=\oint_{b_i} \Omega_k
\eeq
which is the Riemann matrix. Let $z^{-1}$ be a local parameter in a neighborhood
of the marked point $\infty$. Consider the meromorphic differentials
$\Omega ^{(k)}$ which have the only high order 
pole on ${\cal C}$ at $\infty$ with the principle parts
$\Omega ^{(k)}=dz^k + O(z^{-2})$, $z\to \infty$. Let us normalize them by the 
conditions $\oint_{a_i}\Omega ^{(k)}=0$. 

\noindent
{\bf Proposition.} {\it
If the matrix $T$ in (\ref{ag2}) is the 
matrix of periods of holomorphic dif\-fe\-ren\-ti\-als on ${\cal C}$ (\ref{ag3})
and the vectors $\vec U_k=(U_{k,1}, \ldots , U_{k, g})^T$ are given by
\beq\label{ag4}
U_{k, i}=\oint_{b_i}\Omega ^{(k)},
\eeq
then $\tau ({\bf t})$ given by 
(\ref{ag2}) is the tau-function of the KP hierarchy.}

\noindent
The proof is based on properties of Baker-Akhiezer functions on Riemann surfaces.

The set of Riemann matrices has $\frac{1}{2}\, g(g+1)$ complex parameters. 
A fundamental result in the theory of Riemann surfaces states that the set of 
period matrices for $g\geq 2$ has $3g-3$ complex parameters. At $g=2$ and $g=3$ 
these two numbers are equal but for $g\geq 4$ $3g-3$ is strictly less than 
$\frac{1}{2}\, g(g+1)$. The famous 
Schottky problem is to characterize the matrices of periods 
of holomorphic dif\-fe\-ren\-ti\-als on Riemann surfaces among all Riemann 
matrices. A solution to the Schottky problem is suggested by the Novikov's conjecture
(proved by Shiota in 1986): matrices of periods (coming from a Riemann surface) 
are precisely those Riemann matrices for which
the expression (\ref{ag2}) gives a solution to the KP equation. 

For the KdV equation $\vec U_2=0$. In this case the curve $\Gamma$ should be 
hyperelliptic of genus $g$ with a ramification point at $\infty$, 
i.e. given by an equation of the form $y^2=R_{2g+1}(x)$ with a polynomial 
$R_{2g+1}(x)$ of degree $2g+1$. 
The functions $u=2\p_x^2 \log \Theta 
\Bigl (\vec Z_0+ 
x\vec U_1  +t_3 \vec U_3 +\ldots \Bigl | \, T\Bigr )+c$ constructed from
hyperelliptic curves as functions of $x$ 
have a rather special property:
the corresponding Schrodinger operator with the potential $-u$ has only finite number
of unstable bands (gaps) in the spectrum. That is why the algebro-geometric solutions 
are sometimes called finite-gap solutions. 

\subsubsection{General solution to the KP hierarchy: non-local $\bar \p$-problem}

A general method to define the function 
$\psi$ in such a way that the asymptotic equalities
(\ref{lemma2}) would imply the exact ones consists in posing the so-called
non-local
$\bar \p$-problem ($\bar \p \equiv \p_{\bar z}$). Let the function
$\psi =\psi (z)$ be of the form
\beq\label{dbar0}
\psi (z)=e^{\xi ({\bf t},z)}w(z),
\quad \quad 
w(z)=1+\frac{\xi_1}{z}+\frac{\xi_2}{z^2}+\ldots
\eeq
as $z\to \infty$. Impose the equation
\beq\label{dbar1}
\p_{\bar z}\psi (z)= \int K(z, \zeta ) \psi (\zeta )
d^2 \zeta ,
\eeq
where the integration in general goes over all complex plane
with the measure
$d^2 z \equiv d({\cal R}e \, z) d({\cal I}m \, z)$,
and $K(z, \zeta )$ is some kernel function which does not depend on the times $t_i$.
Assume that the function $K(z, \zeta )$ in each of the variables
does not vanish only in some compact domain, then outside this domain, when
$|z|$ is sufficiently large, one can require the holomorphic asymptotics
(\ref{dbar0}).  
The function $K(z, \zeta )$ can be a generalized function 
(a distribution), i.e. it may contain delta-functions concentrated in points
or on contours. 
Equation (\ref{dbar1}) is called the non-local
$\bar \p$-problem (non-local because the right hand side contains the integral).
Assume that this problem has a unique solution
(up to a constant factor). Then we are in the situation discussed above
and $u=-2\xi_1'$ will satisfy the KP equation. It is hard to formulate 
general conditions on the kernel when there is a unique solution. This problem
should be solved by a separate investigation in each concrete case. 

The integro-differential equation
(\ref{dbar1}) can be reduced to an integral equation for the function 
$w=\psi e^{-\xi ({\bf t},z)}$.
Let us represent the $\bar \p$-problem in the form
\beq\label{dbar2}
\p_{\bar z}w (z)= \int K_t(z, \zeta ) w (\zeta )
d^2 \zeta ,
\eeq
where the kernel $K_t(z, \zeta )$ depends on times:
$
K_t(z, \zeta )=e^{\xi ({\bf t}, \zeta )-\xi ({\bf t}, z)}
K(z, \zeta ).
$
Using the formal equality $\p_{\bar z}(1/z)=\pi
\delta ^{(2)}(z)$, we can write 
$$
w(z)=1+\frac{1}{\pi}\int 
\frac{\p_{\bar \zeta}w(\zeta )d^2 \zeta}{z-\zeta}
$$
and then (\ref{dbar2}) becomes the integral equation
\beq\label{dbar3}
w(z)=1+ \frac{1}{\pi}\int \int \frac{K_t(\zeta , \xi )
w(\xi )}{z-\zeta}\, d^2 \zeta d^2 \xi .
\eeq

\noindent
{\bf Remark.} In the case $K(z, \zeta )=K(z)\delta ^{(2)}(z+\zeta )$
one gets solutions of the KdV equation; the $\bar \p$-problem acquires the form
\beq \label{dbar4}
\p_{\bar z}\psi (z)= K(z) \psi (-z).
\eeq

As an example, consider the soliton-like solutions. 
Let the kernel be concentrated at a finite number of points:
\beq\label{dbar5}
K(z, \zeta )=-\pi \sum_{j=1}^{N}\tilde \beta_j 
\delta ^{(2)}(z-q_j)\delta ^{(2)}(\zeta -p_j),
\eeq
then the integral equation (\ref{dbar3}) will be of the form
$$
w(z)=1-\sum_{j=1}^{N}\tilde \beta_j \frac{w(p_j)}{z-q_j}\,
e^{\xi ({\bf t}, p_j)-\xi ({\bf t}, q_j)}.
$$
The right hand side defines the function $w(z)$ as a rational function 
with poles at $q_j$. Denoting $w_j = w(p_j)$,
we have a system of linear equations
\beq\label{dbar6}
w_i =1-\sum_{j=1}^{N}\frac{\tilde \beta _j 
e^{\xi ({\bf t}, p_j)-\xi ({\bf t}, q_j)}}{p_i -q_j}\, w_j
\eeq
from which we should find
$$
\xi_1 =-\sum_{j=1}^{N}\tilde \beta _j
e^{\xi ({\bf t}, p_j)-\xi ({\bf t}, q_j)}\, w_j.
$$
The solution consists in application of the Kramer's rule. It gives
$u=2\p_{x}^{2}\log \tau$ with the tau-function (\ref{kptauN1})
and $\tilde \beta _j =(p_j-q_j)\beta_j$.

\noindent
{\bf Problem.} Find the kernel of the $\bar \p$-problem corresponding to the
solutions depending on functional parameters.

\subsection{Additional (nonabelian) symmetries of the KP hierarchy}

Nonabelian symmetries of the KP hierarchy are richer than in the case of KdV.
In order to define them in a general form, we need an extended Lax formalism.
Let us introduce a new operator,
$M$, which satisfies the same Lax equations and forms a canonical pair 
with $L$, i.e. their commutator is equal to 1.  
(This is the so-called Orlov-Shulman operator.)
The simplest way to define it is to use dressing by the operator $K$.

Note that besides $\p$ there is yet another operator 
commuting with $\p_{t_n}-\p^n$; it is
$$
\Gamma = \sum_{k=1}^{\infty}kt_k \p^{k-1} = x + 2t_2 \p +
3t_3 \p^2 + \ldots
$$
The operator $\p$ acts to the exponential function 
$e^{\xi ({\bf t},z)}$ as multiplication by $z$ while
$\Gamma$ acts as $z$-derivative:
\beq\label{dop0}
\p e^{\xi ({\bf t},z)} =z e^{\xi ({\bf t},z)}, \quad \quad
\Gamma e^{\xi ({\bf t},z)} =\p_z e^{\xi ({\bf t},z)}.
\eeq
We recall that dressing of the obvious commutation relation
$[\p , \p_{t_n}-\p^n ]=0$
by the operator $K$ gives the ``equations of motion'' for the Lax operator. 
In a similar way, dressing of the relation
$[\Gamma , \p_{t_n}-\p^n ]=0$ gives $[M, \p_{t_n}-A_n]=0$ or
\beq\label{dop1}
\p_{t_n}M =[A_n , M]\,, \quad \quad A_n =(L^n)_{+},
\eeq
where $M=K\Gamma K^{-1}$ is the Orlov-Shulman operator. 

\noindent
{\bf Exercise.}
Check (\ref{dop1}) by a direct calculation.

\noindent
The Orlov-Shulman operator can be represented as a series
in powers of the Lax operator:
\beq\label{dop2}
M=\sum_{k\geq 2}kt_k L^{k-1}+x + \sum_{k\geq 1}v_k L^{-k-1},
\eeq
where the ``tail'' of the expansion (negative powers of 
$L$) is obtained from the dressing $KxK^{-1}$ and
$v_k$ depend on $x$ and all higher times. It is clear that
the Lax equation of the type (\ref{dop1}) holds for any function of
$L$ and $M$, in particular:
\beq\label{dop1a}
\p_{t_n}(M^m L^l) =[A_n , M^m L^l].
\eeq
Acting by the dressing operator from the left to both sides 
of relations
(\ref{dop0}), we find how $L$ and $M$ act to the Baker-Akiezer function:
\beq\label{dop0a}
L\psi =z \psi, \quad \quad
M\psi =\p_z \psi .
\eeq
Finally, dressing of the obvious relation
$[\p , \Gamma ]=1$ yields
\beq\label{dop3}
[L, M]=1,
\eeq
so $L$ and $M$ is indeed the canonical pair. 
The same is seen from (\ref{dop0a}).

Now everything is ready for introducing of the additional symmetries.
Let $s_{lm}$ be parameters of the symmetries (now they are numbered by
two indices) and $\p_{s_{lm}}$ be the corresponding vector fields. 
Consider the equations
\beq\label{dop4}
\p_{s_{lm}}L=-[(M^m L^l)_{-}, L],
\eeq
which define flows on the space of operators
$L$. Obviously, 
$\p_{s_{n0}}=\p_{t_n}$.

\noindent
{\bf Proposition.}
{\it Equations (\ref{dop4}) are symmetries
of the KP hierarchy.}

\noindent
It is easy to check that both sides of
(\ref{dop4}) are pseudo-differential operators of order $-1$,
and thus the flows are well-defined. To see that they are symmetries, 
one should calculate $X:=[\p_{s_{lm}}, \p_{t_n}]L =
\p_{s_{lm}}(\p_{t_n}L)-\p_{t_n}(\p_{s_{lm}}L)$ and show that
$X=0$. We have:
$$
\begin{array}{lll}
X&=& \p_{s_{lm}}(\p_{t_n}L)-\p_{t_n}(\p_{s_{lm}}L)
\\ && \\
&=&-\p_{s_{lm}}[(L^n)_- , L] +\p_{t_n}[(M^m L^l )_- , L]
\\ && \\
&=&-[(L^n)_- , \p_{s_{lm}}L]-[\p_{s_{lm}}(L^n)_-, L]+
[\p_{t_n}(M^m L^l )_- , L] +[(M^m L^l )_- , \p_{t_n}L]
\\ && \\
&=& \left [(L^n)_- ,[(M^m L^l )_- , L]\right ]\! -\!
\left [\p_{s_{lm}}(L^n)_- \! -\!
\p_{t_n}(M^m L^l )_- ,  L \right ]-\left [(M^m L^l )_-, [(L^n)_-, L]\right ]
\\ && \\
&=& \left [\p_{t_n}(M^m L^l )_- - \p_{s_{lm}}(L^n)_- +
[(L^n)_- , (M^m L^l )_-], \, L\right ].
\end{array}
$$
When passing to the last line, we have used the Jacobi identity for the double 
commutator. 
Calculate now $\p_{s_{lm}}(L^n)_-$ with the hep of the definition (\ref{dop4})
and show that the operator inside the commutator is equal to 0, i.e.
\beq\label{dop5}
\p_{t_n}(M^m L^l )_- - \p_{s_{lm}}(L^n)_- +
[(L^n)_- , (M^m L^l )_-]=0.
\eeq
Indeed, 
$$
\begin{array}{lll}
\p_{s_{lm}}(L^n)_-&=&-\left ( [(M^m L^l )_-, L^n]\right )_-
\\ && \\
&=&-[(M^m L^l )_-, (L^n)_- ]-\left ( [M^m L^l , (L^n)_+]\right )_-
\\ && \\
&=&-[(M^m L^l )_-, (L^n)_- ]+\p_{t_n}(M^mL^l),
\end{array}
$$
which is equivalent to (\ref{dop5}).

We have shown that the vector fields given by (\ref{dop4}) are symmetries of the
KP hierarchy. However, they do not commute and thus they are not symmetries
for each other. 
In general these symmetries are non-local, i.e. contain
integrals of $u_i$ over $x$.

The doubly infinite family of symmetries constructed above includes
three-\-pa\-ra\-met\-ric set of symmetries which generalize the KdV symmetries 
(\ref{ns1}). For the KP equation they are as follows.
If $u(x,y,t)$ satisfies the KP equation
(\ref{kp0}), then $\tilde u$,
defined as a function of $\tilde x , \tilde y , \tilde t$
by the equalities
\beq\label{dop6}
\left \{
\begin{array}{l}
\tilde u = \,\, \lambda^{-2}u -2a \lambda^{-1},
\\ \\
\tilde x = \,\, \lambda x +2b\lambda y +(3a \lambda^2
+3b^2 \lambda )t,
\\ \\
\tilde y = \lambda ^2 y +3 b\lambda ^2 t,
\\ \\
\tilde t = \,\, \lambda^3 t
\end{array}\right.
\eeq
satisfies the same equation. 
In other words, the transformation
$$
u(x,y,t)\rightarrow \,\,
\lambda^{-2}u\left (\lambda^{-1}x\! -\!
2b \lambda ^{-2}y \! - \!
(3a\lambda^{-2}\! - \! 3b^2 \lambda ^{-3})t, \,
\lambda ^{-2} y \! - \! 3b \lambda ^{-3}t, \,
\lambda^{-3}t\right ) -2a\lambda^{-1}
$$
sends a solution to another solution. (This can be checked by a direct
substitution.) 
At $\lambda =1$, $b=0$, $a\neq 0$ we have the Galilean transformation.
It generates by the vector field
$$
\p_{s_{-1,1}}L=-[(ML^{-1})_- , \, L].
$$

\noindent
{\bf Problem.} Find vector fields of the type (\ref{dop4}) which generate
transformations (\ref{dop6})
with $\lambda =1$, $b\neq 0$, $a=0$ and
$\lambda \neq 1$, $b=0$, $a=0$.

\subsection{Tau-function of the KP hierarchy}

So far the tau-function occasionally appeared as a convenient auxiliary
object in the dis\-cus\-sion of soliton solutions. In fact the tau-function plays
a fundamental role in the theory of the KP hierarchy (and other integrable
hierarchies). 
Passing to the tau-function, regarded as a dependent variable, allows one to
formulate the KP hierarchy as an infinite set of compatible equations for
{\it one} function rather than an infinite number of them, as in the original formulation. 
In terms of the tau-function, all equations of the KP hierarchy become bilinear
and they can be encoded in {\it one} ``generating equation'', which is known as
the difference Hirota equation. However, it is not an easy task to derive all that
starting from the Lax representation. Introducing the very concept of the
tau-function requires some preparation.

\subsubsection{The bilinear identity}

We start from a reformulation of the KP hierarchy in terms of the Baker-Akhiezer
function and its adjoint. 

As before, together with the operator residue
$\mbox{res}=
\mbox{res}_{\p}$ we will consider the ordinary
residue $\mbox{res}_z$ of the Laurent series $\sum_j a_j z^j$ defined as
$\mbox{res}_z (\sum_j a_j z^j) =a_{-1}$.

\noindent
{\bf Proposition.} {\it For the $\psi$-function of any solution to the KP hierarchy
the following bilinear identity holds:
\beq\label{bi101}
\mbox{{\rm res}}_z \Bigl ( \psi ({\bf t},z)\psi ^* ({\bf t}',z)\Bigr )=0,
\eeq
where ${\bf t}=\{t_j\}$, ${\bf t}'=\{t'_j\}$ are two arbitrary sets
of times.}

\noindent
For the proof we use the previously proved lemma which states that
for any pseudo-differential operators
$P$, $Q$ the equality $$\mbox{res}_z \left [ (Pe^{xz})\, 
(Qe^{-xz})\right ]=
\mbox{res}_{\p} (PQ^{\dag})$$
holds true. Assume that the function 
$\psi ({\bf t}', z)$ can be obtained from $\psi ({\bf t}, z)$ as a Taylor series in
$t'_i -t_i$. Then it is enough to prove
(\ref{bi101}) for any terms of this series. 
Note that the $t_i$-derivatives with $i\geq 2$ are expressed linearly
through derivatives with respect to $t_1=x$ by virtue of (\ref{kplin1}).
Hence it is enough to show that for all $n\geq 0$
$$\mbox{res}_z \! \Bigl ( (\p^n \psi ({\bf t},z))\, \psi ^* ({\bf t},z)\Bigr )=0.$$
The latter is easily checked using the above lemma:
$$
\mbox{res}_z \! \left ( \p^n \psi \, \psi ^* \right )=
\mbox{res}_z \! \left ( \p^n K e^{\xi ({\bf t},z)}(K^{\dag})^{-1}
e^{-\xi ({\bf t},z)}\right )=
\mbox{res}_{\p} \left ( \p^n \! K \, K^{-1}\right )
= \mbox{res}_{\p} \, \p^n =0.
$$
The inverse statement is also true:
let $\psi$, $\psi^*$ be the series of the form
$$
\psi =e^{\xi ({\bf t},z)}
\left (1+\xi _1 z^{-1} +\xi_2 z^{-2}+\ldots \right ),
\quad
\psi^* =e^{-\xi ({\bf t},z)}\left (1+\xi^*_1 z^{-1} +
\xi^*_2 z^{-2}+\ldots \right ),
$$
where the coefficients $\xi_j , \xi^*_j$ are some functions of $t_i$,
and for all sets of times $t'_i$, $t_i$ (\ref{bi101}) holds; then
there exists the $L$-operator of the form (\ref{kp1}) satisfying all the Lax equations
and the Baker-Akhiezer function for it coincides with
$\psi$ (and the adjoint function coincides with $\psi ^*$).

\noindent
{\bf Problem.} For the tau-function $\tau =t_2+\frac{1}{2}\, t_1^2$ find $\psi$,
$\psi^*$ and check the bilinear identity.

\noindent
{\bf Problem.} Check the bilinear identity for the functions
$\psi$, $\psi^*$ corresponding to the one-soliton solution. 

Let us explain in more detail how one should understand the residue
in the bilinear identity. We have
$\psi ({\bf t},z)=e^{\xi ({\bf t},z)}w ({\bf t},z)$, 
$\psi^* ({\bf t},z)=e^{-\xi ({\bf t},z)}w^* ({\bf t},z)$, 
where the functions $w$, $w^*$ are expanded in series in inverse powers of
and thus they are regular at $\infty$. The bilinear identity acquires the form
\beq\label{bi1a}
\mbox{res}_z \left ( e^{\xi ({\bf t}-{\bf t}',z)}w({\bf t},z)
w^*({\bf t}',z)\right )=0.
\eeq
One should expand the function $e^{\xi ({\bf t}-{\bf t}',z)}$
in positive powers of $z$, the functions $w({\bf t},z)$, $w^*({\bf t}',z)$ in negative
powers of $z$, multiply these series, extract the coefficient in
front of
$z^{-1}$ and equate it to zero. This is equivalent to vanishing of the contour integral
\beq\label{bi1b}
\oint_{{\sf C}} e^{\xi ({\bf t}-{\bf t}',z)}w({\bf t},z)
w^*({\bf t}',z)dz =0,
\eeq
where the contour ${\sf C}$ should encircle all singularities 
of the function  
$e^{\xi ({\bf t}-{\bf t}',z)}$ and should not encircle singularities of the function 
$w({\bf t},z)w^*({\bf t}',z)$. In particular, if only a finite number 
of the times are not equal to $0$, then the function
$e^{\xi ({\bf t}-{\bf t}',z)}$ is regular everywhere except $\infty$, where it has an
essential singularity. In this case the contour
${\sf C}$ is a circle of radius
$R$ for sufficiently large $R$.

\subsubsection{``Japanese'' formula for the Baker-Akhiezer function}

One of the most important results of the Japanese school (Sato, Jimbo, Miwa and others)
is the remarkable formula for the Baker-Akhiezer function through the tau-function.
(This formula simultaneously serves as a definition of the latter.)

\noindent
{\bf Theorem.} {\it Let $\psi$ be the Baker-Akhiezer function of the KP hierarchy,
then there exists a function
$\tau (t_1 , t_2 , t_3 , \ldots )$ such that
\beq\label{jap1}
\mbox{\fbox{$\displaystyle{\phantom{\int ^{A}_{B}}
\psi ({\bf t},z)=e^{\xi ({\bf t},z)}\,  \frac{\tau (t_1 - \frac{1}{z}, \,
t_2 - \frac{1}{2z^2}, \,
t_3 - \frac{1}{3z^3}, \, \ldots )}{\tau (t_1 , t_2 , t_3 , \ldots )}.
\phantom{\int ^{A}_{B}}}$}}
\eeq
There is also a similar formula for the adjoint Baker-Akhiezer function  through the 
same tau-function:
\beq\label{jap2}
\mbox{\fbox{$\displaystyle{\phantom{\int ^{A}_{B}}
\psi^* ({\bf t},z)=e^{-\xi ({\bf t},z)}\,  \frac{\tau (t_1 + \frac{1}{z}, \,
t_2 + \frac{1}{2z^2}, \,
t_3 + \frac{1}{3z^3}, \, \ldots )}{\tau (t_1 , t_2 , t_3 , \ldots )}.
\phantom{\int ^{A}_{B}}}$}}
\eeq
}

\noindent
Equation (\ref{jap1}) can be written in a different form. Write
$\psi ({\bf t},z)=e^{\xi ({\bf t},z)}w ({\bf t},z)$ and take the logarithmic derivative
with respect to $z$ of both sides of (\ref{jap1}). We get
$$
\p_z \log w({\bf t},z)=\sum_{m\geq 1}\frac{\p \log w({\bf t},z)}{\p t_m}
\, z^{-m-1} +\sum_{m\geq 1}\frac{\p \log \tau}{\p t_m}z^{-m-1}
$$
which is equivalent to the relations
\beq\label{jap3}
\frac{\p \log \tau}{\p t_n}=\mbox{res}_z \Bigl (
z^n \left (\p_z -\p (z)\right )
\log w({\bf t},z)\Bigr ),
\eeq
where we use the notation
$$
\p (z):= \sum_{j\geq 1}z^{-j-1}\frac{\p}{\p t_j}.
$$
Therefore, for the proof of existence of the tau-function it is enough to prove that
the expression
$$
\mbox{res}_z \Bigl (
z^n \left (\p_z -\p (z)\right )
\p_{t_m}\! \log w({\bf t},z)\Bigr )
$$
is symmetric under permutation of $m,n$.

The proof is based on the bilinear identity. Below we present its main points.

Putting  $t'_j =t_j -\zeta^{-j}/j$ in the bilinear identity,
we write it in the form
\beq\label{jap4a}
\mbox{res}_z \left ( \psi ({\bf t},z)
\psi^* ({\bf t} \! -\! [\zeta ^{-1} ],  z)\right )=0,
\eeq
where we have introduced the convenient notation
$$
F({\bf t}\pm 
[z])\equiv F(t_1 \pm z, \, t_2 \pm z^2/2, \,  t_3 \pm z^3 /3, \, \ldots ).
$$
After the substitutions
$\psi ({\bf t},z)=e^{\xi ({\bf t} ,z)}w ({\bf t},z)$,
$\psi^* ({\bf t} ,z)=e^{-\xi ({\bf t},z)}w^* ({\bf t},z)$ equation (\ref{jap4a}) reads
$$
\mbox{res}_z \left (\frac{w ({\bf t},z)\, 
w^* ({\bf t} \! -\! [\zeta ^{-1} ], z)}{1-z/\zeta}\right )=0.
$$
It is easy to check that for any series $\displaystyle{f(z)=1+\sum_{j\geq 1}f_j z^{-j}}$
the identity
\beq\label{jap4}
\mbox{res}_z \left (\frac{f(z)}{1-z/\zeta}\right )=
\sum_{j\geq 1}f_j \, \zeta^{1-j}=\zeta (f(\zeta )-1)
\eeq
holds. Applying it to the previous equality, we get the relation
between 
$w$ and $w^*$:
\beq\label{jap5}
w({\bf t},z)\, w^* ({\bf t}-[z^{-1}], z)=1.
\eeq
Similarly, from the biliniear identity
$
\mbox{res}_z \Bigl ( \psi ({\bf t} ,z)\, \psi^* ({\bf t} \! -\! 
[\zeta _1^{-1}]
\! -\! [\zeta _2^{-1}],  z)\Bigr )=0
$
rewritten in the form
$$
\mbox{res}_z \left (\frac{w ({\bf t},z)
\, w^* ({\bf t}  \! -\! [\zeta _1^{-1}]
\! -\! [\zeta _2^{-1}], z)}{(1-z/\zeta _1)(1-z/\zeta _2)}\right )=0
$$
it follows that
$$
w({\bf t},\zeta _1)\, w^*({\bf t}  \! -\! [\zeta _1^{-1}]
\! -\! [\zeta _2^{-1}], \zeta_1)=
w({\bf t},\zeta _2)\, w^*({\bf t}  \! -\! [\zeta _1^{-1}]
\! -\! [\zeta _2^{-1}], \zeta_2),
$$
where we have used the identity
$$
\frac{1/\zeta_1 -1/\zeta_2}{(1-z/\zeta _1)(1-z/\zeta _2)}
=\frac{1}{\zeta_1 (1-z/\zeta _1)}-\frac{1}{\zeta_2 (1-z/\zeta _2)}
$$
and equation (\ref{jap4}).
Now one can express $w^*$ through
$w$ with the help of (\ref{jap4}) and put $\zeta_1 =z$,
$\zeta_2 =\zeta$. The result is
\beq\label{jap6}
\frac{w({\bf t}, z)}{w({\bf t}\! -\! [\zeta ^{-1}], z)}=
\frac{w({\bf t}, \zeta )}{w({\bf t}\! -\! [z^{-1}], \zeta)}.
\eeq
Let us take logarithm of this equality and apply the operator
$\p_z - \p (z)$. We get:
\beq\label{jap7}
(\p_z - \p (z))\log w({\bf t},z)-(\p_z - \p (z))
\log w({\bf t}\! -\! [\zeta ^{-1}],z)
=-\p (z)\log w({\bf t},\zeta ).
\eeq
For brevity denote
$
Y_n ({\bf t}):= \mbox{res}_z \Bigl ( z^n (\p_z - \p (z))
\log w({\bf t},z)\Bigr )
$.
Multiplying both sides of (\ref{jap7}) by $z^n$ and taking the residue, we have
$$
Y_n ({\bf t})-Y_n ({\bf t}\! -\! [\zeta ^{-1}])=-
\p_{t_n}\log w({\bf t}, \zeta ).
$$
After differentiating with respect to $t_m$ and subtracting the similar equality
with in\-ter\-chan\-ged $m$, $n$, this yields
$$
\p_{t_m}Y_n ({\bf t})-\p_{t_n}Y_m ({\bf t})=
\p_{t_m}Y_n ({\bf t}\! -\! [\zeta ^{-1}])-
\p_{t_n}Y_m ({\bf t}\! -\! [\zeta ^{-1}]).
$$
Let us denote the left hand side by $F_{mn}({\bf t})$. Expand
$F_{mn}({\bf t}-[\zeta ^{-1}])$ 
in powers of $\zeta$: $F_{mn}({\bf t}-[\zeta ^{-1}])=
F_{mn}({\bf t})-\zeta^{-1}\p_{t_1}F_{mn}({\bf t})+\frac{1}{2}\, 
\zeta^{-2}(\p^2_{t_1}F_{mn}({\bf t})-\p_{t_2}F_{mn}({\bf t}))+\ldots $.
Since $F_{mn}({\bf t}-[\zeta ^{-1}])=F_{mn}({\bf t})$ for all $t_k$
and for all solutions, the comparison of coefficients at
$\zeta^{-1}$ implies that $\p_{t_1}F_{mn}({\bf t})=0$
for all ${\bf t}$, i.e. $F_{mn}$ does not depend on $t_1$. Next, comparing the 
coefficients at
$\zeta^{-2}$ we similarly find that 
$F_{mn}$ does not depend on $t_2$. Repeating this argument, we conclude that 
$F_{mn}$ does not depend on all times, i.e. it is a constant. 
For the trivial solution $u_i=0$ the constant is $0$.
Since $F_{mn}$ is a differential polynomial of $u_i$, this constant is equal to
$0$ for any solution. Thus we have shown that
$\p_{t_m}Y_n ({\bf t})=\p_{t_n}Y_m ({\bf t})$, which implies existence of the 
tau-function and validity of equations (\ref{jap1}),
(\ref{jap3}). Equation (\ref{jap2}) is proved in a similar way.

\subsubsection{Equations of the KP hierarchy in the Hirota form}

One can express $\psi$ and $\psi^*$ through the tau-function using the 
``Japanese'' formulas. This yields the bilinear relation for the tau-function:
\beq\label{hir1}
\mbox{res}_z \left ( \tau ({\bf t}-[z^{-1}])\, \tau ({\bf t}' +[z^{-1}])\,
e^{\xi ({\bf t}-{\bf t}',z)}\right )=0,
\eeq
or
\beq\label{hir1a}
\mbox{\fbox{$\displaystyle{\phantom{\int ^{A}_{B}}
\oint_{{\sf C}} e^{\xi ({\bf t}-{\bf t}',z)}\tau ({\bf t}-[z^{-1}])\, 
\tau ({\bf t}' +[z^{-1}])
dz =0
\phantom{\int ^{A}_{B}}}$}}
\eeq
with the convention about the integration contour discussed above. 
This relation is equivalent to an infinite system of bilinear differential equations
for the tau-function. They are obtained by equating to 0 the expansion 
coefficients in the Taylor series for the left hand side in ${\bf t}'-{\bf t}$.
Technically this expansion is conveniently made if one substitutes
$t_i -T_i$ and $t_i +T_i$ for
$t_i$ and $t_i'$ respectively:
$$
\begin{array}{ll}
&\mbox{res}_z \left [ \tau ({\bf t}-{\bf T}-[z^{-1}])\, 
\tau ({\bf t}+{\bf T} +[z^{-1}])\,
e^{-2\xi ({\bf T},z)}\right ]
\\ & \\
=& \mbox{res}_z \left [ e^{\xi (\tilde \p _{\bf T}, z^{-1})} 
(\tau ({\bf t}-{\bf T})\, \tau ({\bf t}+{\bf T}))\,
e^{-2\xi ({\bf T},z)}\right ]
\\ & \\
=& \displaystyle{
\mbox{res}_z \left [\sum_{j\geq 0} z^{-j}h_j(\tilde \p _{\bf T})
(\tau ({\bf t}-{\bf T})\, \tau ({\bf t}+{\bf T}))\,
\sum_{l\geq 0} z^l h_l(-2{\bf T})\right ]}
\\ & \\
=& \displaystyle{\sum_{j\geq 0}h_j(-2{\bf T})h_{j+1}(\tilde \p _{\bf T})\,
\tau ({\bf t}-{\bf T})\, \tau ({\bf t}+{\bf T})\, =\, 0}.
\end{array}
$$
In the second line, the shift by $[z^{-1}]$ is represented as action of
the exponential function of the differential operator
$$
\xi (\tilde \p _{\bf T} , z^{-1})= \sum_{j\geq 1}
\frac{z^{-j}}{j}\, \p_{T_j}
$$
(the notation $\tilde \p _{\bf T} =
\{ \p_{T_1}, \, \frac{1}{2}\, \p_{T_2}, \, \frac{1}{3}\, \p_{T_3},
\ldots \}$ is used). In the third line, the exponential function is expanded 
with the help of the Schur polynomials (\ref{schur}).
The latter equality can be also written in the form
$$
\left. \sum_{j\geq 0}h_j(-2{\bf T})h_{j+1}(\tilde \p _{\bf X})\,
e^{\sum_{l\geq 1}T_l \p_{X_l}}
\tau ({\bf t}-{\bf X})\, \tau ({\bf t}+{\bf X})\right |_{X_m=0} =\, 0.
$$
Using the symbol $D_i$ of the ``Hirota derivative'' defined by the rule
$$
\left. \phantom{\int}
P({\bf D}) f({\bf t})\cdot g({\bf t}) :=
P(\p_{\bf X})(f({\bf t}-{\bf X})g({\bf t}+
{\bf X}))\right |_{{\bf X}=0}
$$
for any polynimial $P({\bf D})$ of $D_i$, we can write it in the form
\beq\label{hir2}
\sum_{j\geq 0}h_j(-2{\bf T})h_{j+1}(\tilde {\bf D})\,
e^{\sum_{l\geq 1}T_l D_{l}}
\tau ({\bf t}) \cdot \tau ({\bf t}) =0.
\eeq
This relation contains, in the encoded form, all bilinear Hirota 
equations for the KP hierarchy.
The first non-trivial equation obtained as a result of 
expanding in the series in $T_i$ has the form
\beq\label{hir3}
\Bigl (D_{1}^4 +3D_{2}^2 -4D_1 D_3 \Bigr )\, \tau \cdot \tau =0.
\eeq

\noindent
{\bf Exercise.} Check that equation (\ref{hir3}) after the substitution
$u=2\p^2 \log \tau$ turns into the KP equation
(\ref{kp0}).

\noindent
{\bf Problem.} Prove that if $\tau ({\bf t})$ is the tau-function 
of the KP hierarchy (a solution to all bilinear Hirota equations), then 
$\tau (-{\bf t})$ is also the tau-function.

\subsection{Bilinear difference Hirota equation}

\subsubsection{The Hirota-Miwa equation}

In the bilinear relation (\ref{hir1}) put ${\bf t}'={\bf t}-
[\lambda _1^{-1}]-[\lambda _2^{-1}]-[\lambda _3^{-1}]$, 
where $\lambda_{1,2,3}$ are arbitrary compllex parameters, i.e.,
put
$$
t_{k}'=t_k - \frac{\lambda_{1}^{-k}}{k}
- \frac{\lambda_{2}^{-k}}{k}
- \frac{\lambda_{3}^{-k}}{k}.
$$
Then the bilinear relation reads
$$
\mbox{res}_z \left (\frac{\tau ({\bf t}-[z^{-1}])\, \tau ({\bf t} \! -\! 
[\lambda _1^{-1}]
\! -\! [\lambda _2^{-1}]\! -\! [\lambda _3^{-1}] + [z^{-1}])}{(1-z/\lambda _1)
(1-z/\lambda _2)(1-z/\lambda _3)}\right )=0.
$$
To use identity
(\ref{jap4}), we represent the product of the pole
factors as a sum of poles:
$$
\frac{(1/\lambda_1 -1/\lambda_2)(1/\lambda_1 -1/\lambda_3)
(1/\lambda_2 -1/\lambda_3)}{(1-z/\lambda _1)(1-z/\lambda _2)(1-z/\lambda _3)}
=\frac{(1/\lambda_2 -1/\lambda_3)/\lambda_1^2}{1-z/\lambda _1} +
(231)+(312),
$$
where the last two terms are obtained from the first one by cyclic permutations
of indices $(1\to 2,\, 2\to 3, \, 3\to 1)$ and 
$(1\to 3, \, 2\to 1, \, 3\to 2)$. Using (\ref{jap4}), we obtain 
the fundamental equation
\beq\label{hir4}
\mbox{\fbox{$\displaystyle{\phantom{\int ^{A}_{B}}
\begin{array}{l}
(\lambda_2 -\lambda_3) \, \tau ({\bf t}-[\lambda_1^{-1}]) \,
\tau ({\bf t}-[\lambda_2^{-1}]-[\lambda_3^{-1}]) 
\\ \\
\phantom{aaaa}+(\lambda_3 -\lambda_1) \, \tau ({\bf t}-[\lambda_2^{-1}]) \,
\tau ({\bf t}-[\lambda_3^{-1}]-[\lambda_1^{-1}]) 
\\ \\
\phantom{aaaaaaaa}+(\lambda_1 -\lambda_2) \, \tau ({\bf t}-[\lambda_3^{-1}]) \,
\tau ({\bf t}-[\lambda_1^{-1}]-[\lambda_2^{-1}])=0
\end{array}
\phantom{\int ^{A}_{B}}}$}}
\eeq
satisfied by the tau-function of the KP hierarchy.
This is the Hirota-Miwa equation.

\noindent
{\bf Problem.} From the bilinear relation derive the equation
\beq\label{hir4a}
(\lambda_0-\lambda_1)(\lambda_2 -\lambda_3) \, \tau ({\bf t}-
[\lambda_0^{-1}]-[\lambda_1^{-1}]) \,
\tau ({\bf t}-[\lambda_2^{-1}]-[\lambda_3^{-1}])
+(231)+(312)=0
\eeq
and show that it is equivalent to (\ref{hir4}).

\noindent
{\bf Problem.} From the bilinear relation derive the equation
\beq\label{hir4b}
\p_{t_1}\log \frac{\tau ({\bf t}+[\lambda_1^{-1}])}{\tau ({\bf t}+[\lambda_2^{-1}])}
=(\lambda_2-\lambda_1)\left (\frac{\tau ({\bf t})\tau ({\bf t}+[\lambda_1^{-1}]+
[\lambda_2^{-1}])}{\tau ({\bf t}+[\lambda_1^{-1}])\tau ({\bf t}+[\lambda_2^{-1}])}-1
\right )
\eeq
and show that it is equivalent to (\ref{hir4}).

\noindent
{\bf Problem.} Prove that the tau-function of the KP hierarchy satisfies the 
equation
\beq\label{hir4c}
\prod_{1\leq i<j\leq m}(\lambda_j -\lambda_i)\,
\tau \Bigl ({\bf t}+\sum_{\alpha =1}^{m}[\lambda_{\alpha}^{-1}]\Bigr )
\tau^{m-1}({\bf t})=\det_{1\leq j,k\leq m}\Bigl ((\lambda_j-\p_{t_1})^{k-1}
\tau ({\bf t}+[\lambda_j^{-1}])\Bigr ).
\eeq
(Hint: use induction in $m$.)

\subsubsection{The $T$-system and the $Y$-system}

The Hirota-Miwa equation can be read as a difference equation for the function
$$
\tau (p_1, p_2 , p_3):= \tau \Bigl ({\bf t}-p_1 [\lambda _1^{-1}]-
p_2 [\lambda_2^{-1}]-p_3 [\lambda_3^{-1}]\Bigr )
$$
(the arguments in the right hand side are 
$t_k - (p_1 \lambda_1^{-k}+p_2 \lambda_2^{-k}+p_3 \lambda_3^{-k})/k$). In this form,
it is called the bilinear difference Hirota equation:
\beq\label{hir5}
\mbox{\fbox{$\displaystyle{\phantom{\int ^{A}_{B}}
\begin{array}{l}
(\lambda_2 -\lambda_3) \, \tau (p_1+1, p_2, p_3) \,
\tau (p_1, p_2 +1, p_3+1) 
\\ \\
\phantom{aaaa} +(\lambda_3 -\lambda_1) \, \tau (p_1, p_2+1, p_3) \,
\tau (p_1+1, p_2 , p_3+1) 
\\ \\
\phantom{aaaaaaaa} +(\lambda_1 -\lambda_2) \, \tau (p_1, p_2, p_3+1) \,
\tau (p_1+1, p_2+1 , p_3)=0
\end{array}
\phantom{\int ^{A}_{B}}}$}}
\eeq
If one forgets about the origin of the function
$\tau (p_1, p_2, p_3)$, then this equation can be understood as
an equation for a function of discrete variables
$p_j$ assuming integer values or as a difference equation for a function
of real or complex variables $p_j$. It is the integrable difference analogue
of the KP equation. One can see that it is much more symmetric than the KP equation 
itself.

Equation (\ref{hir5}) can be represented  
in different equivalent forms. Let us point out two of them.

The linear change of variables $x_1 =p_2+p_3$,
$x_2 =p_1+p_3$, $x_3 =p_1+p_2$ brings equation
(\ref{hir5}) to the form
\beq\label{hir6}
z_1 T(x_1+1)T(x_1-1)+z_2 T(x_2+1)T(x_2-1)
+z_3 T(x_3+1)T(x_3-1)=0,
\eeq
where
$T(x_1, x_2 , x_3)=\tau (p_1, p_2, p_3)$ provided that 
$x_i$ and $p_i$ are connected by the above linear relations.
For brevity in (\ref{hir6}) the arguments which are not shifted are not
written explicitly, i.e., for example, 
$T(x_1+1)=T(x_1+1, x_2, x_3)$ and so on. The constants $z_i$ are subject to the relation
подчинить условию $z_1+z_2+z_3=0$ (then $T=1$ is the simplest solution);
however, one can consider them as arbitrary or even equal to
$1$, which is achieved by the simple transformation
of the $T$-function
$$
T(x_1, x_2, x_3)\longrightarrow z_1^{-x_{1}^{2}/2}\, 
z_2^{-x_{2}^{2}/2}\, z_3^{-x_{3}^{2}/2}\, 
T(x_1, x_2, x_3).
$$
It is this form in which this equation was introduced by Hirota in 1981.
It was suggested as a universal discretization of soliton equations. 
Sometimes it is called the $T$-system (in analogy with  
independently suggested 
$Y$-system, which is a consequence of (\ref{hir6}), see below). 
The majority (if not all) of known soliton equations are obtained form
(\ref{hir6}) by various reductions, limits, changes of variables.

Note that if $T(x_1, x_2, x_3)$ is a solution to 
(\ref{hir6}), then
$$
f_0(x_1+x_2+x_3)f_1(-x_1+x_2+x_3)f_2(x_1-x_2+x_3)f_3(x_1+x_2-x_3)
T(x_1, x_2, x_3),
$$
where $f_i$ are arbitrary functions, is also a solution. 
Introduce the new function
\beq\label{Y}
Y(x_1, x_2, x_3)=\frac{T(x_1, x_2, x_3+1)
T(x_1, x_2, x_3-1)}{T(x_1+1, x_2, x_3)T(x_1-1, x_2, x_3)}
\eeq
which remains invariant under multiplication of the 
$T$-function by $f_i$, as above. The Hirota equation for $T$ implies 
the following equation for $Y$: 
\beq\label{hir7}
Y(x_1, x_2+1, x_3)Y(x_1, x_2-1, x_3)=
\frac{(1+Y(x_1, x_2, x_3+1))
(1+Y(x_1, x_2, x_3-1))}{(1+Y^{-1}(x_1+1, x_2, x_3))
(1+Y^{-1}(x_1-1, x_2, x_3))}
\eeq
which is called the $Y$-system.

\subsubsection{Auxiliary linear problems for the Hirota equation}

Let us come back to equation (\ref{hir5}) and denote for brevity
$$
\tau (p_1 +1, p_2 , p_3):= \tau_1,\quad  \tau (p_1, p_2 +1 , p_3):=
\tau_2, \quad \tau (p_1 +1, p_2 +1 , p_3):= \tau_{12}, \quad
\ldots
$$
and $\tau (p_1+2, p_2, p_3):=\tau_{11}$ and so on.
Let $\alpha \beta \gamma $ be any cyclic permutation of 
$123$. Consider the following system of three 
linear equations for the function 
$\psi =\psi (p_1 , p_2 , p_3)$:
\begin{equation}\label{A1}
\left ( e^{\p_{\alpha}}+z_{\gamma}\, \frac{\tau \,
\tau_{\alpha \beta}}{\tau_{\alpha}\tau_{\beta}} \right ) \psi =
e^{\p_{\beta}}\psi\,, \quad \alpha \beta \gamma = 123, \; 231, \;
312,
\end{equation}
where $z_{\alpha}$ are some parameters and $\p_{\alpha}\equiv \p/
\p_{p_{\alpha}}$. It is not difficult to see that the compatibility 
of these equations implies the Hirota equation for $\tau$: 
\begin{equation}\label{AH}
z_1 \tau_1 \tau_{23}+z_2 \tau_2 \tau_{13}+ z_3
\tau_3 \tau_{12}=0\,.
\end{equation}
Indeed, consider, for  example, the second and the third equations and rewrite
them in the form
$$
\begin{array}{l}
\displaystyle{
\psi (p_1+1)=\psi(p_3+1)+z_2 \frac{\tau \tau_{13}}{\tau_1 \tau_3}}\, \psi ,
\\ \\
\displaystyle{
\psi (p_2+1)=\psi(p_3+1)-z_1 \frac{\tau \tau_{23}}{\tau_2 \tau_3}}\, \psi .
\end{array}
$$
These equations allow one to represent the function 
$\psi (p_1+1, p_2+1)$ as a linear combination of
$\psi (p_3)$, $\psi (p_3+1)$, $\psi (p_3+2)$ in two different ways. 
Compatibility of the linear problems means that the results must coincide.  
Equating to each other the expressions obtained in this way, we see that
the terms proportional to 
$\psi (p_3)$ and $\psi (p_3+2)$ cancel identically while the terms
proportional to $\psi (p_3+1)$ give a non-trivial relation
(provided that $\psi (p_3+1)$ is not identically zero):
$$
z_2\frac{\tau_{133}\tau_{3}}{\tau_{13}\tau_{33}} -
z_1\frac{\tau_{123}\tau_{1}}{\tau_{12}\tau_{13}}=
z_2\frac{\tau_{123}\tau_{2}}{\tau_{12}\tau_{23}}-
z_1\frac{\tau_{233}\tau_{3}}{\tau_{23}\tau_{33}},
$$
which means that
$$
\frac{z_1\tau_{1}\tau_{23}+z_2\tau_{2}\tau_{13}}{\tau_{12}\tau_3}
$$
is a periodic function of
$p_3$ with period $1$ and an arbitrary function of 
$p_1, p_2$. Since we do not imply any special
periodicity properties, we assume that this function
does not depend on $p_3$. Therefore, we come to the relation
$$
z_1 \tau (p_1+1)\tau (p_2 +1, p_3+1)+z_2
\tau (p_2+1)\tau (p_1 +1, p_3+1)=
h(p_1, p_2) \tau (p_3+1)\tau (p_1 +1, p_1+1),
$$
where $h$ may be an arbitrary function of $p_1, p_2$.
The compatibility with the third linear problem implies that
$h$ must be a constant equal to 
$-z_3$, hence the Hirota equation follows. In the cases when 
$h$ can be fixed in some other way, (for example, from boundary conditions),
two liner problems are enough. 

\noindent
{\bf Remark.} Generally speaking, compatibility of linear problems
follows from existence of a continuous family of common solutions. In our case
the coefficient functions in the difference operators
(\ref{A1}) are such that the compatibility is equivalent to the presence 
of {\it at least one} 
non-trivial solution.

Introduce the function $\varphi = \psi \tau$, then the linear problems
acquire the form
\begin{equation}\label{A2}
\tau_{\gamma}\varphi_{\beta}- \tau_{\beta}\varphi_{\gamma}
+z_{\alpha}\tau_{\beta \gamma}\varphi =0 \,, \quad \quad
\alpha \beta \gamma = 123, \; 231, \; 312\,.
\end{equation} 
From the first and the second equations we get
$$
\tau_2 = \frac{\tau_3 \varphi_2 + z_1
\tau_{23}\varphi}{\varphi_3}\,, \quad \tau_1 = \frac{\tau_3
\varphi_1 - z_2 \tau_{13}\varphi}{\varphi_3}\,.
$$
Substituting this into the Hirota equation, we obtain yet another linear problem
com\-pa\-ti\-ble with the other three:
\begin{equation}\label{A3}
z_1 \tau_{23}\varphi_1 + z_2 \tau_{13}\varphi_2 +
z_3 \tau_{12}\varphi_3 =0\,.
\end{equation}
All four linear problems can be unified into one matrix equation
\begin{equation}\label{A4}
\left (
\begin{array}{cccc}
0 & \tau_3 & -\tau_2 & z_1 \tau_{23}\\ &&&\\
-\tau_3 &0& \tau_1 &  z_2 \tau_{13}\\ &&&\\
\tau_2 & -\tau_1 & 0 & z_3 \tau_{12}\\ &&&\\
-z_1 \tau_{23} & -z_2 \tau_{13} & -z_3 \tau_{12} &
0
\end{array} \right )
\left (
\begin{array}{c}
\varphi_1 \\ \\    \varphi_2 \\ \\ \varphi_3 \\ \\
\varphi
\end{array} \right ) =0\,.
\end{equation}
The determinant of the antisymmetric matrix in the left hand side is equal to
$(z_1 \tau_1 \tau_{23} +  z_2 \tau_2 \tau_{13}
+z_3 \tau_3 \tau_{12} )^2$. It vanishes if $\tau$ satisfies the Hirota equation,
with the rank of the matrix in this case being equal to 2, hence only two 
of the four equations are linearly independent.

The system
(\ref{A2}) can be treated as a system of linear equations 
for $\tau$ with the coefficients
$\varphi$. Shifting the variables $p_{\beta} \to
p_{\beta}-1$, $p_{\gamma} \to p_{\gamma}-1$ aand changing their sign, 
$p_{1,2,3}\to -p_{1,2,3}$, we see that the form of this system is the same.
Since the Hirota equation is invariant under simultaneous change of signs 
of all variables, the compatibility condition gives the Hirota equation for
$\varphi$ of the same form:
\begin{equation}\label{A5}
z_1 \varphi_{23}\varphi_1 + z_2 \varphi_{13}\varphi_2 +
z_3 \varphi_{12}\varphi_3 =0\,.
\end{equation}

Let us pass to the new variables $x_1, x_2 , x_3$ according to the formulas
$$
\begin{array}{l}
p_1 =\frac{1}{2}(- \varepsilon_1 x_1 \! +\! \varepsilon_2
x_2 \! +\! \varepsilon_3 x_3), \\ \\
p_2 =\frac{1}{2}(\, \varepsilon_1 x_1 -\varepsilon_2 x_2 +
\varepsilon_3 x_3), \\ \\
p_3 =\frac{1}{2}(\, \varepsilon_1 x_1 +\varepsilon_2 x_2 -
\varepsilon_3 x_3)\,.
\end{array}
$$
Here $\varepsilon_{\alpha}=\pm 1$ is some fixed set of signs 
(there are $2^3 = 8$ possible choices). The inverse transformation is given by
$$
x_1 =\varepsilon_1 (p_2 + p_3)\,, \quad x_2 =\varepsilon_2 (p_1 +
p_3)\,, \quad x_3 =\varepsilon_3 (p_1 + p_2) \,.
$$
Introduce the functions $T(x_1, x_2 , x_3)= \tau (p_1 , p_2 , p_3)$ and 
$F(x_1, x_2 , x_3)= \varphi (p_1 , p_2 , p_3 )$, with the variables
$x_{\alpha}$ and $p_{\alpha}$ being connected by the above formulas. 
In the new variables, the system of four linear problems has the form
\begin{equation}\label{A6}
\left (
\begin{array}{cccc}
0 & T_{12} & -T_{13} & z_1 T_{1123}\\ &&&\\
-T_{12} &0& T_{23} &  z_2 T_{1223}\\ &&&\\
T_{13} & -T_{23} & 0 & z_3 T_{1233}\\ &&&\\
-z_1 T_{1123} & -z_2 T_{1223} & -z_3 T_{1233} & 0
\end{array} \right )
\left (
\begin{array}{c}
F_{23} \\ \\  F_{13} \\ \\ F_{12} \\ \\ F
\end{array} \right ) =0\,,
\end{equation}
where $T_1 \equiv T(x_1 \! +\! \varepsilon_1 , x_2 ,
x_3)$, $T_{12} \equiv T(x_1 \! +\! \varepsilon_1 , x_2 \! +\!
\varepsilon_2, x_3)$, $T_{1123} \equiv T(x_1 \! +\! 2\varepsilon_1 ,
x_2 \! +\! \varepsilon_2, x_3\! +\! \varepsilon_3)$, and so on (and
similarly for $F$).

Compatibility of these linear problems implies the Hirota equation
$$
z_1 T_{1123}T_{23} + z_2 T_{1223}T_{13} + z_3
T_{1233}T_{12} =0\,.
$$
After the shift of the variables $x_{\alpha}\to
x_{\alpha}-\varepsilon_{\alpha}$ we get the equation
$$
\begin{array}{l}
z_1 T(x_1 \! +\! \varepsilon_1, x_2 , x_3) T(x_1 \! -\!
\varepsilon_1, x_2 , x_3) + z_2 T(x_1, x_2 \! +\!
\varepsilon_2 , x_3) T(x_1, x_2 \! -\!\varepsilon_2, x_3) 
\\ \\
\phantom{aaa}+ \,\,
z_3 T(x_1, x_2, x_3 \! +\! \varepsilon_3) T(x_1, x_2, x_3 \!
-\! \varepsilon_3) =\, 0.
\end{array}
$$
Note that its form does not depend on the choice of signs $\varepsilon_{\alpha}$
and coincides with (\ref{hir6}).
The systems of linear equations
(\ref{A6}) give difference Backlund transformations for it.
However, only four of them
(correspnding, for example, to the choices
$\varepsilon_1 = \varepsilon_2 =\varepsilon_3
=1$, $-\varepsilon_1 = \varepsilon_2 =\varepsilon_3 =1$,
$\varepsilon_1 = -\varepsilon_2 =\varepsilon_3 =1$, $\varepsilon_1 =
\varepsilon_2 =-\varepsilon_3 =1$) are really different because
the simultaneous change of signs means passing to the conjugate
system of linear problems in which the roles of 
$T$ and $F$ are interchanged.

\subsubsection{Continuous (dispersionless) limit of the Hirota
equation}

The bilinear difference Hirota equation admits different continuous limits.
As we know, one of them leads to the KP equation with the whole hierarchy.
Here we describe another continuous limit which leads to the dispersionless
analogue of the KP hierarchy. 

We begin with introducing a small parameter
$\epsilon$ and redefine the times and the tau-function 
in the following way:
\beq\label{displess0}
t_k = \frac{T_k}{\epsilon}\,, \quad \quad
\tau ({\bf T})=\tau_{\epsilon}({\bf T}).
\eeq
Introduce also the differential operator
\beq\label{displess1}
D(z)=\sum_{k\geq 1}\frac{z^{-k}}{k}\, \p_{T_k}
\eeq
which is a ``generating function'' of the vector fields
$\p _{T_k}$. Equation (\ref{hir4}) will acquire the form
\beq\label{displess2}
(\lambda_2-\lambda_3)\left (e^{-\epsilon D(\lambda_1)}\tau_{\epsilon}\right )
\left (e^{-\epsilon (D(\lambda_2)+D(\lambda_3))}\tau_{\epsilon}\right )
+(231)+(312)=0.
\eeq

The parameter $\epsilon$ plays the role of the lattice spacing.
The dispersionless limit is well-defined on the class of solutions
for which a finite limit
предел
\beq\label{displess3}
F({\bf T})=\lim_{\epsilon \to 0}\epsilon^2 \log \tau_{\epsilon}({\bf T})
\eeq
exists, i.e., such that the tau-function $\tau_{\epsilon}({\bf T})$ 
behaves in the limit
$\epsilon \to 0$ as
$e^{F({\bf T})/\epsilon^2}$, where $F({\bf T})$ is some function of the variables
$T_k$. If the limit exists, it is sometimes called the
dispersionless tau-function although actually it is the limit
of its re-scaled logarithm. The tau-function itself in this limit does not make sense
because the expression 
$\tau_{\epsilon}=e^{F/\epsilon^2}$ is not defined at $\epsilon = 0$. Note that
the soiton-like solutions do not have the dispersionless limit
(the very existence of a soliton is the effect due to non-zero dispersion). 

\noindent
{\bf Problem.} Give an example of a solution
which has the dispersionless limit. 

Substituting $\tau_{\epsilon}=e^{F/\epsilon^2}$ into equation (\ref{displess2}),
we get
$$
(\lambda_2 -\lambda_3)e^{\epsilon^{-2}e^{-\epsilon D(\lambda_1)}F}
e^{\epsilon^{-2}e^{-\epsilon (D(\lambda_2)+D(\lambda_3)}F}
+(231)+(312)=0.
$$
The limit $\epsilon \to 0$
yields the following equation for $F$:
\beq\label{displess4}
(\lambda_1 -\lambda_2)e^{D(\lambda_1)D(\lambda_2)F}+
(\lambda_2 -\lambda_3)e^{D(\lambda_2)D(\lambda_3)F}+
(\lambda_3 -\lambda_1)e^{D(\lambda_1)D(\lambda_3)F}=0,
\eeq
which is called the dispersionless analogue of the KP hierarchy in the Hirota form.

\noindent
{\bf Exercise.} Give a detailed derivation of equation (\ref{displess4}).

\noindent
Tending $\lambda_3 \to \infty$ we will have:
$$
(\lambda_1 -\lambda_2)\left (1-e^{D(\lambda_1)D(\lambda_2)F}\right )
=\Bigl (D(\lambda_1) -D(\lambda_2)\Bigr )\p_{t_1}F
$$
or
\beq\label{displess5}
D(\lambda_1)D(\lambda_2)F=\log 
\frac{p(\lambda_1)-p(\lambda_2)}{\lambda_1 -\lambda_2},
\eeq
where
$$
p(\lambda )=\lambda -\sum_{k\geq 1}\frac{\lambda^{-k}}{k}\,
F_{1k}\,,       \quad F_{ik}:=\p_{t_i}\p_{t_k}F.
$$
In fact the relation (\ref{displess5}) is equivalent to (\ref{displess4}), 
since the latter is obtained from the former by applying $\exp$, 
multiplying by $\lambda_1 -\lambda_2$ and summing three such equations
(for the pairs
$\{\lambda_1, \lambda_2\}$,
$\{\lambda_2, \lambda_3\}$ and $\{\lambda_3, \lambda_1\}$).
It is seen from (\ref{displess5}) that these relations 
express 
$F_{ij}$ with $i,j\geq 2$ through $F_{1j}$, $j\geq 1$.

\newpage

\section{Toda lattice hierarchy}

The Toda lattice hierarchy, or, more precisely, 2D Toda lattice hierarchy (2DTL)
is ``twice as large'' as the KP hierarchy. It has the double infinite set of times
$$\{ \ldots t_{-3}, t_{-2}, t_{-1}, t_0, t_1, t_2, t_3, \ldots \},$$
with the zeroth time, $t_0=n$, taking integer values. The KP hierarchy can be 
embedded into the 2DTL one in the sense that when all the non-positive times
are frozen, the dependent variables as functions of the other times satisfy the equations
of the KP hierarchy. In this section we briefly discuss the main points of the theory
of the 2DTL hierarchy.

\subsection{Commutation representations of the 2DTL hierarchy}

The 2DTL hierarchy has two Lax operators, $L$ and $\bar L$ (we use the same notation
as for the Lax operator of the KP hierarchy and hope that this will not lead to a confusion 
because they do not mix). They are {\it pseudo-difference} operators, i.e. infinite series in
powers of the shift operator $e^{\p_n}$ acting on a function $f(n)$ 
as $e^{\pm \p_n}f(n)=f(n\pm 1)$:
\beq\label{tl1}
\begin{array}{l}
L=e^{\p_n}+u_0 +u_1e^{-\p_n}+ u_2e^{-2\p_n}+\ldots \, ,
\\ \\
\bar L = ce^{-\p_n} +\bar u_0 +\bar u_1e^{\p_n}+ \bar u_2e^{2\p_n}+\ldots \,.
\end{array}
\eeq
Here the coefficients $c, u_i, \bar u_i$ are functions of $n$ and all the higher times
($\bar u_i$ is {\it not} a complex conjugate of $u_i$). 

On the algebra of pseudo-difference 
operators we have two truncation operations, $(\ldots )_{\geq 0}$ and 
$(\ldots )_{<0}$ defined by
$$
\Bigl (\sum_{k\in \z}a_k e^{k\p_n}\Bigr )_{\geq 0}=
\sum_{k\geq 0}a_k e^{k\p_n}, \quad
\Bigl (\sum_{k\in \z}a_k e^{k\p_n}\Bigr )_{< 0}=
\sum_{k<0}a_k e^{k\p_n}.
$$
The 2DTL hierarchy is the system of differential-difference equations 
for the coefficient functions of the Lax operators that follow
from the Lax equations
\beq\label{tl2}
\p_{t_j}L=[B_j, L], \quad \p_{t_j}\bar L=[B_j, \bar L], \quad j=\pm 1, \pm 2, \ldots ,
\eeq
where the difference operators $B_j$ are
\beq\label{tl3}
B_j = (L^{j})_{\geq 0}, \quad B_{-j}=(\bar L^{j})_{<0}, \quad j\geq 1.
\eeq
For example, $B_1=e^{\p_n}+u_0(n)$, $B_{-1}=c(n)e^{-\p_n}$.

As in the KP case, the Lax representation is equivalent to the zero curvature
equations
\beq\label{tl4}
\p_{t_j}B_k - \p_{t_k}B_j -[B_j, B_k]=0, \quad j,k=\pm 1, \pm 2, \ldots .
\eeq

\noindent
{\bf Exercise.} Show that the zero curvature equation at $j=1$, $k=-1$ gives the system
of equations
\beq\label{tl5}
\left \{ \begin{array}{l}
\p_{t_1}\log c(n)=u_0(n)-u_0(n-1),
\\ \\
\p_{t_{-1}} u_0(n)=c(n)-c(n+1).
\end{array} \right.
\eeq

\noindent
Excluding $u_0$ from (\ref{tl5}), one obtains the following closed second order 
dif\-fe\-ren\-ti\-al-\-dif\-fe\-ren\-ce equation for $c(n)$:
\beq\label{tl6}
\p_{t_1}\p_{t_{-1}}\log c(n)=2c(n)-c(n+1)-c(n-1)
\eeq
which is one of the forms of the 2D Toda equation. In terms of the function
$\varphi_n$ introduced through the relation $c(n)=e^{\varphi_n -\varphi_{n-1}}$ it
acquires the most familiar form
\beq\label{tl7}
\p_{t_1}\p_{t_{-1}}\varphi_n=e^{\varphi_n -\varphi_{n-1}}-
e^{\varphi_{n+1}-\varphi_n}.
\eeq

Let us point out two important reductions of the 2DTL equation.
The Toda chain equation is the 1D reduction of the 2DTL such that
$(\p_{t_1}+\p_{t_{-1}})c(n)=(\p_{t_1}+\p_{t_{-1}})u_0(n)=0$. The Toda chain equation reads
\beq\label{tl7a}
\p^2_{t_1}\varphi_n=e^{\varphi_{n+1}-\varphi_n}-e^{\varphi_n -\varphi_{n-1}}.
\eeq
The 2-periodic constraint $\varphi_{n+2}=\varphi_n$ leads to the sine-Gordon equation
in ``light cone coordinates''
\beq\label{tl17b}
\p_{t_1}\p_{t_{-1}}\varphi =4\sin \varphi
\eeq
for $\varphi =i(\varphi_0-\varphi_1)$. 

The form of $L, \bar L$ can be made more symmetric if one applies a ``gauge transformation''
$$
L\mapsto L^G =G^{-1}LG, \quad \bar L\mapsto \bar L^G =G^{-1}\bar LG,
$$
$$
B_j \mapsto B_j^G=G^{-1}B_jG-G^{-1}\p_{t_j}G
$$
with a function $G(n)=e^{\alpha \varphi_n}$. At $\alpha =\frac{1}{2}$ we have the 
gauge-transformed operators
$$
\begin{array}{l}
L=u_{-1}e^{\p_n}+u_0 +u_1e^{-\p_n}+ u_2e^{-2\p_n}+\ldots \, ,
\\ \\
\bar L = \bar u_{-1}
e^{-\p_n} +\bar u_0 +\bar u_1e^{\p_n}+ \bar u_2e^{2\p_n}+\ldots 
\end{array}
$$
with $u_{-1}(n)=\bar u_{-1}(n+1)$ and
$$
\begin{array}{l}
B_j =  (L^{j})_{>0}+\frac{1}{2}(L^{j})_{0}, \quad
B_{-j}  = (\bar L^{j})_{<0}+\frac{1}{2}(\bar L^{j})_{0},
\end{array}
$$
with the obvious definition of the truncation operations 
$(\ldots )_{> 0}$ and 
$(\ldots )_{0}$.
In what follows we use the original $\alpha =0$ gauge. 

\subsection{Conserved quantities}

We call {\it the residue} $\mbox{res}_{e^{\p}}$ 
of a pseudo-difference operator $\displaystyle{P=\sum_k p_k(n)e^{k\p_n}}$ the
coefficient $p_0$: $\mbox{res}_{e^{\p}}P=(P)_0=p_0$. The following proposition is a difference
counterpart of the corresponding property of the pseudo-differential operators.

\noindent
{\bf Proposition.} {\it The residue of the commutator of pseudo-difference operators
is a ``full difference'':
$$
\mbox{res}_{e^{\p}}[P, Q]=\Delta C, \quad \Delta :=e^{\p_n}-1,
$$
where $C$ is a polynomial of coefficients of the operators $P$, $Q$.}

\noindent
The proof is elementary. 

The densities of conserved quantities for the 2DTL hierarchy are 
$\mbox{res}_{e^{\p}}L^k =(L^k)_0$, $k\geq 1$, i.e., the conserved quantities
are
\beq\label{t17b}
J_k=\sum_{n=-\infty}^{\infty}\mbox{res}_{e^{\p}}L^k.
\eeq
Indeed,
$$
\p_{t_m}J_k=\p_{t_m}\sum_{n=-\infty}^{\infty}\mbox{res}_{e^{\p}}(L^k)=
\sum_{n=-\infty}^{\infty}\mbox{res}_{e^{\p}} [B_k, L^k]=\sum_{n=-\infty}^{\infty}\Delta C(n)
$$
which is zero for rapidly decreasing and periodic solutions. A similar series of conserved
quantities exists for the second Lax operator $\bar L$.

\subsection{Dressing operators and $\psi$-functions}

Similarly to the KP case, the dressing operators are pseudo-difference
operators of the form
\beq\label{tl8}
\begin{array}{l}
W=1+ w_1e^{-\p_n}+ w_2e^{-2\p_n}+\ldots \, ,
\\ \\
\bar W = \bar w_0 +\bar w_1e^{\p_n}+ \bar w_2e^{2\p_n}+\ldots 
\end{array}
\eeq
such that
\beq\label{tl9}
L=We^{\p_n}W^{-1}, \quad \bar L = \bar W e^{-\p_n}\bar W^{-1}.
\eeq
The equations of motion for the dressing operators are
\beq\label{tl10}
\begin{array}{l}
\p_{t_j}W=B_jW-We^{j\p_n}, \quad \p_{t_{-j}}W=B_{-j}W, \quad j\geq 1,
\\ \\
\p_{t_{-j}}\bar W=B_{-j}\bar W-\bar We^{-j\p_n}, \quad \p_{t_{j}}\bar W=B_{j}\bar W, 
\quad j\geq 1,
\end{array}
\eeq
Accordingly, there are two $\psi$-functions, $\psi$ and $\bar \psi$, 
depending on the spectral parameter $z$, which are obtained by
applying the dressing operators to the functions $z^n e^{\xi ({\bf t}_+, z)}$ and
$z^n e^{\xi ({\bf t}_-, z^{-1})}$:
\beq\label{tl11}
\begin{array}{l}
\psi = Wz^n e^{\xi ({\bf t}_+, z)}=z^n e^{\xi ({\bf t}_+, z)}\Bigl (
1+w_1 z^{-1} +w_2 z^{-2} +\ldots \Bigr ),
\\ \\
\bar \psi =\bar Wz^n e^{\xi ({\bf t}_-, z^{-1})}=
z^n e^{\xi ({\bf t}_-, z^{-1})}\Bigl (\bar w_0+ \bar w_1 z +\bar w_2 z^2 +\ldots \Bigr ),
\end{array}
\eeq
where ${\bf t}_{\pm}=\{t_{\pm 1}, t_{\pm 2}, t_{\pm 3}, \ldots \}$.
Here $\psi$ should be understood as a series around $z=\infty$ while $\bar \psi$
as a series around $z=0$. In fact for algebro-geometric solutions $\psi$ and $\bar \psi$
are expansions around $z=\infty$ and $z=0$ of one and the same function on a Riemann
surface (the Baker-Akhiezer function). The functions $\psi$, $\bar \psi$ satisfy
the linear equations
\beq\label{tl12}
\p_{t_j}\psi =B_j \psi , \qquad \p_{t_j}\bar \psi =B_j \bar \psi , \qquad
j=\pm 1, \pm 2, \ldots
\eeq
and
\beq\label{tl13}
L\psi =z\psi , \qquad \bar L\bar \psi =z^{-1}\bar \psi .
\eeq
Their compatibility is equivalent to the zero curvature equations and the Lax equations.
The simplest linear equations are
\beq\label{tl14}
\p_{t_1}\psi (n) =\psi (n+1)+u_0(n)\psi (n), \qquad
\p_{t_{-1}}\psi (n) =c(n) \psi (n-1).
\eeq

Along with the $\psi$-functions $\psi$, $\bar \psi$, one may introduce the adjoint
functions $\psi^*$, $\bar \psi^*$:
\beq\label{tl11a}
\psi^* =(W^{\dag})^{-1}z^{-n} e^{-\xi ({\bf t}_+, z)}, \quad
\bar \psi^* =(\bar W^{\dag})^{-1}z^{-n} e^{-\xi ({\bf t}_+, z)},
\eeq
where $(f(n)e^{k\p_n})^{\dag}=e^{-k\p_n}f(n)$.

\subsection{The tau-function of the 2DTL hierarchy}

The tau-function $\tau_n ({\bf t}_+, {\bf t}_-)$ of the 2DTL hierarchy 
can be introduced by the formulas similar to (\ref{jap1}):
\beq\label{tl15}
\psi (n)=z^n e^{\xi ({\bf t}_+, z)}\, 
\frac{\tau_n ({\bf t}_+-[z^{-1}], {\bf t}_-)}{\tau_n ({\bf t}_+, {\bf t}_-)}, \quad
\psi ^* (n)=z^{-n} e^{-\xi ({\bf t}_+, z)}\, 
\frac{\tau_n ({\bf t}_++[z^{-1}], {\bf t}_-)}{\tau_n ({\bf t}_+, {\bf t}_-)},
\eeq
\beq\label{tl16}
\bar \psi (n)=z^n e^{\xi ({\bf t}_-, z^{-1})}\, 
\frac{\tau_{n+1} ({\bf t}_+, {\bf t}_- -[z])}{\tau_n ({\bf t}_+, {\bf t}_-)}, \quad
\bar \psi ^*(n)=z^{-n} e^{-\xi ({\bf t}_-, z^{-1})}\, 
\frac{\tau_{n-1} ({\bf t}_+, {\bf t}_- +[z])}{\tau_n ({\bf t}_+, {\bf t}_-)}.
\eeq
All equations of the 2DTL hierarchy are encoded in the bilinear relation
\beq\label{tl17}
\begin{array}{c}
\displaystyle{\oint_{{\sf C}} z^{n-n'}e^{\xi ({\bf t}_+ - {\bf t}_+', z)}
\tau_n ({\bf t}_+ -[z^{-1}], {\bf t}_-)\tau_{n'} ({\bf t}_+' +[z^{-1}], {\bf t}_-')\, dz}
\\ \\
=\, \displaystyle{\oint_{{\sf C}} z^{-n+n'}e^{\xi ({\bf t}_- - {\bf t}_-', z)}
\tau_{n+1} ({\bf t}_+, {\bf t}_- -[z^{-1}], )\tau_{n'-1} ( {\bf t}_+', 
{\bf t}_-' +[z^{-1}])\, z^{-2}\, dz}
\end{array}
\eeq
valid for any sets ${\bf t}_{+}, {\bf t}_{-}, {\bf t}_{+}', {\bf t}_{-}'$ and integers 
$n , n'$. Note that at $n=n'$, ${\bf t}_{-}={\bf t}_{-}'$ the right hand side vanishes
and the bilinear relation reduces to the one for the KP hierarchy (\ref{hir1a}) for the set
of ``positive'' times. Similarly, at $n-n'=-2$,  ${\bf t}_{+}={\bf t}_{+}'$ the left
hand side vanishes and we get the bilinear relation for the KP hierarchy
for the set of ``negative'' times. 

Taking $n=n'$, ${\bf t}_+'={\bf t}_+ -[a^{-1}]$, ${\bf t}_-'={\bf t}_- -[b^{-1}]$ and
noting that $\displaystyle{e^{\xi ({\bf t}_+ - {\bf t}_+', z)}=
\Bigl (1\! -\! \frac{z}{a}\Bigr )^{-1}}$,
$\displaystyle{e^{\xi ({\bf t}_- - {\bf t}_-', z)}=\Bigl (1-\frac{z}{b}\Bigr )^{-1}}$,
one finds from (\ref{tl17}) with the help of residue calculus:
\beq\label{tl18}
\begin{array}{c}
\tau_n ({\bf t}_+-[a^{-1}], {\bf t}_-)\tau_n({\bf t}_+, {\bf t}_- -[b^{-1}])-
\tau_n ({\bf t}_+, {\bf t}_-)\tau_n ({\bf t}_+-[a^{-1}], {\bf t}_- -[b^{-1}])
\\ \\
=\, (ab)^{-1}\tau_{n+1}({\bf t}_+, {\bf t}_- -[b^{-1}])
\tau_{n-1} ({\bf t}_+-[a^{-1}], {\bf t}_-).
\end{array}
\eeq
The simplest tau-function obeying this equation is
$$
\tau_n ({\bf t}_+, {\bf t}_-)=\exp \Bigl (-\sum_{k\geq 1}kt_kt_{-k}\Bigr ).
$$
It corresponds to the trivial solution. 

The 2D Toda equation itself is obtained from
(\ref{tl18}) as a coefficient at $(ab)^{-1}$ in the expansion as $a,b \to \infty$:
\beq\label{tl19}
\p_{t_1}\p_{\bar t_1}\log \tau_n =-\frac{\tau_{n+1}\tau_{n-1}}{\tau^2_n}.
\eeq
The original variables $c(n)$, $u_0(n)$ are connected with the tau-funcntion by
the formulas
\beq\label{tl20}
c(n)=\frac{\tau_{n+1}\tau_{n-1}}{\tau^2_n}, \qquad
u_0(n)=\p_{t_1}\log \frac{\tau_{n+1}}{\tau_n}.
\eeq

\subsection{Solutions to the 2DTL hierarchy}

The 2DTL hierarchy has different classes of solutions which are analogues 
of the corresponding classes for the KP hierarchy. Below we give some details
on the soliton solutions and elliptic solutions.

\subsubsection{Soliton solutions}

The soliton solutions to the KP hierarchy can be extended to the 2DTL hierarchy.
Here we present the 2DTL analogues of the determinant 
representations (\ref{kptauN}) and (\ref{kptauN1}).
It is convenient to redefine the tau-function:
$$
\tau '_n({\bf t}_+, {\bf t}_-)=\exp \Bigl (\sum_{k\geq 1}kt_kt_{-k}\Bigr )
\tau _n({\bf t}_+, {\bf t}_-).
$$
The analogue of (\ref{kptauN}) is
\beq\label{tl21}
\tau_n' ({\bf t}_+, {\bf t}_-)=
\det_{N\! \times \! N}\Bigl (e^{\xi ({\bf t}_+, q_i)+\xi ({\bf t}_-, q_i^{-1})}
q_i^{n-j}\! +\! b_i e^{\xi ({\bf t}_+, p_i)+\xi ({\bf t}_-, p_i^{-1})}p_i^{n-j}\Bigr ).
\eeq
The analogue of (\ref{kptauN1}) is
\beq\label{tl22}
\tau_n' ({\bf t}_+, {\bf t}_-)=
\det_{N\! \times \! N}
\left (\delta_{ij}+\frac{a_kq_k}{q_k\! -\! p_i}\Bigl (\frac{p_i}{q_k}\Bigr )^n
e^{\xi ({\bf t}_+, p_i)-\xi ({\bf t}_+, q_k)+
\xi ({\bf t}_-, p_i^{-1})-\xi ({\bf t}_-, q_k^{-1})}
\right ).
\eeq
These two forms are in fact equivalent (i.e. differ by an exponential function of a linear
form in times and by a constant factor). 
The structure of the right hand side of (\ref{tl22}) is as follows:
\beq\label{tl23}
\tau_n' ({\bf t}_+, {\bf t}_-)
=1+\sum_i e^{\eta_i}+\sum_{i<j}c_{ij} e^{\eta_i +\eta_j}+
\sum_{i<j<k} c_{ij}c_{ik}c_{jk}e^{\eta_i +\eta_j+\eta_k}+\ldots ,
\eeq
where
$$
e^{\eta_i}=\frac{a_iq_i}{q_i-p_i}\Bigl (\frac{p_i}{q_k}\Bigr )^n
e^{\xi ({\bf t}_+, p_i)-\xi ({\bf t}_+, q_k)+
\xi ({\bf t}_-, p_i^{-1})-\xi ({\bf t}_-, q_k^{-1})},
$$
$$
c_{ij}=\frac{(p_i-p_j)(q_i-q_j)}{(p_i-q_j)(q_i-p_j)}.
$$
The more general soliton-like solutions like the ones discussed in section 3.6.2
also exist.

\subsubsection{Elliptic solutions and the Ruijsenaars-Schneider system}

In this section we study solutions to the Toda lattice equation 
for which $c(n), u_0(n)$ are elliptic (double-periodic) functions of $x=n\eta$
($\eta$ is an arbitrary parameter). We denote them by $c(x), u_0(x)$. 
For such solutions the tau-function has the form
\beq\label{te1}
\tau (x)=Ce^{cx^2 +rxt_1 +\bar rx\bar t_1 +\gamma t_1 \bar t_1}\prod_{i=1}^{N}\sigma (x-x_i),
\eeq
then
$$
c(x)=e^{2\eta ^2c}\prod_k \frac{\sigma (x-x_k+\eta )\sigma (x-x_k-\eta )}{\sigma ^2(x-x_k)},
$$
$$
u_0(x)=\sum_k \dot x_k \Bigl (\zeta (x-x_k)-\zeta (x-x_k+\eta )\Bigr )+r\eta ,
$$
where $\dot x_k=\p_{t_1}x_k$. 

We will investigate the dynamics of poles
as functions of the time $t_1$. 
To this end, 
it is enough to solve the first linear problem in (\ref{tl14})
$$
\p_{t_1}\psi (x)=\psi (x+\eta ) +u_0(x)\psi (x)
$$
with $u_0(x)$ as above and the following pole ansatz for the $\psi$-function
similar to (\ref{ell4}):
\beq\label{te2}
\psi =z^{x/\eta}e^{t_1z}\sum_{i=1}^N c_i \Phi (x-x_i, \lambda ).
\eeq
Here $\Phi$ is the same function (\ref{Phi}) as in section 3.6.6.
Substituting (\ref{te2}) into
(\ref{tl14}), we get:
$$
z\sum_i c_i \Phi (x-x_i)+\sum_i \dot c_i \Phi (x-x_i)-\sum_i c_i \dot x_i \Phi '(x-x_i)=
z\sum_i c_i \Phi (x-x_i+\eta )
$$
$$+\left (
\sum_k \dot x_k \Bigl (\zeta (x-x_k)-\zeta (x-x_k+\eta )\Bigr )+r\eta \right )
\sum_i c_i \Phi (x-x_i).
$$
The second order poles at $x=x_i$ cancel identically. 
The cancellation of first order poles at $x=x_i$ and $x=x_i -\eta$ 
leads to the conditions
$$
\left \{ \begin{array}{l}
\displaystyle{zc_i+\dot c_i=r\eta c_i +\dot x_i \sum_{k\neq i}c_k \Phi (x-x_k)
+c_i\sum_{k\neq i}\dot x_k \Bigl (\zeta (x_i-x_k)-\zeta (x_i-x_k+\eta )\Bigr )}
\\ \\
\displaystyle{zc_i-\dot x_i \sum_k c_k \Phi (x_i-x_k-\eta )=0}.
\end{array}
\right.
$$
In the matrix form they read
\beq\label{te3}
\left \{ \begin{array}{l}
{\cal L}{\bf c}=z{\bf c}
\\ \\
\dot {\bf c}={\cal M}{\bf c}
\end{array}
\right.
\eeq
with the $N\! \times \! N$ matrices
$
{\cal L}=\dot X A^{-}$, ${\cal M}=r\eta I+\dot X A -\dot X A^{-}+D^{0}-D^{+},
$
where $A$ is the off-diagonal matrix $A_{ij}=(1-\delta_{ij})\Phi (x_i-x_j)$ and
$$
A^{-}_{ij}=\Phi (x_i-x_j-\eta ), \quad
D^{\pm}_{ij}=\delta_{ij}\sum_{k\neq i}\dot x_k \zeta (x_i-x_k\pm \eta ), 
\quad
D^{0}_{ij}=\delta_{ij}\sum_{k\neq i}\dot x_k \zeta (x_i-x_k ).
$$
The compatibility condition of the linear system (\ref{te3}) is
$\dot {\cal L}+[{\cal L}, {\cal M}]=0$. 
A direct calculation shows that
$$
\dot {\cal L}+[{\cal L}, {\cal M}]=\Bigl (\ddot X \dot X^{-1}+D^++D^--2D^0\Bigr ){\cal L},
$$
so the compatibility condition is
$\ddot X \dot X^{-1}\! +\! D^+ \! +\! D^- \! -\! 2D^0=0$,
which implies equations of motion
\beq\label{te4}
\begin{array}{lll}
\ddot x_i &=&\displaystyle{-\sum_{k\neq i}\dot x_i\dot x_k \Bigl (
\zeta (x_i-x_k+\eta )+\zeta (x_i-x_k-\eta )-2\zeta (x_i-x_k)\Bigr )}
\\ && \\
&=&\displaystyle{\sum_{k\neq i}\dot x_i\dot x_k\frac{\wp '(x_i-x_k)}{\wp (\eta )-
\wp (x_i-x_k)}}
\end{array}
\eeq
together with their Lax representation. These are equations of motion for the 
elliptic Ruijsenaars-Schneider $N$-body system (a relativistic generalization of the 
Calogero-Moser system). 

It can be directly verified that the Ruijsenaars-Schneider system is Hamiltonian
with the Hamiltonian
\beq\label{te5}
{\cal H}=\sum_i e^{p_i}\prod_{k\neq i}\frac{\sigma (x_i-x_k+\eta )}{\sigma (x_i-x_k)},
\eeq
where $p_i, x_i$ are canonical variables. Clearly,
\beq\label{te8a}
{\cal H}=\sum_i \dot x_i =\mbox{const}\, \mbox{tr}\, {\cal L}.
\eeq

It can be also shown that the $t_{-1}$-dynamics of poles leads to the same equations
of motion (\ref{te4}). 

\newpage

\section{Tau-functions as vacuum expectation values
of fer\-mi\-onic operators}

The tau-functions among all functions of infinitely many variables 
are characterized by the property that they satisfy the set of bilinear
Hirota equations. An outstanding discovery of Japanese school 
(Sato, Jimbo, Miwa and others) is another,
equivalent, characterization of tau-functions
as vacuum expectation values of special quantum field operators 
composed of free fermions. 

\subsection{Fermionic operators}

Let us introduce the fermionic operators 
$\psi_n , \psistar_{n}$, $n\in \ZZ$ with the standard (anti)com\-mu\-ta\-tion relations
$[\psi_n , \psi_m ]_+ = [\psistar_n, \psistar_m]_+=0$,
$[\psi_n , \psistar_m]_+=\delta_{mn}$. They generate the infinite dimensional
Clifford algebra. We will also use their Fourier transforms
\beq\label{ferm0}
\psi (z)=\sum_{k\in \z}\psi_k z^k, \quad \quad
\psistar (z)=\sum_{k\in \z}\psistar_k z^{-k}
\eeq
which have the meaning of Fermi fields in the complex plane of the variable
$z$. (We hope that the standard notation 
$\psi (z)$ for the Fermi field will not lead to a confusion with the Baker-Akhiezer 
function $\psi ({\bf t},z)$.)

From the fact that the anti-commutator of any linear combinations of the fermionic
operators is a number, it follows that the commutator of any bilinear expressions in
$\psi_n$ and $\psistar_{n}$ is again bilinear in
$\psi_n$ and $\psistar_{n}$. For example, 
$$
[\psi_m \psistar _n , \, \psi_{m'} \psistar _{n'}]=
\delta _{nm'} \psi_{m} \psistar _{n'}-
\delta _{mn'} \psi_{m'} \psistar _{n}.
$$
We see that the expressions $\psi_m \psistar _n$ commute in the same way as
matrices $E^{(mn)}$ with matrix elements
$E^{(mn)}_{ij}=\delta_{im}\delta_{jn}$, which are generators of the algebra
$gl (\infty )$ of infinite matrices with only finite
(but arbitrary) number of nonzero elements.
More generally, consider a bilinear expression 
\beq\label{XA}
X_A =\sum_{ij}A_{ij}\psi _i \psistar_j
\eeq
with some matrix $A$, 
then $[X_A , \, X_B ]=X_{[A, B]}$, and
$$
[X_A , \, \psi _n ]=\sum_i A_{in}\psi _i \,, \quad \quad
[X_A , \, \psistar _n ]=-\sum_i A_{ni}\psi^*_i .
$$
In order to find the adjoint action of 
$e^{X_A}$ on fermions, we apply the useful formula
\beq\label{ABA}
e^A B e^{-A}=B + [A, B] +\frac{1}{2!}[A, [A,B]] +\ldots
\eeq
which is valid for any two operators $A$, $B$. 

\noindent
{\bf Problem.} Prove (\ref{ABA}). (Hint: 
consider $C(t) = e^{tA} B e^{-tA}$ and show that
the Taylor expansions in $t$ of the both sides
of (\ref{ABA}) coincide.)

\noindent
Using (\ref{ABA}), we get: 
\beq\label{ferm1}
\begin{array}{l}
\displaystyle{
e^{X_A}\psi_n e^{-X_A}=\sum_i \psi _i \, R_{in}},
\\ \\
\displaystyle{
e^{X_A}\psistar_n e^{-X_A}=\sum_i (R^{-1})_{ni}\psistar _i },
\end{array}
\eeq
where the matrix $R$ is connected with $A$ by the relation $R=e^A$.

Thus exponential functions of expressions bilinear in 
$\psi_n$ and $\psistar_{n}$ possess a rather special property: 
the result of their adjoint action on linear combinations of 
$\psi_n$'s (or $\psistar_n$'s) are again linear.
One can say that elements
$G$ of the form
\beq\label{Cl}
G=\exp \left (\sum_{ij}A_{ij}\psi _i \psistar_j \right )
\eeq
perform a linear transformation in the space of fermionic operators:
\beq\label{fermrot}
\begin{array}{l}
\displaystyle{
G\psi_n =\sum_i \psi _i G R_{in}},
\\ \\
\displaystyle{
\psistar_n G=\sum_i G\psistar _i R_{ni} }.
\end{array}
\eeq
Note that the operator $\psistar_n$ transforms with the matrix which is
inverse to the transposed matrix of the transformation for 
$\psi_n$.

\noindent
{\bf Problem.} Show that the operators (\ref{Cl}) obey the group composition law:
$e^{X_A}e^{X_B}=e^{X_C}$, where the matrix
$C$ is defined by the relation $e^C=e^A e^B$.

We will call $G$ of the form (\ref{Cl}) an element of the Clifford group
${\cal G}_{\rm Cliff}$ (this name is not quite precise because 
the Clifford algebra consists of all, not only bilinear, combinations of the 
fermionic operators; however, this name was used in the original papers by
Sato, Jimbo, Miwa and others). The Clifford group is isomorphic to
the infinite dimensional group
$GL(\infty )$. With some reservations, the definition of the Clifford group
can be ex\-ten\-ded 
to matrices $A_{ij}$ in (\ref{Cl}) having infinite number of nonzero elements
(but such that
$A_{ij}=0$ for sufficiently large $|i-j|$).

\subsection{The space of states and the basis}

Introduce the vacuum state $\left |0\rbr$ 
(``Dirac sea''), in which all one-fermion states 
with negative (positive) $n$ are free (occupied): 
$$
\psi_n \rvac =0, \quad n< 0; \quad \quad \quad
\psistar_n \rvac =0, \quad n\geq 0
$$
(for brevity we call indices 
$n\geq 0$ positive).
With respect to this vacuum, $\psi_n$ with
$n<0$ and $\psistar_n$ with $n\geq 0$ are annihilation operators
while 
$\psistar_n$ with $n<0$ and
$\psi_n$ with $n\geq 0$ are creation operators of a particle and a hole
respectively. One can also define
the ``shifted'' Dirac vacuum
$\rvacn$:
$$
\rvacn = \left \{
\begin{array}{l}
\psi_{n-1}\ldots \psi_1 \psi_0 \rvac , \,\,\,\,\, n> 0
\\ \\
\psistar_n \ldots \psistar_{-2}\psistar_{-1}\rvac , \,\,\,\,\, n<0.
\end{array} \right.
$$
The dual vacuum state
(vector from the dual Hilbert space) is such that
$$
\lvac \psistar_n  =0, \quad n< 0; \quad \quad \quad
\lvac \psi_n  =0, \quad n\geq 0
$$
and
$$
\lvacn = \left \{
\begin{array}{l}
\lvac \psistar_{0}\psistar_{1}\ldots \psistar_{n-1} , \,\,\,\,\, n> 0
\\ \\
\lvac \psi_{-1}\psi_{-2}\ldots \psi_{n} , \,\,\,\,\, n<0.
\end{array} \right.
$$
We have:
$$
\begin{array}{l}
 \psi_m \rvacn =0, \quad m < n; 
\qquad 
\psistar_m \rvacn =0, \quad m \ge n, \\ \\
\lvacn  \psi_{m}=0 , \quad m \ge n; 
\qquad 
\lvacn  \psistar_{m}=0 , \quad m < n 
\end{array}
$$
and also
$$
\begin{array}{l}
\psi_n \rvacn = \left|n+1 \rbr,
\qquad  \psistar_n \left|n+1 \rbr = \left|n \rbr,
\\ \\
 \lbr n+1 \right|\psi_n = \lbr n \right|
\qquad  
\lbr n \right|\psistar_n = \lbr n+1 \right| .
\end{array}
$$

As a basis of the Hilbert space
${\cal H}_F$ of states in the theory of free fermions one can take
all states which are obtained form the vacuum
$\left |0\rbr$ by application of a finite number of creation operators
(of particles and holes). 
Let a particle
(created by $\psistar_n$)
carry the charge $-1$, and a hole (created by $\psi_n$) carry the charge
$+1$. Then all basis states have a definite charge equal to the difference 
between the numbers of holes and particles.

Let us give a more precise definition. 
The basis states $\left |\lambda , n\rbr$ are parametrized by the integer number
$n$ and the Young diagram $\lambda$ in the following way.
Let $\lambda =
(\lambda_1 , \ldots , \lambda_{\ell})$ с $\ell =\ell (\lambda )$
be a Young diagram with nonzero rows $\lambda_1, \ldots , \lambda_{\ell}$.
In the Frobenius notation
$\lambda =
(\vec \alpha |\vec \beta )=(\alpha_1, \ldots , \alpha_{d(\lambda )}|
\beta_1 , \ldots , \beta_{d(\lambda )})$, where 
$d(\lambda )$ is the number of boxes on the main diagonal and
$\alpha_i =\lambda_i -i$,
$\beta_i =\lambda'_i -i$. Here $\lambda'$ is the transposed diagram 
(reflected with respect to the main diagonal). 
Then the basis states and their dual are defined as
\beq\label{lambda1}
\begin{array}{l}
\left |\lambda , n\rbr :=
\psistar_{n-\beta_1 -1}\ldots \psistar_{n-\beta_{d(\lambda )}\! -1}\,
\psi_{n+\alpha_{d(\lambda )}}\ldots \psi_{n+\alpha_1}\rvacn ,
\\ \\
\lbr \lambda , n \right |:=
\lvacn \psistar_{n+\alpha_1}\ldots \psistar_{n+\alpha_{d(\lambda )}}\,
\psi_{n-\beta_{d(\lambda )}\! -1}\ldots \psi_{n-\beta_1 -1} .
\end{array}
\eeq
The state $\left |\lambda , n\rbr$ has charge $n$.
For the empty diagram we put
$\left < \emptyset , n\right |=\lvacn$, 
$\left | \emptyset , n\right >=\rvacn$.

\subsection{Vacuum expectation values}

The vacuum expectation value $\lvac \ldots \rvac$ is the Hermitean linear form
on the Clifford algebra. It is defined by the following properties:
$\left. \lvac \! 0\right > =1$,
$\lvac \psi_n \psi_m\rvac = \lvac \psistar_n \psistar_m \rvac =0$
for all $m,n$, and
$$
\lvac \psi_n \psistar_m\rvac =\delta_{mn}\quad
\mbox{for $m<0$}, \quad \quad
\lvac \psi_n \psistar_m\rvac =0 \quad \mbox{for $m\geq 0$}.
$$
The vacuum expectation value of an operator with nonzero charge is equal to 0.

\noindent
{\bf Exercise.} Using the commutation relations for the fermionic operators and the
definition of the shifted vacuum, show that 
$\lvacn \! \left. n\right >=1$ and
$$
\lvacn \psi_i \psistar_j\rvacn =\delta_{ij}\quad
\mbox{at $j<n$}, \quad \quad
\lvacn \psi_i \psistar_j\rvacn =0 \quad \mbox{at $j\geq n$}
$$
for all $i,j$ and $n$.

The scalar product in the fermionic Fock space
${\cal H}_{F}$ is introduce as the vacuum expectation value
of product of the operators creating the states from the vacuum.
The basis vectors (\ref{lambda1})
are then orthonormal:
$$
\lbr \lambda , n\right | \left. \mu , m \rbr =
\delta_{mn}\delta_{\lambda \mu}.
$$
This can be seen moving the operators 
$\psi_{n-\beta_i -1}$ to the right, taking onto account that
the sequences 
$\alpha_1, \alpha_2, \ldots , \alpha_d$ and
$\beta_1, \beta_2, \ldots , \beta_d$ are strictly decreasing.

In general vacuum expectation values of products of fermion operators 
are given by the Wick theorem.
Let $v_i =\sum_j v_{ij}\psi_j$ (respectively, 
$w_i^*=\sum_j w^{*}_{ij}\psistar_j$)
be arbitrary linear combinations of the operators
$\psi_j$ (respectively, $\psistar_j$). 
In the simplest form the Wick theorem states that
$$
\lvacn v_1 \ldots v_m w_m^* \ldots w_1^* \rvacn =
\det_{i,j =1,\ldots , m}\lvacn v_i w_j^* \rvacn ,
$$
$$
\lvacn  w_1^*  \ldots w_m^* v_m  \ldots  v_1  \rvacn =
\det_{i,j =1,\ldots , m}\lvacn w_i^* v_j  \rvacn .
$$
This can be proved by induction. We will not prove this now because later
we will prove a more general statement.

For the fermion fields $\psi (z)$, $\psistar (z)$ we have:
$$
\lvacn \psistar (\zeta ) \psi (z)  \rvacn =
\sum_{j,k}\zeta^{-j}z^k \lvacn \psistar_j \psi_k \rvacn =
\sum_{k\geq n}(z/\zeta )^k =
\frac{z^n\zeta^{1-n}}{\zeta -z}
$$
(assuming that $|\zeta |>|z|$) and
$$
\lvacn \psi (z) \psistar (\zeta )   \rvacn =
\sum_{j,k}\zeta^{-j}z^k \lvacn \psi_k \psistar_j  \rvacn =
\sum_{k< n}(z/\zeta )^k =\frac{z^n\zeta^{1-n}}{z-\zeta}
$$
(assuming that $|z|>|\zeta |$).

\noindent
{\bf Problem.}
Prove the formula
\beq\label{ferm5}
\begin{array}{c}
\displaystyle{
\lvacn \psistar (\zeta_1)\ldots \psistar (\zeta_m)
\psi (z _m)\ldots \psi (z _1)\rvacn =
\prod_{l=1}^{m} (z_l / \zeta_l)^n \cdot \, \det_{i,j}
\frac{\zeta _i}{\zeta_i-z_j}}
\\  \\
\displaystyle{
\phantom{aaaaaaaaaaaaaaaaaaaa}
=\,\, \frac{\prod\limits_{i<i'}(z_i-z_{i'})
\prod\limits_{j>j'}(\zeta _j-\zeta _{j'})}{\prod\limits_{i,j}(\zeta_i-z_j)}
\prod_l z_{l}^{n}\zeta_{l}^{1-n}}.
\end{array}
\eeq

\subsection{Normal ordering}

Here we introduce the useful operation of normal ordering. 
For this, it is necessary to fix a vacuum state. 
The normal ordering $\normord (\ldots )\normord $
with respect to the Dirac vacuum $\rvac$ is defined as follows:
all annihilation operators are moved to the right, all creation operators
are moved to the left, taking into account that each time the positions of two neighboring 
fermionic operators are interchanged the sign factor $(-1)$ appears.  
For example:
$\normord  \psistar_{1}\psi_{1}\normord =
-\psi_{1} \psistar_{1}$, 
$\normord  \psi_{-1}\psi_{0}\normord =
-\psi_{0} \psi_{-1}$, 
$\normord  \psi_{2}\psistar_{1}\psi_1 \psistar_{-2}\normord =
\psi_{2}\psi_1 \psistar_{-2}\psistar_{1}$,
and so on. 

\noindent
{\bf Problem.}
Prove the identity
$
e^{\alpha \psi_k \psistar_k}=1+(e^{\alpha}-1)\psi_k \psistar_k=
\normord e^{(e^{\alpha}-1)\psi_k \psistar_k}\normord \,,
\quad k\geq 0.
$

\noindent
Under the sign of normal ordering, all fermionic operators, both
$\psi_j$ and  $\psistar_j$, anti-commute. We stress that the relations of the 
Clifford algebra are {\it not valid} under the sign of normal ordering, i.e., for example,
$\normord \psistar_{1}\psi_{1}\normord 
\neq \normord (1-\psi_{1} \psistar_{1})\normord$.

Using the normal ordering, one can introduce the charge operator
$Q$:
\beq\label{chargeoper}
Q=\sum_{k\in \z}\normord \psi_k \psistar _k\normord .
\eeq
This operator counts the charge of a state:
$Q\left |\lambda , n\right > =n \left |\lambda , n\right >$, and thus
$\left <\mu , m\right |Q\left |\lambda , n\right >=
n\delta_{nm}\delta_{\lambda \mu}$ (without normal ordering
this matrix element is ill-defined!). The operator
$Q$ has commutation relations
$
[Q, \psi_n ]=\psi_n$, 
$[Q, \psistar _n]=-\psistar_n$ which mean that
$\psi_n$, $\psistar_n$ have charges $\pm 1$. More generally, we say that
an element $X$ of the Clifford algebra
has charge $q$ if $[Q, X]=qX$. 

The definition of normal ordering is closely connected with vacuum expectation value.

\noindent
{\bf Exercise.}
Check that
$\normord \psistar_k \psi_l\normord  =
\psistar_k \psi_l -\lvac \psistar_k \psi_l \rvac$.

\noindent
More generally, for any linear combinations
$f_0, f_1,
\ldots , f_m$ of fermionic operaators $\psi_i, \psistar_j$
the recurrence formula
\beq\label{normord}
f_0 \normord  f_1 f_2 \ldots
f_m  \normord  =\normord
f_0 f_1 f_2 \ldots  f_m \normord 
+\sum_{j=1}^{m}(-1)^{j-1}
\lvac f_0 f_j \rvac \normord f_1 f_2 \ldots
\not{\!\!f}_{\!\!j} \ldots f_m\normord 
\eeq
holds, where
$\not{\!\!f}_{\!\!j}$ means that this operator 
should be omitted. This relation allows one to express normally ordered 
monomials with any number of fermionic operators as linear combinations of monomials
without normal ordering and vice versa. 

\noindent
{\bf Problem.} Prove the identity $e^{\alpha \psi_k \psistar_k}=
\normord e^{(e^{\alpha}-1)\psi_k \psistar_k} \normord$, $k\geq 0$.

In a similar way, one can define normal ordering with respect to any vacuum.
For example, one can consider the empty vacuum
$\left |\infty \right >$. With respect to this vacuum, all
$\psi_j$'s are annihilation operators and all
$\psistar_j$s are creation operators. The corresponding normal ordering will be 
denoted by
$\normordbare (\ldots )\normordbare$.
Examples: 
$\normordbare  \psistar_{m}\psi_{n}\normordbare =
\psistar_{m}\psi_{n}$, $\normordbare \psi_{n}\psistar_{m}\normordbare =
-\psistar_{m}\psi_{n}$ 
and
\beq\label{Bex}
\normordbare
\exp \Bigl (\sum_{ik}B_{ik}\psistar_i \psi_k\Bigr )
\normordbare =
1+ \sum_{i,k}B_{ik}\psistar_i \psi_k +\frac{1}{2!}
\sum_{i,i'k,k'}B_{ik}B_{i'k'}\psistar_i \psistar _{i'}
\psi_{k'} \psi_k +\ldots
\eeq

\subsection{Group and quasigroup elements of the Clifford algebra}

\subsubsection{Group elements}

Bilinear combinations
$\sum_{mn} b_{mn}\psistar_m \psi_n$
with certain conditions on the matrix
$b = (b_{mn})$ form an infinite dimensional Lie algebra. 
Exponentiating them, one gets an infinite dimensional group,
one of the versions of
$GL(\infty )$. Elements o this group can be re\-pre\-sen\-ted in the form
\begin{equation}\label{gl}
G=\exp \Bigl (\sum_{i, k \in {\z}}b_{ik}\psistar_i \psi_k\Bigr ).
\end{equation}
The inverse element has the same form with the matrix 
$(-b_{ik} )$. 

As it was already mentioned, the group elements
(\ref{gl}) have a rather special property:
the adjoint action of such element preserves the linear space spanned by
the fermionic operators
$\psi _n$ and the same is true for $\psistar _n$. More precisely, we have:
$$
G\psistar _n G^{-1} = \sum_l \psistar_l R_{ln}\,, \quad
G\psi _n G^{-1} = \sum_l  (R^{-1})_{nl}\psi_l
$$
or
\beq\label{rotation}
G\psistar_n =\sum_{l} R_{ln} \psistar_l  G\,, \quad \quad
\psi_n G =  \sum_l R_{nl} G\psi_l
\eeq
with some matrix $R = (R_{nl})$ .

\noindent
{\bf Problem.} Prove that $R=e^b$. 

\noindent
Hence it is clear that the product of group elements is an element of
the same form:
\beq\label{comp1}
\exp \Bigl (\sum_{i, k \in {\z }}b'_{ik}\psistar_i \psi_k\Bigr )
\exp \Bigl (\sum_{i, k \in {\z }}b_{ik}\psistar_i \psi_k\Bigr )=
\exp \Bigl (\sum_{i, k \in {\z }}b''_{ik}\psistar_i \psi_k\Bigr ),
\eeq
where $e^{b'} e^b =e^{b''}$. 
It is also clear that the multiplication of
$G$ of the form
(\ref{gl}) by arbitrary complex number 
preserves the characteristic property (\ref{rotation}). From the fact that
the center of the Clifford algebra is the field of complex numbers
$\CC$ it follows that two group elements 
$G, G'$ with the same matrix of linear transformation 
$R$ can differ by a $c$-number $c\in \CC$ only: $G'=c G$. 

Group elements can be also represented as
{\it normally ordered} exponential functions of bilinear forms. 
For example, it is easy to check that
$G=\normordbare e^{ B_{ik}\psistar_i \psi_k}\normordbare$
(here and below the summation over repeated indices 
is implied) 
satisfies the first relation in (\ref{rotation})
with $R_{ln}=\delta_{ln}+B_{ln}$:
$$
\begin{array}{c}
\normordbare e^{ B_{ik}\psistar_i\psi_k}\normordbare \psistar_n =
\left ( 1+ B_{a_1 b_1}\psistar_{a_1}\psi _{b_1}+\frac{1}{2!}
B_{a_1 b_1}B_{a_2 b_2}\psistar_{a_1}\psistar_{a_2}\psi _{b_2}\psi _{b_1}+
\ldots  \right ) \psistar_n
\\  \\
=\psistar_n \normordbare e^{ B_{ik}\psistar_i\psi_k}\normordbare
\! +\! B_{a_1 n}\psistar_{a_1}\! +\! 
B_{a_1 n}B_{a_2 b_2}\psistar_{a_1}\! \psistar_{a_2}
\psi _{b_2}\! +\! \frac{1}{2!}B_{a_1 n}B_{a_2 b_2}B_{a_3 b_3}
\psistar_{a_1}\! \psistar_{a_2}\! \psistar_{a_3}\psi _{b_3}\psi _{b_2}+\ldots 
\\ \\
=\psistar_n \normordbare e^{ B_{ik}\psistar_i\psi_k}\normordbare
+B_{an}\psistar_a \normordbare e^{ B_{ik}\psistar_i\psi_k}\normordbare \, =\,
(\delta_{an}+B_{an})\psistar_a 
\normordbare e^{ B_{ik}\psistar_i\psi_k}\normordbare .
\end{array}
$$

\noindent
{\bf Exercise.}
Prove the second relation in (\ref{rotation}). 

\noindent
In a similar way, one can show that
\beq\label{ferm1a}
\exp \Bigl (b_{ik}\psistar_i \psi_k\Bigr )=
\normordbare \exp \Bigl ((e^b -I)_{ik}\psistar_i \psi_k\Bigr )
\normordbare ,
\eeq
where $I$ is the unity matrix. The composition law has the form
\beq\label{comp2}
\normordbare \exp \Bigl (
B'_{ik}\psistar_i \psi_k\Bigr )
\normordbare
\normordbare \exp \Bigl (
B_{ik}\psistar_i \psi_k\Bigr )
\normordbare =
\normordbare \exp \Bigl (
(B\! +\! B'\! +\! B'B)_{ik}\psistar_i \psi_k\Bigr ).
\normordbare 
\eeq
This directly follows from the composition law (\ref{comp1})
and the relation $B=e^b -I$.

Let us prove another useful formula which allows one to 
represent the group element as a normal ordered exponent with respect to
different vacua: 
\begin{equation}\label{no1}
\normordbare \exp \Bigl (
B_{ik}\psistar_i \psi_k\Bigr )
\normordbare
=\det (I\! +\! P_{+} B)\,
\normord \exp \Bigl (
A_{ik}\psistar_i \psi_k\Bigr )\normord
\end{equation}
or, equivalently, 
\begin{equation}\label{no1a}
\normord \exp \Bigl (
A_{ik}\psistar_i \psi_k\Bigr ) 
\normord =\det (I\! -\! P_{+} A)\,
\normordbare \exp \Bigl (
B_{ik}\psistar_i \psi_k\Bigr ).
\normordbare
\end{equation}
Here $P_{+}$ is the projector to the space of positive modes 
($(P_{+})_{ik}=\delta_{ik}$ at $i,k\geq 0$ and 0 otherwise),
and the matrices $A,B$ are connected by the relations
\beq\label{no2}
B-A =AP_{+} B\,, \quad \mbox{i.e.,}\quad
B=(I\! - \! AP_+)^{-1}A \quad \mbox{or}\quad 
A=B(I\! + \! P_+B)^{-1}.
\eeq
For the proof we first notice that we can write
$\normord e^{A_{ik}\psistar_i \psi_k} 
\normord$ as a composition of three operators: 
$$
\normord e^{A_{ik}\psistar_i \psi_k} 
\normord =\underbrace{\phantom{\normord}
e^{A_{\bar a b}\psistar_{\bar a} \psi_b}\phantom{\normord}}_{G_1}\,\cdot \,
\underbrace{ \, \normord
e^{A_{a b}\psistar_{a} \psi_b}\cdot
e^{A_{\bar a \bar b}\psistar_{\bar a} \psi_{\bar b}}
\normord }_{G_2} \, \cdot \,
\underbrace{\phantom{\normord}
e^{A_{a \bar b}\psistar_{a} \psi_{\bar b}}\phantom{\normord} }_{G_3},
$$
where in the right hand side the summation over repeated {\it positive}
indices $a,b$ (and {\it negative} $\bar a, \bar b$) is implied. 
(summation over repeated
$i,k$ in the left hand side is over all integers). 
The operator $G_1$ contains only creation operators while  
$G_3$ contains only annihilation operators. Note also that the two operators
under the sign of normal ordering commute with each other. 
It is not difficult to find the linear transformations performed by
$G_1, G_2, G_3$. For $G_{1}, G_{3}$ this is especially easy:
$$
\psi_n G_1 =G_1 \left \{
\begin{array}{l}\psi_n +A_{nb}\psi_b \,, \quad n<0
\\ \psi_n \,, \quad \quad \quad \quad \quad \! n\geq 0,
\end{array}\right.
\quad \quad 
\psi_n G_3 =G_3 \left \{
\begin{array}{l}
\psi_n \,, \quad \quad \quad \quad \quad \! n< 0
\\
\psi_n +A_{n\bar b}\psi_{\bar b} \,, \quad n\geq 0.
\end{array}\right.
$$
For $G_2$ we write $G_2 = \normord G_2^+ G_2^-\normord$
with $G_2^+=e^{A_{a b}\psistar_{a} \psi_b}$,
$G_2^-=e^{A_{\bar a \bar b}\psistar_{\bar a}\psi_{\bar b}}$.
Moving $\psi_n$ through this element, one can ignore either
$G_2^+$ or $G_2^-$, depending on whether $n$ is negative or positive. 
The remaining calculations are similar to the above calculation with the normally
ordered expression
$\normordbare (\ldots )\normordbare$.
It yields:
$$
\left \{ \begin{array}{l}
\psi_n G_2 =G_2 (\psi_n +A_{n\bar b}\psi_{\bar b}),
\quad n<0,
\\  \\
G_2\psi_n =(\psi_n -A_{n b}\psi_b )G_2 , \quad n\geq 0.
\end{array} \right.
$$
It is useful to rewrite these transformations in the block-matrix form:
$$
\left ( \begin{array}{c}\psi_{-}G_1\\ \psi_+ G_1\end{array}\right )
=\left (\begin{array}{cc} I&A^-_{+}\\0& I\end{array}\right )
\left ( \begin{array}{c}G_1\psi_{-}\\ G_1\psi_+ \end{array}\right ),
$$
$$
\left ( \begin{array}{c}\psi_{-}G_3\\ \psi_+ G_3\end{array}\right )
=\left (\begin{array}{cc} I&0\\A^+_{-}& I\end{array}\right )
\left ( \begin{array}{c}G_3\psi_{-}\\ G_3\psi_+ \end{array}\right ),
$$
$$
\left ( \begin{array}{c}\psi_{-}G_2\\ \psi_+ G_2\end{array}\right )
=\left (\begin{array}{cc} I\! +\! A^-_{-}&0\\0& (I\! -\! A^+_{+})^{-1}
\end{array}\right )
\left ( \begin{array}{c}G_2\psi_{-}\\ G_2\psi_+ \end{array}\right )
$$
(assuming that the matrix $I\! -\! A^+_{+}$ is invertible).
In this notation
$\displaystyle{
P_+ = \left (\begin{array}{cc}0&0\\ 0&I\end{array}
\right )}
$. The full transformation matrix
is obtained as the product of these three:
$$
\begin{array}{lll}
R&=&\displaystyle{\left (\begin{array}{cc} I&A^-_{+}\\0& I
\end{array}\right )
\left (\begin{array}{cc} I\! +\! A^-_{-}&0\\0& (I\! -\! A^+_{+})^{-1}
\end{array}\right )
\left (\begin{array}{cc} I&0\\A^+_{-}& I\end{array}\right )}
\\ &&\\
&=& \left (\begin{array}{cc}I&0\\ 0&I\end{array}
\right )+
\left (\begin{array}{cc} A_{-}^{-}\! +\!  A_{+}^{-}(I\! -\! A^+_{+})^{-1}
 A_{-}^{+}
&A^-_{+} (I\! -\! A^+_{+})^{-1}\\(I\! -\! A^+_{+})^{-1} A^+_-& 
(I\! -\! A^+_{+})^{-1}A^+_+
\end{array}\right ).
\end{array}
$$
It can be checked that the second matrix in the last line
(which is $R-I=B$) is precisely $(I\! -\! AP_+)^{-1}A$
in agreement with (\ref{no2}). It remains to find the scalar 
multiplier in 
(\ref{no1a}). Let us find the vacuum expectation value 
of the both sides with respect to the bare (empty) vacuum.
Then we should show that
$$
\left <\infty \right |\normord \exp \Bigl (
A_{ik}\psistar_i \psi_k\Bigr ) 
\normord \left |\infty \right > =\det (I\! -\! P_{+} A).
$$
Using the representation of the operator in the left hand side as composition
of three, we have:
$$
\begin{array}{c}
\left <\infty \right |\normord e^{
A_{ik}\psistar_i \psi_k}
\normord \left |\infty \right >=
\left <\infty \right |\normord e^{-A_{ab}\psi_b \psistar_a}
\normord \left |\infty \right >
\\ \\
=\, \displaystyle{
\sum_{k\geq 0}\frac{(-1)^k}{k!}A_{a_1 b_1}\ldots 
A_{a_k b_k}\left <\infty \right | \psi_{b_1}\ldots \psi_{b_k}
\psistar_{a_k}\ldots \psistar_{a_1}\left |\infty \right >}
\\ \\
=\, \displaystyle{
\sum_{k\geq 0}\frac{(-1)^k}{k!}
\sum_{a_1, \ldots , a_k \geq 0}
\left |\begin{array}{cccc}A_{a_1 a_1}&A_{a_1 a_2} & \ldots &A_{a_1 a_k}
\\  
A_{a_2 a_1}&A_{a_2 a_2} & \ldots &A_{a_2 a_k}
\\  
\ldots &  \ldots & \ldots & \ldots 
\\  
A_{a_k a_1}&A_{a_k a_2} & \ldots &A_{a_k a_k}
\end{array}\right |}
\\ \\
=\, \det (I-A_+^+)=\det (I-P_+A).
\end{array}
$$

\subsubsection{Quasigroup elements}

The normal ordering allow one to represent in the form
(\ref{ferm1}) not only group elements but also some not invertible
elements of the Clifford algebra which obey the property
\beq\label{rotation-a}
G\psistar_n =\sum_{l} R_{ln} \psistar_l  G\,, \quad \quad
\psi_n G =  \sum_l R_{nl} G\psi_l\,,
\eeq
or
\beq\label{rotation-b}
\psistar_n G=\sum_{l} R'_{ln} G \psistar_l  \quad \quad
G \psi_n  =  \sum_l R'_{nl} \psi_l \, G
\eeq
with some (not necessarily invertible)
matrices $R, R'$ (for non-invertible elements only one pair
of these relations is satisfied). 
We call an element 
$G$ of the Clifford algebra a quasigroup element 
if the commutation relations (\ref{rotation-a}) or
(\ref{rotation-b}) hold. If the matrix 
$R$ is not invertible, then $G$ is also not invertible.
In this case it can not be written in the exponential form
(\ref{gl}) but can be written as a normally ordered exponent. 

\noindent
{\bf Example.} Let $\Psi$, $\Phi^*$ be arbitrary linear combinations
of the fermionic operators $\psi_n$,
$\psistar_n$ respectively. Consider the element
$$
G=e^{\beta \Phi^* \Psi}=
\normordbare e^{\alpha \Phi^* \Psi}\normordbare
= 1+\alpha \Phi^* \Psi = 1+\alpha \gamma -\alpha \Psi \Phi^* ,
$$
where
$\gamma := \lbr \infty \right | \Psi \Phi^* \left |\infty \rbr $
and $\alpha $, $\beta $ are connected by the relation
$e^{\gamma \beta} =1+\gamma \alpha $.
For almost all $\alpha $ the element $G$ is invertible and
the two representations (with and without normal ordering)
are absolutely equivalent. 
However, at $ \alpha =-1/\gamma $ (assumig that
$\gamma \neq 0$) 
$G=\normordbare e^{\alpha \Phi^* \Psi}\normordbare$ 
becomes non-invertible and acquires the form
$$
G= \frac{\Psi \, \Phi^*}{\lbr \infty
\right | \! \Psi \Phi^* \! \left |\infty \rbr}.
$$

\subsection{The basic bilinear relation}

It easily follows from (\ref{rotation-a}) or (\ref{rotation-b}) 
that any quasigroup element satisfies the com\-mu\-ta\-tion relation
\beq\label{commute}
\mbox{\fbox{$\displaystyle{\phantom{\int^A}
\sum_{k \in {\z}} 
\psi_{k} G \otimes  \psi_{k}^{*} G =
\sum_{k \in {\z}}G\psi_{k} \otimes G \psi_{k}^{*}\phantom{\int^A}}$}}
\eeq
which we call the basic bilinear relation (BBR). 
It means that
$G\otimes G$ commutes with
$\sum_{k}\psi_k \otimes \psistar_k$. In terms of matrix elements,
the BBR states that
\begin{equation}
\sum_{k \in {\z}} \lbr U \right|  \psi_{k} G \left|V \rbr
 \lbr U^{\prime} \right|  \psistar_{k} G \left|V^{\prime} \rbr =
\sum_{k \in {\z}} \lbr U \right| G \psi_{k} \left|V \rbr
 \lbr U^{\prime} \right| G \psistar_{k} \left|V^{\prime} \rbr 
 \label{bilinear-fermi}
\end{equation}
for any states $\left|V \rbr , \left|V^{\prime} \rbr$,
$\lbr U \right|,  \lbr U^{\prime} \right|$
from the spaces
${\cal H}_{F}$ and its dual. 
Indeed, substituting (\ref{rotation-a}) 
(or (\ref{rotation-b})) instead of $\psi_k G$ and $G\psistar_k$ 
(or $\psistar_k G$ and $G\psi_k$) 
in the right and left hand sides of (\ref{commute}), we get an identity.

It turns out that the elements
$G$ satisfying the BBR, are not exhausted by normally ordered exponents. 
For example, it is easy to check that
$G=\psi_n$ and
$G=\psistar_n$ satisfy (\ref{commute}) but they are not normally ordered exponents
of anything.  
Besides, the element
$G=\psi_n$ does not perform a linear transformation
of the space with basis $\psistar_k$
neither of the form (\ref{rotation-a}) nor (\ref{rotation-b}).
The class of such examples can be significantly enlarged. 

\noindent
{\bf Exercise.} Let $\Psi$, $\Phi^*$ be arbitrary linear combinations
of $\psi_n$,
$\psistar_n$ respectively. Check that
$G=\Psi$ and $G=\Phi^*$ satisfy the condition (\ref{commute}).

Let us point out two general properties of elements satisfying the BBR.

\noindent
{\bf Proposition.} {\it  Elements of the Clifford algebra satisfying the BBR
(\ref{commute}) form a se\-mi\-group: if $G$ and $G'$ satisfy it, then
$GG'$ does.}

\noindent
This is obvious from (\ref{commute}). 

\noindent
{\bf Proposition.} {\it All solutions to the BBR
(\ref{commute}) have a definite charge, i.e., 
$[Q, G]=qG$ with some integer $q$.}

\noindent
The proof is omitted.

The BBR in the form
(\ref{commute}) or (\ref{bilinear-fermi}) is the basis for what follows.
We will extend the notion of quasigroup elements calling quasigroup elements
{\it all solutions to the BBR}. 
For simplicity we will work, as a rule, with elements
$G$ having zero charge (for example, with normally ordered exponents) but all
main statements can be easily extended to the general case.

\subsection{The generalized Wick theorem}

Vacuum expectation values of products of fermionic operators with insertions of
qu\-asi\-group elements obey some special properties which are described by the 
generalized Wick theorem. 
Let $v_i =\sum_j v_{ij}\psi_j$ be an arbitrary linear combination of
the operators $\psi_j$ and
$w_i^*=\sum_j w^{*}_{ij}\psistar_j$ be an arbitrary linear combination of the operators
$\psistar_j$.

\noindent
{\bf Proposition} (the generalized Wick theorem). {\it Let
$G, G'$ be any two quasigroup elements with zero charge.
Then for any
$v_j, w^*_i$ and arbitrary $n$ such that $\lvacn G'G\rvacn \neq 0$ 
the following identity holds:
\begin{equation}\label{Wick1}
\frac{\lvacn G' v_1 \ldots v_m w^{*}_m \ldots w^{*}_1 G\rvacn }{\lvacn
G'G\rvacn }=\det_{i,j =1,\ldots , m}
\frac{\lvacn G'v_j w^{*}_iG\rvacn }{\lvacn
G'G\rvacn }.
\end{equation}
}

\noindent
This statement can be proved by induction. Assume that
thhe identity
(\ref{Wick1}) holds for some
$m \geq 1$ (obviously, this is true for 
$m=1$).
Set
$$
\lbr U\right |=\lvacn G' w_{1}^{*}, \quad
\lbr U'\right |=\lvacn G' v_1 v_2
\ldots v_{m+1} w_{m+1}^{*}w_{m}^{*}\ldots
w_{2}^{*},
\quad \left |V\rbr = \left |V'\rbr =\rvacn .
$$
Substituting this into the BBR (\ref{bilinear-fermi}), we see that
its right hand side vanishes identically since
either $\psi_k \rvacn =0$ or
$\psistar_k \rvacn =0$, and thus
$$
\sum_k \lvacn G' w_{1}^* \psi_k G\rvacn \lvacn G' v_1 \ldots 
v_{m+1} w^*_{m+1}\ldots w^*_2 \psistar_k G\rvacn =0.
$$
Substitute 
$w_{1}^* \psi_k = w^{*}_{1k}-\psi_kw_{1}^*$ in the first multiplier
and move $\psistar_k$ in the second multiplier through the chain of operators
$w_j^*$'s. In the left hand side, we get 
$$
\begin{array}{c}
\lvacn G'G\rvacn \lvacn G'v_1 \ldots 
v_{m+1} w^*_{m+1}\ldots w^*_1 G\rvacn
\\ \\\displaystyle{
-\, (-1)^m \sum_k \lvacn G' \psi_k w_{1}^*  G\rvacn
\lvacn G' v_1 \ldots 
v_{m+1} \psistar_k w^*_{m+1}\ldots w^*_2  G\rvacn .
}
\end{array}
$$
Now move $\psistar_k$ to the left through the chain of operators
$v_j$ and use at each step the relation $v_j\psistar_k=
v_{jk}-\psistar_k v_j$. As a result, we obtain
$$
\begin{array}{c}
\lvacn G'G\rvacn \lvacn G'v_1 \ldots 
v_{m+1} w^*_{m+1}\ldots w^*_1 G\rvacn
\\ \\ \displaystyle{
+\, \sum_{j=1}^{m+1}(-1)^j \lvacn G' v_j w_{1}^*  G\rvacn
\lvacn G' v_1 \ldots \not \! v_j \ldots 
v_{m+1} w^*_{m+1}\ldots w^*_2  G\rvacn }
\\ \\
\displaystyle{+\, \sum_k \lvacn G' \psi_k w_{1}^*  G\rvacn
\lvacn G' \psistar_k v_1 \ldots 
v_{m+1} w^*_{m+1}\ldots w^*_2  G\rvacn }.
\end{array}
$$
In the last line we can again use the BBR, to bring the last line to the form 
$$
\sum_k \lvacn \psi_k G'  w_{1}^*  G\rvacn
\lvacn \psistar_k G'  v_1 \ldots 
v_{m+1} w^*_{m+1}\ldots w^*_2  G\rvacn .
$$
This is equal to 0 since again either $\lvacn \psi_k =0$ or
$\lvacn \psistar_k =0$. Therefore, we come to the relation
$$
\begin{array}{c}
\displaystyle{\lvacn G'G\rvacn \lvacn G'v_1 \ldots 
v_{m+1} w^*_{m+1}\ldots w^*_1 G\rvacn}
\\ \\ \displaystyle{
+\, \sum_{j=1}^{m+1}(-1)^{j} \lvacn G' v_j w_{1}^*  
G\rvacn
\lvacn G' v_1 \ldots \not \! v_j \ldots 
v_{m+1} w^*_{m+1}\ldots w^*_2  G\rvacn =0}
\end{array}
$$
or
$$
\begin{array}{c}
\displaystyle{\frac{\lvacn G'v_1 \ldots 
v_{m+1} w^*_{m+1}\ldots w^*_1 G\rvacn}{\lvacn G'G\rvacn}}
\\ \\ \displaystyle{
=\, \sum_{j=1}^{m+1}(-1)^{j-1} \frac{\lvacn G' v_j w_{1}^*  
G\rvacn}{\lvacn G'G\rvacn}
\, \frac{\lvacn G' v_1 \ldots \not \! v_j \ldots 
v_{m+1} w^*_{m+1}\ldots w^*_2  G\rvacn}{\lvacn G'G\rvacn} }.
\end{array}
$$
By the assumption of induction, the second ratio in the right hand side
is given by the determinant
of the $m\times m$ matrix written above.
Now notice that the expression in the right hand side is the expansion 
of the determinant of the corresponding $(m+1)\times (m+1)$ matrix
in the first column and thus the proposition is proved.

\subsection{The operators $e^{J_{\pm}}$}

For applications to integrable hierarchies, especially important are 
group elements of a special form, which we introduce in this section.

Consider the operators
\beq\label{Jk}
J_k =\sum_{j\in \z}\normord \psi_j \psistar_{j+k}\normord
=\mbox{res}_z \Bigl ( \normord \psi (z)
z^{k-1}\psistar (z)\normord \Bigr )
\eeq
which are Fourier modes of the current operator
$J(z)=\normord \psi (z)\psistar (z)\normord $. The normal ordering in 
(\ref{Jk}) is essential only at $k=0$, and in this case
$J_0$ coincides with the charge operator
$Q$ (\ref{chargeoper}). If $k\neq 0$, then the normal ordering  
can be omitted:
\beq\label{Jka}
J_k =\sum_{j\in \z}\psi_j \psistar_{j+k}\,, \quad \quad
k\neq 0.
\eeq
These operators have the form (\ref{XA})
with the matrix $A_{ij}=\delta_{i, j-k}$. This matrix has the infinite nonzero
diagonal, and one should be careful 
when working with formal expressions containing infinite sums (see the example
of calculation of the commutator below). 

\noindent
{\bf Exercise.} Show that
$[J_k , \psi _m]=\psi_{m-k}$, $[J_k , 
\psistar _m]=-\psistar_{m+k}$.

First consider the operators $J_k$ with positive $k$. Put
\beq\label{ferm2}
J_+ =J_+({\bf t}_+) = \sum_{k\geq 1}t_k J_k
\eeq
with arbitrary parameters $t_k$ (they will be identified with the times of the KP  
hierarchy), the whole set of which will be denoted as
${\bf t}_+ = \{ t_1, t_2, \ldots \}$ or simply ${\bf t}$ if this does not lead to 
a confusion. 

\noindent
{\bf Exercise.} Verify that $J_+({\bf t}_+)$ annihilates the right vacuum: 
$J_+({\bf t}_+)\rvac =0$.

\noindent
It is easy to check that all $J_k$ with $k\geq 1$
commute with each other and the fermionic fields
$\psi (z), \, \psistar (z)$ transform diagonally under the action of
$e^{J_+}$:
\beq\label{ferm3}
\begin{array}{l}
e^{J_+({\bf t})}\psi (z)e^{-J_+({\bf t})}=e^{\xi ({\bf t} ,\, z)}\psi (z),
\\ \\
e^{J_+({\bf t})}\psistar (z)e^{-J_+({\bf t})}=
e^{-\xi ({\bf t} , \, z)}\psistar (z).
\end{array}
\eeq
Here we use the previously introduced notation 
$\xi ({\bf t}, z)=\displaystyle{\sum_{j\geq 1}t_j z^j}$. Then it is obvious that
the operators 
$\psi_n$, $\psistar_{n}$ transform as follows:
\beq\label{ferm4}
\begin{array}{l}
\displaystyle{
e^{J_+({\bf t})}\psi _n e^{-J_+({\bf t})}=
\sum_{k\geq 0}\psi_{n- k}h_k({\bf t})},
\\ \\
\displaystyle{
e^{J_+({\bf t})}\psistar _n e^{-J_+({\bf t})}=
\sum_{k\geq 0}\psistar_{n+ k}h_k(-{\bf t})},
\end{array}
\eeq
where the Schur polynomials $h_k$ are defined by (\ref{schur}).

In a similar way, introduce the operator
\beq\label{ferm201}
J_- =J_-({\bf t}_-) = \sum_{k\geq 1}t_{-k} J_{-k}
\eeq
with arbitrary parameters $t_{-k}$. In what follows, to avoid many signs
$\pm$, we will write simply ${\bf t}$ for any half-infinite set of times
(implying ${\bf t}_+$ or ${\bf t}_-$).

As in the case of $J_+$, it is easy to check that all $J_k$ with $k\leq  -1$
commute with each other and the fermionic fields
$\psi (z), \, \psistar (z)$ transform diagonally:
\beq\label{ferm3a}
\begin{array}{l}
e^{J_{-}({\bf t})}\psi (z)e^{-J_{-}({\bf t})}=
e^{\xi ( {\bf t} ,\, z^{-1})}\psi (z),
\\ \\
e^{J_{-}({\bf t})}\psistar (z)e^{-J_{-}( {\bf t})}=e^{-\xi ({\bf t} , \,
z^{-1})}\psistar (z),
\end{array}
\eeq
which is equivalent to the following action on
$\psi_n$,
$\psistar_n$:
\beq\label{ferm4a}
\begin{array}{l}
\displaystyle{
e^{J_-({\bf t})}\psi _n e^{-J_-({\bf t})}=
\sum_{k\geq 0}\psi_{n+ k}h_k({\bf t})},
\\ \\
\displaystyle{
e^{J_-({\bf t})}\psistar _n e^{-J_-({\bf t})}=
\sum_{k\geq 0}\psistar_{n- k}h_k(-{\bf t})}.
\end{array}
\eeq

Let us find the commutator $[J_k, J_l]$:
$$
[J_k, J_l]=\sum_j [J_k, \psi_j \psistar_{j+l}]=
\sum_j \left ( [J_k, \psi_j]\psistar_{j+l}+
\psi_j [J_k, \psistar_{j+l}]\right ).
$$
Using the formulas obtained above, we will have:
\beq\label{JkJl1}
[J_k, J_l]=\sum_j \left (\psi_{j-k}\psistar_{j+l}-\psi_j
\psistar_{j+l+k}\right )\stackrel{?}{=}\, 0.
\eeq
Formally one could shift the summation index in the first sum
$j\to j+k$ and get 0. 
To check the result, calculate $\lvac [J_k, J_{-k}]\rvac$
with $k>0$ ``by hands'':
$$
\lvac [J_k, J_{-k}]\rvac =\lvac J_k J_{-k}\rvac
=\sum_{-k\leq j <0}\sum_{0\leq l<k}
\lvac \psi_j \psistar_{j+k}\psi_l \psistar_{l-k}\rvac =
\sum_{l=0}^{k-1} 1 = k\neq 0. \,\,\,\, (!)
$$
Where is the mistake? The matter is that the shift of the summation index is legal
only if $k+l\neq 0$. In this case the sum of operators in the right hand side is well-defined
(and is indeed equal to 0) because its matrix elements between any basis states
contain only finite number of terms. If
$k+l=0$, then an infinite sum appears which requires
an additional definition. In this case it is necessary first to rewrite 
the right hand side in terms of normally ordered expressions:
$$
[J_k, J_{-k}]=\sum_j \normord 
\left (\psi_{j-k}\psistar_{j-k}-\psi_j
\psistar_{j}\right ) \normord +\sum_j \Bigl (\theta (j<k)-\theta (j<0)
\Bigr ).
$$
Here $\theta (j<k)=1$ if $j<k$ and $0$ otherwise. The normally ordered expressions
are well-defined and now the summation index in the first sum can be safely shifted. 
The remaining terms give the commutation rule for the Fourier 
modes of the current operator: 
\beq\label{JkJl2}
[J_k, J_l]=k\delta_{k+l, 0}.
\eeq
It is identical to the commutation rule of bosonic operators.

\noindent
{\bf Exercise.} Prove the formulas
\beq\label{JkJl3}
\begin{array}{l}\displaystyle{
[J_+({\bf t}_+), \, J_-({\bf t}_-)]=\sum_{k\geq 1}kt_kt_{-k}},
\\ \\
\displaystyle{
e^{J_+({\bf t}_+)}e^{J_-({\bf t}_-)}=
\exp \Bigl (\sum_{k\geq 1}kt_kt_{-k}\Bigr )
e^{J_-({\bf t}_-)}
e^{J_+({\bf t}_+)}}.
\end{array}
\eeq

\noindent
{\bf Exercise.} Calculate $\lvac e^{J_+({\bf t})} \normordbare
e^{(\psi_{-1}-\psi_1)(\psistar_{-1}-\psistar_1)}\normordbare \rvac$.

\noindent
{\bf Problem.} Find $e^{tH_1}\psi (z)e^{-tH_1}$ and $e^{tH_1}\psistar (z)e^{-tH_1}$,
where $\displaystyle{H_1=\sum_{k\in \z}k\normord \psi_k\psistar_k \normord }$ and
calculate $\lvacn e^{J_+({\bf t}_+)}e^{tH_1}e^{-J_-({\bf t}_-)}\rvacn$.

\subsection{Boson-fermion correspondence}

\subsubsection{Bosonization rules}

As we have just seen, the Fourier modes $J_k$ of the current operator
have the same commutation relations as the oscillator modes of operators 
of free bosonic field:
$[J_k, J_l]=k\delta_{k+l,0}$, i.e. $J_k$ and $J_{-k}/k$ 
are canonically conjugated. 
the operator $J_0=Q$ plays a special role. Introduce the operator $P$
canonically conjugated to $Q$; the operator
$e^P$ is the operator of shift of the index: 
$$
e^P \psi_n e^{-P}=\psi_{n+1}\,, \quad \quad
e^P \psistar_n e^{-P}=\psistar_{n+1}.
$$
On the vacuum states 
$e^{\pm P}\rvacn = \left |n\pm 1\right >$. One can check that this definition
is indeed equivalent to the commutation relation 
$[Q,P]=1$.  

Introduce the chiral bosonic field
\beq\label{bos1}
\begin{array}{lll}
\phi (z)&=& \displaystyle{\sum_{k>0}\frac{J_{-k}}{k}\, z^k +P +
Q \log z -\sum_{k>0}\frac{J_{k}}{k}\, z^{-k}}
\\ &&\\
&=& \displaystyle{J_{-}([z])+P +J_0 \log z -J_{+}([z^{-1}])}.
\end{array}
\eeq
In the last line we use the notation
$J_{\pm}([z])=J_{\pm 1}z+\frac{1}{2}J_{\pm 2}z^2 +
\frac{1}{3}J_{\pm 3}z^3 +\ldots$.
We have $z\p_z \phi (z)=J(z)$ or
$$
\phi (z_2)-\phi(z_1)=\int_{z_1}^{z_2}\!\! J(z)\frac{dz}{z}.
$$
The operators $J_{+k}$ с $k>0$ annihilate the right vacuum
(they are bosonic annihilation operators) 
while $J_{-k}$ with $k>0$ annihilate the left vacuum (they are bosonic creation
operators).
Introduce the bosonic normal ordering
$\normordboson (\ldots )\normordboson$ with the same rule that
all creation operators are moved to the left and all annihilation operators
are moved to the right and consider normally ordered exponents of the bosonic fields:
\beq\label{bos2}
\begin{array}{l}
\normordboson e^{\phi (z)}\normordboson =
e^{J_{-}([z])} \, e^P z^Q \, e^{-J_{+}([z^{-1}])},
\\ \\
\normordboson e^{-\phi (z)}\normordboson =
e^{-J_{-}([z])} \, z^{-Q}  e^{-P} \, e^{J_{+}([z^{-1}])}
\end{array}
\eeq
(the normal ordering affects only the operators with
$J_{k}$ с $k\neq 0$).
From (\ref{JkJl3}) it follows that
$$
\begin{array}{l}
e^{J_{\pm}({\bf t})}\, \normordboson e^{\phi (z)}\normordboson \, 
e^{-J_{\pm}({\bf t})}
=e^{\xi({\bf t},z^{\pm 1})}\normordboson e^{\phi (z)}\normordboson ,
\\ \\
e^{J_{\pm}({\bf t})}\, \normordboson e^{-\phi (z)}\normordboson \, 
e^{-J_{\pm}({\bf t})}
=e^{-\xi({\bf t},z^{\pm 1})}\normordboson e^{-\phi (z)}\normordboson .
\end{array}
$$
These formulas tell us that  
$\normordboson e^{\pm \phi (z)}\normordboson$ behave as the fermionic fields
$\psi (z), \psistar (z)$. Moreover, it can be shown that all matrix elements
of the operators
$\normordboson e^{\pm \phi (z)}\normordboson$ and 
$\psi (z), \psistar (z)$ between all basis states coincide. 
Therefore, one can identify
\beq\label{bos3}
\psi (z) =\normordboson e^{\phi (z)}\normordboson 
\quad \quad
\psistar (z) =\normordboson e^{-\phi (z)}\normordboson .
\eeq
We will not give a complete proof and will just compare the matrix elements
between the states 
$\lvacn e^{J_{+}({\bf t}_+)}$ and $e^{J_{-}({\bf t}_-)}
\left |n-1\right >$ (obviously, the matrix elements of the operators in question
are nonzero only if the charge of the right state is less than the charge of the 
left one by 1).  
In fact this is enough if to know that these states are ``generating functions''
of the basis states (the so called coherent states) but we will not prove this. 
We want to show that
$$
\lvacn e^{J_{+}({\bf t}_+)}\normordboson e^{\phi (z)}\normordboson
e^{J_{-}({\bf t}_-)}
\left |n-1\right > =
\lvacn e^{J_{+}({\bf t}_+)} \psi (z)e^{J_{-}({\bf t}_-)}
\left |n-1\right >
$$
for all sets of times ${\bf t}_+ , {\bf t}_-$. A simple direct calculation
using the definition
(\ref{bos2}), the commutation relation  (\ref{JkJl3}) 
and the relations (\ref{ferm3}), (\ref{ferm3a}) shows that the both sides
are equal to
$$
z^{n-1}\exp \Bigl (\xi ({\bf t}_+, z)-\xi ({\bf t}_-, z^{-1})
+\sum_{k\geq 1}kt_k t_{-k}\Bigr ).
$$

Let us check that the bosonization rules (\ref{bos3}) imply that
\beq\label{bos4}
\normord \psi (z) \psistar (z)\normord = z\p_{z}\phi (z).
\eeq
In this sense they agree with (\ref{bos1}).
We write
$
\psi (z_2) \psistar (z_1)=\normordboson e^{\phi (z_2)}\normordboson
\normordboson e^{-\phi (z_1)}\normordboson
$
and transform both sides so that they would contain normally ordered 
expressions: 
$$
\psi (z_2) \psistar (z_1)=\normord \psi (z_2) \psistar (z_1)\normord
+\frac{z_1}{z_2 -z_1},
$$
$$
\normordboson e^{\phi (z_2)}\normordboson
\normordboson e^{-\phi (z_1)}\normordboson =
\frac{z_1}{z_2 -z_1}\, \normordboson 
e^{\phi (z_2)-\phi (z_1)}\normordboson .
$$
Putting $z_1 =z$, $z_2=z_1+\varepsilon$, we get
$$
\frac{z}{\varepsilon}\Bigl (\normordboson 
e^{\phi (z+\varepsilon )-\phi (z)}\normordboson \, - 1\Bigr )
=\normord \psi (z+\varepsilon ) \psistar (z)\normord
$$
which coincides with (\ref{bos4}) in the limit $\varepsilon \to 0$. 

When applied to the vacuum states, the operators 
(\ref{bos2}) simplify because either
$e^{J_-([z])}$ or $e^{J_+([z^{-1}])}$ acts to vacuum trivially. 
This leads to the following simplified bo\-so\-ni\-za\-tion rules:
\beq\label{bf3}
\begin{array}{l}
\lvacn \psi (z)= z^{n-1}\left <n\! -\! 1\right | 
e^{-J_+([z^{-1}])},
\\ \\
\lvacn \psi^{*} (z)= z^{-n}\left <n\! +\! 1\right | 
e^{J_+([z^{-1}])}
\end{array}
\eeq
for the left vacuum and
\beq\label{bf3a}
\begin{array}{l}
\psi (z)\rvacn = z^{n}\,
e^{J_-([z])}\, \left |n\! +\! 1\right >
\\ \\
\psi^{*} (z)\rvacn = z^{-n+1}
e^{-J_-([z])}\left |n\! -\! 1\right >
\end{array}
\eeq
for the right vacuum.

\noindent
{\bf Exercise.} Show that
\beq\label{bf3b}
\lvacn \psistar (\zeta )\psi (z) =\frac{z^n \zeta^{1-n}}{\zeta -z}
\lvacn e^{J_+([\zeta ^{-1}]-[z^{-1}])}.
\eeq

\noindent
{\bf Problem.} With the help of bosonization rules
prove that 
\beq\label{bf4}
\begin{array}{l}\displaystyle{
\lvacn \psi (z_1)\ldots \psi (z_m)=(z_1 \ldots z_m)^{n-m}\,
\prod_{i<j}
(z_i-z_j)\,
\left <n\! -\! m \right | e^{-J_+([z_1^{-1}])-\ldots -J_+([z_m^{-1}])}},
\\ \\
\displaystyle{
\lvacn \psistar (z_1)\ldots \psistar (z_m)=(z_1 \ldots z_m)^{-n-m+1}\!
\prod_{i<j}
(z_i\! -\! z_j)
\left <n\! +\! m \right | e^{J_+([z_1^{-1}])+\ldots +J_+([z_m^{-1}])}}
\end{array}
\eeq
and derive similar formulas for the right vacuum.

\subsubsection{Vertex operators}

The bosonization rules can be represented in a different form which is based on
the explicit realization of the bosonic Fock space
${\cal H}_B$ as the space of polynomials of 
infinite number of variables 
$t_1, t_2, t_3, \ldots$. More precisely, consider the space
$$
{\cal H}_B = \CC [w,w^{-1}, \, t_1, t_2, t_3, \ldots ]
=\bigoplus_{l\in \z}w^l \,
\CC [t_1, t_2, t_3, \ldots ],
$$
where we have added an extra variable $w$ to take into account fermionic states
with different charge, and construct the map
$\Phi : {\cal H}_F \rightarrow {\cal H}_B$ defined for an arbitrary vector
$\left |U\right > \in {\cal H}_F$
as follows:
\beq\label{bf1}
\Phi (\left |U\right >)=\sum_{l\in \z} w^l  \left <l\right |
e^{J_+({\bf t})}\left |U\right >.
\eeq
If the state $\left |U\right >$ has a definite charge $m$, then the sum
contains only one term with $l=m$. The fermionic creation and 
annihilation operators turn into some operators acting in the space off functions of
$w$ and $t_i$.

\noindent
{\bf Proposition.} {\it 
For the modes of the current there is the correspondence
\beq\label{bf102}
\Phi (J_k \left |U\right >)=\left \{\begin{array}{l}
\,\, \p_{t_k}\Phi (\left |U\right >)\,, \quad \,\, \, \, k>0,
\\ \\
\,\, w\p_w \Phi (\left |U\right >)\,, \!\!\!\! \quad \quad k=0,
\\ \\
-kt_{-k}\Phi (\left |U\right >)\,, \quad k<0.
\end{array}\right.
\eeq
}
the first two formulas are obvious from the definition (\ref{bf1}).
The third one follows from the commutation relation
$e^{J_+({\bf t})}J_{-k}=(J_{-k}+kt_k)e^{J_+({\bf t})}$ ($k>0$)
obtained with the help of (\ref{ABA}).

Introduce {\it vertex operators} by the formulas
\beq\label{bf2}
\begin{array}{l}
\displaystyle{
X(z)=\exp \left (\sum_{j\geq 1}t_j z^j\right )
\exp \left (-\sum_{j\geq 1}\frac{1}{jz^j}\, \p_{t_j}\right )
e^P z^Q},
\\ \\
\displaystyle{
X^{*}(z)=\exp \left (-\sum_{j\geq 1}t_j z^j\right )
\exp \left (\sum_{j\geq 1}\frac{1}{jz^j}\, \p_{t_j}\right )
z^{-Q} e^{-P}}.
\end{array}
\eeq
Here the operators $P,Q$ are defined by the action on functions of 
$w$: 
$$
e^P f(w)=wf(w)\,, \quad \quad
z^Q f(w)=f(zw).
$$
(it is easy to check that $[Q,P]=1$). Using the shorthand notation introduced 
above, one can write
\beq\label{bf2a}
\begin{array}{l}
\displaystyle{
X(z)=e^{\xi ({\bf t},z)} e^{-\xi (\tilde \p , z^{-1})}
e^P z^Q},
\\ \\
\displaystyle{
X^{*}(z)=e^{-\xi ({\bf t},z)} e^{\xi (\tilde \p , z^{-1})}
z^{-Q} e^{-P} }.
\end{array}
\eeq

\noindent
{\bf Proposition.} {\it Under the boson-fermion correspondence,
the fermionic fields $\psi (z)$, $\psistar (z)$ are represented by the vertex operators
$X(z), X^{*}(z)$:
$$
\begin{array}{l}
\Phi (\psi (z)\left |U\right >)=X(z)\Phi (\left |U\right >),
\\ \\
\Phi (\psistar (z)\left |U\right >)=X^{*}(z)
\Phi (\left |U\right >).
\end{array}
$$
}

\noindent
For the proof of the first equality we write separately the 
left and right hand sides and compare them: 
$$
\begin{array}{l}\displaystyle{
\Phi (\psi (z)\left |U\right >)=
\sum_n w^n \left <n\right | 
e^{J_+({\bf t})}\psi (z)\left |U\right > = e^{\xi ({\bf t},z)}
\sum_n w^n \left <n\right | \psi (z) e^{J_+({\bf t})}\left |U\right >},
\\ \\
\displaystyle{
X(z)\Phi (\left |U\right >)=e^{\xi ({\bf t},z)}
\sum_n z^{n-1}w^n \left <n\! -\! 1\right | 
e^{-J_+([z])}
e^{J_+({\bf t})}\left |U\right >}.
\end{array}
$$
It remains to use the bosonization rules (\ref{bf3}).
The second equality is proved in a similar way.

\subsection{Tau-functions as vacuum expectation values}

Here is the main statement connecting the theory of free fermions
constructed above with the theory of integrable hierarchies.
If $G$ is an arbitrary quasigroup element 
of the Clifford algebra (for simplicity, with zero charge),
then the vacuum expectation value
\beq\label{vac1}
\mbox{\fbox{$\displaystyle{\phantom{\int ^{A}_{B}}
\tau ({\bf t})=\lvac e^{J_+({\bf t})}G\rvac
\phantom{\int ^{A}_{B}}}$}}
\eeq
is the tau-function of the KP hierarchy and the ratios of the expectation values
\beq\label{vac2}
\mbox{\fbox{$\displaystyle{\phantom{\int ^{A}_{B}}
\begin{array}{l}
\displaystyle{
\psi ({\bf t},z)=\frac{\left < 1\right | e^{J_+({\bf t})}
\psi (z) G\rvac}{\lvac e^{J_+({\bf t})}G\rvac}},
\\ \\
\displaystyle{
\psi^{*} ({\bf t},z)=\frac{\left < -1\right | e^{J_+({\bf t})}
\psi^{*} (z) G\rvac}{z\lvac e^{J_+({\bf t})}G\rvac}}
\end{array}
\phantom{\int ^{A}_{B}}}$}}
\eeq
are the Baker-Akhiezer function and its adjoint.

First of all let us show that these formulas agree with 
(\ref{jap1}), (\ref{jap2}). Indeed, moving the operators 
$\psi$, $\psistar$ in the numerator to the left and using formulas (\ref{bf3}),
we see that the Baker-Akhiezer functions 
$\psi ({\bf t},z)$ and $\psi ^{*}({\bf t},z)$ are connected with $\tau$ by the
``Japanese'' formulas (\ref{jap1}), (\ref{jap2}). It remains to show that
the function $\tau ({\bf t})$ defined by 
(\ref{vac1}) satisfies the bilinear relation (\ref{hir1}),
i.e. 
$$
\mbox{res}_z \left [ e^{\xi ({\bf t}-{\bf t}',z)}
\lvac e^{-J_+([z^{-1}])}e^{J_+({\bf t})} G \rvac
\lvac e^{J_+([z^{-1}])}e^{J_+({\bf t}')} G \rvac \right ]=0.
$$
Using equations (\ref{bf3}), it is easy to see that it is equivalent to 
the identity (\ref{bilinear-fermi}), in which one should put
$\left < U\right | =\left < 1\right | e^{J_+({\bf t})}$,
$\left < U'\right | =\left < -1\right | e^{J_+({\bf t}')}$.

More generally, there are the following classes of tau-functions
(which generalize each other):
\begin{itemize}
\item Tau-functions of the KP hierarchy depending on the times
${\bf t}=\{t_1, t_2, \ldots \}$:
\begin{equation}\label{ferm6}
\tau ({\bf t})=\lvac e^{J_+ ({\bf t})}G\rvac .
\end{equation}
\item
Tau-functions of the modified KP hierarchy (mKP):
\begin{equation}\label{ferm6a}
\tau_n ({\bf t})=\lvacn e^{J_+ ({\bf t})}G\rvacn .
\end{equation}
The equations of the mKP hierarchy are differential-difference equations
including shifts of the variable 
$n=t_0$ (the ``zeroth time'').
At any fixed $n$ the tau-function (\ref{ferm6a})
solves the KP hierarchy.
\item
Tau-functions of the 2DTL hierarchy:
\beq\label{ferm6b}
\tau_n ({\bf t}_+ , {\bf t}_-)=
\lvacn e^{J_+ ({\bf t}_+)}Ge^{-J_- ({\bf t}_-)}\rvacn .
\eeq
This is the most general tau-function wich can be constructed 
using the one-component fermions. 
At fixed $n$, ${\bf t}_-$ this formula gives 
the tau-function of the KP hierarchy
(as a function of ${\bf t}_+$).
\end{itemize}

Consider the tau-function of the mKP hierarchy. Let us show that it satisfies a
bilinear relation which is a direct consequence of the BBR in the form
(\ref{bilinear-fermi}).
Set $\left|V \rbr =\rvacn$,
$\left|V^{\prime} \rbr =\left |n'\right >$ with $n\geq n'$,
where $\rvacn$ and $\left |n'\right >$ are two shifted Dirac vacua. 
We have
\begin{equation}\label{B1}
\sum_{k \in {\z}} \lbr U \right|  \psi_{k} G \left|n \rbr
\lbr U^{\prime} \right|  \psistar_{k} G \left|n' \rbr =0
\end{equation}
since either $\psi_k$ or $\psistar_k$ annihilates the right vacuum in each term
of the right hand side of
(\ref{bilinear-fermi}).
Now set 
$\lbr U \right| = \lbr n+1 \right | e^{J_{+}({\bf t})}$,
$\lbr U' \right| = \lbr n'-1 \right | e^{J_{+}({\bf t'})}$
and write
$$
\begin{array}{lll}
0&=&\displaystyle{\sum_k
\lbr n+1 \right | e^{J_{+}({\bf t})} \psi_k G \rvacn
\lbr n'-1 \right | e^{J_{+}({\bf t'})}\psistar_k G\left |n'\right >}
\\ &&\\
&=& \displaystyle{
\mbox{res}_{z}\left [
z^{-1} \lbr n+1 \right | e^{J_{+}({\bf t})} \psi (z) G \rvacn
\lbr n'-1 \right | e^{J_{+}({\bf t'})}\psistar (z)
G\left |n'\right >\right ]}
\\ &&\\
&=& \displaystyle{
\mbox{res}_{z}\left [
e^{\xi ({\bf t}-{\bf t'},z)}z^{n-n'}
\lbr n \right | e^{J_+({\bf t}-[z^{-1}])}G\rvacn
\lbr n' \right | e^{J_+({\bf t'}+[z^{-1}])}G\left |n' \rbr \right ],}
\end{array}
$$
where we have used the commutation relations of the Fermi operators with
$e^{J_+({\bf t})}$ and the bosonization rules
(\ref{bf3}). 
(As before, $\mbox{res}_{z}$ means the coefficient in front of 
$z^{-1}$ in the Laurent series.)
We thus obtain that the identity
\begin{equation}\label{bi1}
\oint_{{\sf C}} z^{n-n'} e^{\xi ({\bf t}-{\bf t'},z)}
\tau_n ({\bf t}-[z^{-1}])\tau_{n'}({\bf t'}+[z^{-1}])dz =0
\end{equation}
holds for all ${\bf t}, {\bf t'}$
and $n\geq n'$. 
At $n=n'$ (\ref{bi1}) turns into the bilinear relation for the tau-function of the 
KP hierarchy:
\begin{equation}\label{bi1KP}
\oint_{{\sf C}} e^{\xi ({\bf t}-{\bf t'},z)}
\tau ({\bf t}-[z^{-1}])\tau ({\bf t'}+[z^{-1}])dz =0.
\end{equation}

\noindent
At $n=n'+1$ (\ref{bi1}) gives the bilinear relation
for the tau-function of the mKP hierarchy:
\begin{equation}\label{bi1MKP}
\oint_{{\sf C}} ze^{\xi ({\bf t}-{\bf t'},z)}
\tau_{n+1} ({\bf t}-[z^{-1}])\tau _n ({\bf t'}+[z^{-1}])dz =0.
\end{equation}

\noindent
{\bf Example.} The $N$-soliton tau-function (\ref{tl21}) is obtained in the fermionic
formalism as follows. Introduce the fermionic operators
$\Psi_i(p_i, q_i)=\psi (q_i)+b_i \psi (p_i)$. They are group-like elements with charge
$+1$. Therefore, 
$$
\tau_n({\bf t}_+, {\bf t}_-)=\lvacn e^{J_+({\bf t}_+)}\Psi_1(p_1, q_1)\ldots
\Psi_N(p_N, q_N)e^{-J_-({\bf t}_-)}\left |n\! -\! N\right >
$$
is the tau-function of the 2DTL hierarchy. It coincides with (\ref{tl21}).

\noindent
{\bf Problem.} Show that 
$$
G=\normordbare \exp \left (\sum_{k=1}^N a_k \psistar (q_k)\psi (p_k)\right )\normordbare
$$
in (\ref{ferm6b}) leads to the $N$-soliton tau-function in the form (\ref{tl22}).

\section*{Acknowledgments}
\addcontentsline{toc}{section}{\hspace{6mm}Acknowledgments}

The author thanks A.Alexandrov, V.Akhmedova, A.Anokhina, I.Krichever, A.Orlov and T.Takebe
for discussions and help. This work was funded by the Russian Academic Excellence
Project `5-100' and supported in part by RFBR grant
18-01-00461.

\end{document}